\documentclass[12pt,a4paper]{article}
\usepackage{amssymb,amsmath,bm}
\usepackage{graphicx}

\def\simge{\mathrel{\rlap{\raise 0.511ex \hbox{$>$}}{\lower 0.511ex \hbox{$\sim$}}}}
\def\simle{\mathrel{\rlap{\raise 0.511ex \hbox{$<$}}{\lower 0.511ex \hbox{$\sim$}}}}
\def\slash#1{\setbox0=\hbox{$#1$}\dimen0=\wd0
     \setbox1=\hbox{/} \dimen1=\wd1 \ifdim\dimen0>\dimen1
     \rlap{\hbox to \dimen0{\hfil/\hfil}} #1                        \else
     \rlap{\hbox to \dimen1{\hfil$#1$\hfil}}
     /   \fi}

\def\slash#1{\mbox{$\not \!\! #1$}}
\def\Dslash{{\slash {\cal D}}}

\def\lvec#1{\setbox0=\hbox{$#1$}
    \setbox1=\hbox{$\scriptstyle\leftarrow$}
    #1\kern-\wd0\smash{
    \raise\ht0\hbox{$\raise1pt\hbox{$\scriptstyle\leftarrow$}$}}
    \kern-\wd1\kern\wd0}
\def\rvec#1{\setbox0=\hbox{$#1$}
    \setbox1=\hbox{$\scriptstyle\rightarrow$}
    #1\kern-\wd0\smash{
    \raise\ht0\hbox{$\raise1pt\hbox{$\scriptstyle\rightarrow$}$}}
    \kern-\wd1\kern\wd0}

\def\diracstar#1#2{
    \setbox0=\hbox{$\gamma$}\setbox1=\hbox{$\gamma_{#1}$}
    \gamma_{#1}\kern-\wd1\kern\wd0
    \smash{\raise4.5pt\hbox{$\scriptstyle#2$}}}

\newcommand{\pslash}{\not\hspace{-3pt}p}

\newcommand{\lsim}{
\mathrel{\hbox{\rlap{\hbox{\lower4pt\hbox{$\sim$}}}\hbox{$<$}}}}

\newcommand{\gsim}{
\mathrel{\hbox{\rlap{\hbox{\lower4pt\hbox{$\sim$}}}\hbox{$>$}}}}

\newcommand{\gev}{\, {\rm GeV}}
\newcommand{\mev}{\, {\rm MeV}}

\newcommand{\be}{\begin{equation}}
\newcommand{\ee}{\end{equation}}
\newcommand{\bea}{\begin{eqnarray}}
\newcommand{\eea}{\end{eqnarray}}
\newcommand{\nn}{\nonumber}

\begin{document}

\rightline{\small Preprint MITP/14-020, ROM2F/2014/01, RM3-TH/14-4}
\vspace{1.5cm}

\centerline{\huge Up, down, strange and charm quark masses} 
\vspace{0.5cm}
\centerline{\huge with $N_f = 2+1+1$ twisted mass lattice QCD}

\vspace{1.5cm}

\centerline{\large N. Carrasco$^{(a)}$, A. Deuzeman$^{(b,*)}$, P. Dimopoulos$^{(c,d)}$, R. Frezzotti$^{(d,e)}$,}  
\vspace{0.25cm}
\centerline{\large V. Gim\'enez$^{(f)}$, G. Herdoiza$^{(g)}$, P. Lami$^{(h,a)}$, V. Lubicz$^{(h,a)}$, D. Palao$^{(i)}$,}
\vspace{0.25cm}
\centerline{\large  E. Picca$^{(h,a)}$, S. Reker$^{(l,*)}$, L. Riggio$^{(h,a)}$, G.C. Rossi$^{(d,e)}$, F. Sanfilippo$^{(m)}$,}
\vspace{0.25cm}
\centerline{\large L. Scorzato$^{(n)}$, S. Simula$^{(a)}$, C. Tarantino$^{(h,a)}$, C. Urbach$^{(p)}$, U. Wenger$^{(b)}$}

\vspace{1cm}

\centerline{\includegraphics[draft=false]{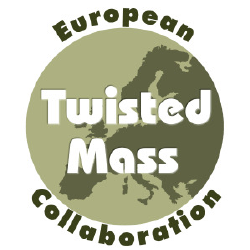}}

\vspace{1cm}

\centerline{\it $^{(a)}$ INFN, Sezione di Roma Tre}
\centerline{\it Via della Vasca Navale 84, I-00146 Rome, Italy}
\vskip 2 true mm
\centerline{\it $^{(b)}$Albert Einstein Center for Fundamental Physics, Institute for Theoretical Physics,}
\centerline{\it  University of Bern, Sidlerstrasse 5, CH-3012 Bern, Switzerland}
\vskip 2 true mm
\centerline{\it $^{(c)}$ Centro Fermi - Museo Storico della Fisica e Centro Studi e Ricerche Enrico Fermi}
\centerline{\it Compendio del Viminale, Piazza del Viminale 1 I–00184 Rome, Italy }
\vskip 2 true mm
\centerline{\it $^{(d)}$ Dipartimento di Fisica, Universit\`a di Roma ``Tor Vergata''}
\centerline{\it Via della Ricerca Scientifica 1, I-00133 Rome, Italy}
\vskip 2 true mm
\centerline{\it $^{(e)}$ INFN, Sezione di ``Tor Vergata"}
\centerline{\it Via della Ricerca Scientifica 1, I-00133 Rome, Italy}
\vskip 2 true mm
\centerline{\it $^{(f)}$  Departament de F\'{\i}sica Te\`orica and IFIC, Univ. de Val\`encia-CSIC}
\centerline{\it Dr.~Moliner 50, E-46100 Val\`encia, Spain,}
\vskip 2 true mm
\centerline{\it $^{(g)}$ PRISMA Cluster of Excellence, Institut f{\"u}r Kernphysik}
\centerline{\it Johannes Gutenberg-Universit{\"a}t, D-55099 Mainz, Germany}
\vskip 2 true mm
\centerline{\it $^{(h)}$ Dipartimento di Matematica e Fisica, Universit\`a  Roma Tre}
\centerline{\it Via della Vasca Navale 84, I-00146 Rome, Italy}
\vskip 2 true mm
\centerline{\it $^{(i)}$ Goethe-Universit\"at, Institut f\"ur Theoretische Physik}
\centerline{\it Max-von-Laue-Stra{\ss}e 1, D-60438 Frankfurt am Main, Germany}
\vskip 2 true mm
\centerline{\it $^{(l)}$ Centre for Theoretical Physics, University of Groningen}
\centerline{\it Nijenborgh 4, 9747 AG Groningen, the Netherlands}
\vskip 2 true mm
\centerline{\it $^{(m)}$ School of Physics and Astronomy, University of Southampton,}
\centerline{\it SO17 1BJ Southampton, United Kingdom}
\vskip 2 true mm
\centerline{\it $^{(n)}$ INFN-TIFPA, Trento Institute for Fundamental Physics and Application,}
\centerline{\it Via Sommarive 14, I-38123 Trento, Italy}
\vskip 2 true mm
\centerline{\it $^{(p)}$ Helmholtz-Institut f{\"u}r Strahlen- und Kernphysik (Theorie) and}
\centerline{\it Bethe Center for Theoretical Physics, Universit{\"a}t Bonn, 53115 Bonn, Germany}
\vskip 2 true mm
\centerline{\it $^{(*)}$ Current address: Shell Global Solutions International}
\centerline{\it Kessler Park 1, 2288 GS Rijswijk, The Netherlands}

\vspace{2cm}

\begin{abstract}
We present a lattice QCD calculation of the up, down, strange and charm quark masses performed using the gauge configurations produced by the European Twisted Mass Collaboration with $N_f = 2 + 1 + 1$ dynamical quarks, which include in the sea, besides two light mass degenerate quarks, also the strange and charm quarks with masses close to their physical values. 
The simulations are based on a unitary setup for the two light quarks and on a mixed action approach for the strange and charm quarks.
The analysis uses data at three values of the lattice spacing and pion masses in the range $210 \div 450$ MeV, allowing for accurate continuum limit and controlled chiral extrapolation.
The quark mass renormalization is carried out non-perturbatively using the RI$^\prime$-MOM method. The results for the quark masses converted to the $\overline{\rm MS}$ scheme are: $m_{ud}(2~\gev) = 3.70 (17) \mev$, $m_s(2~\gev) = 99.6 (4.3) \mev$ and $m_c(m_c) = 1.348 (46) \gev$.
We obtain also the quark mass ratios $m_s / m_{ud} = 26.66 (32)$ and $m_c / m_s = 11.62 (16)$.
By studying the mass splitting between the neutral and charged kaons and using available lattice results for the electromagnetic contributions, we evaluate $m_u / m_d = 0.470 (56)$, leading to $m_u = 2.36 (24)$ MeV and $m_d = 5.03 (26)$ MeV.
\end{abstract}

\newpage

\section{Introduction}
\label{sec:intro}

The precise knowledge of the quark masses and of the hadronic parameters in general plays a fundamental role both in testing the Standard Model (SM) and in the search for new physics (NP). 
Despite its unquestionable successes in describing experimental data the SM does not provide any explanation for the quark masses. 
On the theoretical side, the understanding of the hierarchical pattern of the quark masses remains an open and fascinating challenge.
On the phenomenological side, since several important observables depend on the quark masses, a precise determination of these values is crucial to constrain the SM and through comparisons between theory and experiments to search for NP.

In the determination of the quark masses lattice QCD (LQCD) plays a primary role as it is a non-perturbative approach based on first principles. It consists in simulating QCD by formulating the Lagrangian on a discrete and finite Euclidean space-time which allows for a numerical computation of the path integral via MonteCarlo methods. The finite volume, the lattice spacing and generally the lower bound on the simulated light quark masses, which are limited by the currently available computing power, introduce errors which have to be well under control and accounted for.

Thanks to the increased computational power as well as to the algorithm and action improvements of the last decade, LQCD simulations have made significant progresses reaching a remarkable level of precision. In particular, this is due to the so-called unquenched calculations, where the contribution of loops of dynamical sea quarks is taken into account. As a matter of fact, most of the recent lattice determinations of quark masses have been performed with either two (up and down) \cite{Blossier:2010cr, Fritzsch:2012wq} or three (up, down and strange) \cite{Bazavov:2009bb}-\cite{Blum:2010ym} dynamical sea quarks.

In this paper we present an accurate determination of the up, down, strange and charm quark masses using the gauge configurations produced by the European Twisted Mass (ETM) Collaboration with four flavors of dynamical quarks ($N_f = 2+1+1$), which include in the sea, besides two light mass degenerate quarks, also the strange and charm quarks with masses close to their physical values. 
Such a setup is the closest one to the real world, adopted till now only by the ETM \cite{Baron:2010bv, Baron:2010th, Baron:2011sf, Ottnad:2012fv} and the MILC \cite{Bazavov:2010ru} Collaborations.

The simulations have been carried out at three different values of the inverse bare lattice coupling $\beta$, namely $\beta = 1.90, ~ 1.95 $ and $2.10$, to allow for a controlled extrapolation to the continuum limit. 
For $\beta = 1.90$ and $\beta = 1.95$ two different lattice volumes have been considered. We also used non-perturbative renormalization constants evaluated in the RI$^\prime$-MOM scheme, whose calculation is discussed in \ref{sec:RCs}.
The fermions were simulated using the Wilson Twisted Mass Action \cite{Frezzotti:2000nk, Frezzotti:2003xj} which, at maximal twist, allows for automatic ${\cal{O}}(a)$-improvement \cite{Frezzotti:2003ni, Frezzotti:2004wz}.
In order to avoid the mixing in the strange and charm sectors we adopted the non-unitary set up described in Ref.~\cite{Frezzotti:2004wz}, in which the strange and charm valence quarks are regularized as Osterwalder-Seiler (OS) fermions \cite{Osterwalder:1977pc}.
For the links the Iwasaki action \cite{Iwasaki} was adopted, because it proved to relieve simulations with light quark masses allowing to bring the simulated pion mass down to approximately $210 \mev$.

Since simulations were not performed at the physical point for the up and down quark masses, a chiral extrapolation is needed. In order to estimate the associated systematic error we studied the dependence on the light quark mass by using different fit formulae based on the predictions of Chiral Perturbation Theory (ChPT) as well as on polynomial expressions. 

To account for finite size effects (FSE) we used the resummed asymptotic formulae developed in Ref.~\cite{Colangelo:2010cu} for the pion sector, which include the effects due to the neutral and charged pion mass splitting (present in the twisted mass formulation), and the formulae of Ref.~\cite{CDH05} for the kaon sector.
We checked the accuracy of these predictions for FSE on the lattice data obtained at fixed quark masses and lattice spacings, but different lattice volumes.

As for the continuum limit, in order to lower the impact of discretization effects as much as possible and to keep the continuum extrapolation under control we tried two different procedures, which both use $f_\pi$ to set the scale.
The first one involves the Sommer parameter $r_0$ \cite{Sommer:1993ce} in units of the lattice spacing $a$, i.e.~$r_0/a$, as the intermediate scaling variable, while in the second one we used the mass of a fictitious pseudoscalar (PS) meson made of two strange-like quarks (or a strange-like and a charm-like quark), $a M_{s^\prime s^\prime}$ (or $a M_{c^\prime s^\prime}$), trying to exploit cancellation of discretization effects in ratios like $M_K / M_{s^\prime s^\prime}$ (or $M_{D_s} / M_{c^\prime s^\prime}$).
In particular for the kaon and $D_s (D)$ meson masses these ratios lead to a significant reduction of discretization effects. 
Of course, in order to determine the lattice scale, the continuum limit of $M_{s^\prime s^\prime}$ (or $M_{c^\prime s^\prime}$) has to be performed eventually.
The fact that we obtain compatible predictions from the two procedures strengthens the validity of our results and shows that the impact of the discretization effects is safely kept under control.

As described in \ref{sec:RCs}, by using dedicated ensembles of gauge configurations produced with $N_f = 4$ degenerate flavors of sea quarks \cite{ETM:2011aa}, we computed the quark mass renormalization constants (RCs) $Z_\mu = 1 / Z_P$ in the RI$^\prime$-MOM scheme using two different methods, labelled as M1 and M2.
The first method (M1) aims at removing ${\cal{O}}(a^2p^2)$ effects, while in the second method (M2) the renormalization constants are taken at a fixed reference value of $p^2$.
The use of the two sets of renormalization constants is expected to lead to the same final results once the continuum limit for the physical quantity of interest is performed.

Summarizing, our analysis has followed eight branches differing in the choice of the scaling variable (either $r_0/a$ or $a M_{s^\prime s^\prime}$), the fitting procedures (either ChPT or polynomial expansion) and the method (either M1 or M2) used to determine the values of the RCs $Z_P$.

First we calculated the up/down average quark mass from the analysis of the pion mass and decay constant.
Then, using either $r_0$ or $M_{s^\prime s^\prime}$ ($M_{c^\prime s^\prime}$) as well as the lattice spacing and the light quark mass determined from the pion sector, we extracted the strange and charm quark masses from the analysis of $K$- and $D$-meson correlators, respectively.
The differences among the results obtained within the various branches of the analysis have been used to estimate the systematic uncertainties.
  
The final results obtained for the quark masses in the $\overline{\rm MS}$ scheme are:
 \bea
    m_{ud} (2~\gev) & = & 3.70 ~ (17) \mev ~ , \nn \\
    m_s (2~\gev) & = & 99.6 ~ (4.3) \mev ~ , \nn \\
    m_c (m_c ) & = & 1.348 ~ (46) \gev ~ ,
    \label{eq:qmasses}
 \eea
where the errors are the sum in quadrature of the statistical and systematic uncertainties.

By studying the light-quark mass dependence of the squared kaon mass we calculated also the leading strong isospin breaking (IB) effect on the charged and neutral kaon masses, $\hat{M}_{K^0}$ and $\hat{M}_{K^+}$, which occurs in the pure QCD sector of the SM due to the quark mass difference $(m_d - m_u)$.
Adopting the recent FLAG estimate $\hat{M}_{K^+} - \hat{M}_{K^0} = - 6.1 (4) \mev$ \cite{FLAG}, based on the results for the electromagnetic self-energies in neutral and charged PS mesons obtained in Refs.~\cite{Basak:2008na, Basak:2012zx, Basak:2013iw, Portelli:2010yn, Portelli:2012pn, deDivitiis:2013xla}, we find
  \be
    \frac{m_u}{m_d} = 0.470 ~ (56) ~ ,
    \label{eq:mudratio}
 \ee
which is independent of both the renormalization scheme and scale.
Combining Eqs.~(\ref{eq:qmasses}-\ref{eq:mudratio}) we obtain the following predictions for the up and down quark masses:
 \bea
    m_u  (2~\gev) & = & 2.36 ~ (24) \mev ~ ,\nn \\
    m_d  (2~\gev) & = & 5.03 ~ (26) \mev ~ .
    \label{eq:mu&md}
 \eea

Finally, by introducing suitable ratios of meson masses (see Sections \ref{sec:ratio_sl} and \ref{sec:ratio_cs}) we determined the quark mass ratios $m_s / m_{ud}$ and $m_c / m_s$, obtaining
 \bea
    \frac{m_s}{m_{ud}} & = & 26.66 ~ (32) ~ ,\nn \\[2mm]
    \frac{m_c}{m_s} & = & 11.62 ~ (16) ~ ,
    \label{eq:mratios}
 \eea
which are independent of both the renormalization scheme and scale.
We also quote our results for the ratios
 \bea
     R & \equiv & \frac{m_s - m_{ud}}{m_d - m_u} = 35.6 ~ (5.1) ~ , \nn \\[2mm]
     Q & \equiv & \sqrt{\frac{m_s^2 - m_{ud}^2}{m_d^2 - m_u^2}} =  22.2 ~ (1.6) ~ ,
     \label{eq:R&Q}
 \eea
which provide information on the relative size of SU(3) and SU(2) symmetry breaking effects.

\section{Simulation details}
\label{sec:simulations}

The present work is based on the $N_f = 2+1+1$ gauge field configurations generated by the ETMC \cite{Baron:2010bv, Baron:2011sf}  using the following action
 \be
    S = S_g + S_{tm}^\ell  + S_{tm}^h ~ ,
    \label{eq:totalaction}
 \ee
where the gluon action $S_g$ is the Iwasaki one \cite{Iwasaki}. 
For the fermions we have adopted the Wilson twisted-mass action, given explicitly for the mass-degenerate up/down quark doublet by \cite{Frezzotti:2000nk}
 \be
    S_{tm}^\ell  = a^4 \sum\nolimits_x {\overline{\psi}(x) \left\{ \frac{1}{2} \gamma_\mu (\nabla_\mu + \nabla^*_\mu)  - i \gamma _5 \tau^3 
                           \left[ m_0 - \frac{a}{2} \nabla_\mu \nabla^*_\mu \right] + \mu_\ell \right\} \psi (x)} 
    \label{eq:action_ud}
 \ee
and for the strange and charm doublet by \cite{Frezzotti:2003xj}
 \be
    S_{tm}^h  = a^4 \sum\nolimits_x {\overline{\psi}(x)\left\{ \frac{1}{2} \gamma_\mu  (\nabla_\mu   + \nabla^*_\mu  ) - i \gamma _5 \tau ^1 
                        \left[ m_0  - \frac{a}{2} \nabla_\mu \nabla^*_\mu \right] + \mu_\sigma + \mu_\delta  \tau^3 \right\} \psi (x)},
    \label{eq:action_sc}
 \ee
where $\nabla_\mu$ and $\nabla^*_\mu$ are nearest-neighbor forward and backward covariant derivatives, $\mu_\ell$ is the light quark mass and $m_0$ is the ``untwisted" mass.
The latter is tuned to its critical value $m_{cr}$ as discussed in Ref.~\cite{Baron:2010bv} in order to guarantee the automatic ${\cal{O}}(a) $-improvement at maximal twist \cite{Frezzotti:2003ni, Frezzotti:2004wz}.
Finally in Eq.~(\ref{eq:action_sc}) the twisted masses $\mu_\sigma$ and $\mu_\delta$ are related to the renormalized strange and charm sea quark masses via the relation \cite{Frezzotti:2004wz}
 \be
       m_{c,s}^{sea}  = \frac{1}{Z_P} \left( \mu_\sigma  \pm \frac{Z_P}{Z_S} \mu_\delta \right) 
       \label{eq:mcs}
 \ee
with $Z_P$ and $Z_S$ being the pseudoscalar and scalar renormalization constants, respectively.

The twisted-mass action (\ref{eq:totalaction}) leads to a mixing in the strange and charm sectors \cite{Frezzotti:2003xj, Baron:2010th}. 
In order to avoid the mixing of $K$- and $D$-meson states in the correlation functions, we adopted a non-unitary set up \cite{Frezzotti:2004wz} in which the strange and charm valence quarks are regularized as Osterwalder-Seiler (OS) fermions \cite{Osterwalder:1977pc}.
Thus, while we keep the light sector unitary, the action in the strange and charm sectors ($f = s, ~ c$) reads as
 \be
    S^f_{OS}  = a^4 \sum\nolimits_x {\overline{q}_f (x) \left\{ \frac{1}{2} \gamma_\mu  (\nabla_\mu + \nabla^*_\mu  ) - i \gamma _5 r_f 
                        \left[ m_0  - \frac{a}{2} \nabla_\mu \nabla^*_\mu \right] + \mu _f \right\} q_f (x)} ~ ,
    \label{eq:OS}
 \ee
where $r_f = \pm 1$.
When constructing meson correlation functions (including the pion) the Wilson parameters of the two valence quarks are always chosen to have opposite values.
This choice guarantees that the squared PS meson mass, $M_{PS}^2$, differs from its continuum counterpart only by terms of ${\cal{O}}(a^2 \mu)$ \cite{Frezzotti:2003ni, Frezzotti:2005gi}.

The details of our lattice set up are collected in Table \ref{tab:masses&simudetails}, where the number of gauge configurations analyzed ($N_{cfg}$) corresponds to a separation of $20$ trajectories.
At each lattice spacing, different values of the light sea quark masses have been considered. 
The light valence and sea quark masses are always taken to be degenerate. 
The masses of both the strange and the charm sea quarks are fixed, at each $\beta$, to values close to the physical ones \cite{Baron:2010bv}.
We have simulated three values of the valence strange quark mass and six values of the valence heavy quark mass, which are needed for the interpolation in the physical charm region as well as to extrapolate to the $b$-quark sector for future studies. 
In particular, for the light sector the quark masses were simulated in the range $3 ~ m_\ell^{phys} \lesssim  \mu_\ell \lesssim 12 ~ m_\ell^{phys} $, for the strange sector in $0.7 ~ m_s^{phys} \lesssim  \mu_s \lesssim 1.2 ~ m_s^{phys}$, while for the charm sector in $0.7 ~ m_c^{phys} \lesssim  \mu_c \lesssim 2.5 ~ m_c^{phys} $.
Quark propagators with different valence masses are obtained using the so-called multiple mass solver method \cite{Jegerlehner:1996pm, Jansen:2005kk}, which allows to invert the Dirac operator for several quark masses at a relatively low computational cost.   

\begin{table}[hbt!]
\begin{center}
\footnotesize
\renewcommand{\arraystretch}{1.20}
\begin{tabular}{||c|c|c|c|c|c|c|c|c||}
\hline
ensemble & $\beta$ & $V / a^4$ &$a\mu_{sea}=a\mu_\ell$&$a\mu_\sigma$&$a\mu_\delta$&$N_{cfg}$& $a\mu_s$& $a\mu_c$ \\
\hline
$A30.32$ & $1.90$ & $32^{3}\times 64$ &$0.0030$ &$0.15$ &$0.19$ &$150$&  $0.0145,$& $0.1800, 0.2200,$ \\
$A40.32$ & & & $0.0040$ & & & $90$ &$0.0185,$ &0.2600, 0.3000, \\
$A50.32$ & & & $0.0050$ & & &  $150$ &$ 0.0225$ &  0.3600, 0.4400 \\
\cline{1-7} 
$A40.24$ & $1.90$ & $24^{3}\times 48 $ & $0.0040$  &$0.15$ & $0.19$& $150$ & & \\
$A60.24$ & & & $0.0060$ & & &  $150$ &&  \\
$A80.24$ & & & $0.0080$ & & &  $150$ & &  \\
$A100.24$ &   & & $0.0100$ & & &  $150$ & &  \\
\hline
$B25.32$ & $1.95$ & $32^{3}\times 64$ &$0.0025$&$0.135$ &$0.170$& $150$& $0.0141,$& $0.1750, 0.2140,$ \\
$B35.32$ & & & $0.0035$  & & & $150$ &$  0.0180,$ &0.2530, 0.2920, \\
$B55.32$ & & & $0.0055$ & & & $150$ &$ 0.0219$&  0.3510, 0.4290 \\
$B75.32$ &  & & $0.0075$ & & & $75$ &&   \\
\cline{1-7}
$B85.24$ & $1.95$ & $24^{3}\times 48 $ & $0.0085$ &$0.135$ &$0.170$ & $150$ & & \\
\hline
$D15.48$ & $2.10$ & $48^{3}\times 96$ &$0.0015$&$0.12$ &$0.1385 $& $60$& $0.0118,$& $0.1470, 0.1795, $ \\ 
$D20.48$ & & & $0.0020$  &  &  & $90$ &$0.0151,$ &0.2120, 0.2450, \\
$D30.48$ & & & $0.0030$ & & & $90$ & $  0.0184$&  0.2945, 0.3595 \\
 \hline   
\end{tabular}
\renewcommand{\arraystretch}{1.0}
\end{center}
\caption{\it Values of the simulated sea and valence quark bare masses for each ensemble used in this work.}
\label{tab:masses&simudetails}
\end{table}

We studied the dependence of the PS meson masses and of the pion decay constant on the renormalized light quark mass fitting simultaneously the data at different lattice spacings and volumes. 
In particular, we anticipate that the values of the lattice spacing found in our pion analysis are $a = 0.0885(36), ~ 0.0815(30),  ~ 0.0619(18)$ fm at $\beta = 1.90, ~ 1.95$ and $2.10$, respectively, so that the lattice volume goes from $\simeq 2$ to $\simeq 3$ fm. 
In Table \ref{tab:MpiL} we provide for each ensemble the central values of the pion mass (covering the range $\simeq 210 \div 450 \mev$), of the lattice size $L$ and of the product $M_{\pi}L$.

\begin{table}[hbt!]
\begin{center}
\renewcommand{\arraystretch}{1.20}
\begin{tabular}{||c||c|c|c|c||}
\hline
ensemble & $\beta$ & $L$(fm) & $M_\pi$(MeV)& $M_\pi L$ \\
\hline 
$A30.32$ & $1.90$&$ 2.84$&$245$&$3.53$\\
$A40.32$ &            &            &$282$&$4.06$\\
$A50.32$ &            &            &$314$&$4.53$\\
 \hline
$A40.24$ & $1.90$&$ 2.13$&$282$&$3.05$\\
$A60.24$ &            &            &$344$&$3.71$\\
$A80.24$ &            &            &$396$&$4.27$\\
$A100.24$ &          &            &$443$&$4.78$\\
 \hline
$B25.32$ & $1.95$&$ 2.61$&$239$&$3.16$\\
$B35.32$ &            &            &$281$&$3.72$\\
$B55.32$ &            &            &$350$&$4.64$\\
$B75.32$ &            &            &$408$&$5.41$\\
 \hline
$B85.24$ & $1.95$&$ 1.96$&$435$&$4.32$\\
 \hline
$D15.48$ & $2.10$&$ 2.97$&$211$&$3.19$\\
$D20.48$ &            &            &$243$&$3.66$\\
$D30.48$ &            &            &$296$&$4.46$\\
 \hline
\end{tabular}
\renewcommand{\arraystretch}{1.20}
\end{center}
\caption{\it Central values of the pion mass $M_\pi$, of the lattice size $L$ and of the product $M_\pi L$ for the various ensembles used in this work. The values of $M_\pi$ are extrapolated to the continuum and infinite volume limits, according to the ChPT fit (\ref{eq:cptmpi2Ch}), described in Section \ref{sec:r0}.}
\label{tab:MpiL}
\end{table}
The statistical accuracy of the meson correlators is significantly improved by using the so-called ``one-end" stochastic method \cite{McNeile:2006bz}, which includes spatial stochastic sources at a single time slice chosen randomly.
Statistical errors on the meson masses are evaluated using the jackknife procedure, while statistical errors based on data obtained from independent ensembles of gauge configurations, like the errors of the fitting procedures, are evaluated using a bootstrap sampling with ${\cal{O}}(100)$ events to take properly into account cross-correlations.

In Table \ref{tab:zpr0a} we present the values of the RCs $Z_P$ corresponding to the two methods M1 and M2, described in Section \ref{sec:intro} (see also \ref{sec:RCs}), and the values of $r_0/a$ used to convert the data at different values of lattice spacing to the common scale given by the Sommer parameter $r_0$. 
For each $\beta$ the values of $r_0/a$ have been calculated at the various values of the light quark mass \cite{Baron:2010bv, Baron:2011sf} and then extrapolated to the chiral limit, assuming either a linear or a quadratic dependence in $a \mu_{sea}$. 
Our results for $r_0/a$ are consistent within the errors with the findings of Refs.~\cite{Ottnad:2012fv,Herdoiza:2013sla}, where the extrapolation to the chiral limit was performed using only a linear dependence on $a \mu_{sea}$.
The errors reported in Table \ref{tab:zpr0a} represent the sum in quadrature of the statistical uncertainty and of the systematic error associated to the two different chiral extrapolations. 

\begin{table}[hbt!]
\begin{center}
\renewcommand{\arraystretch}{1.20}
\begin{tabular}{||c|c|c|c||}
\hline
$\beta$ & $Z_P^{\overline{\rm MS}}(2\,\mathrm{GeV})(M_1)$ & $Z_P^{\overline{\rm MS}}(2\,\mathrm{GeV})(M_2)$ & $r_{0}/a$ \\
\hline
$1.90$ & $0.529(7)$ & $0.574(4)$ & $5.31(8)$ \\
$1.95$ & $0.509(4)$ & $0.546(2)$ & $5.77(6)$ \\
$2.10$ & $0.516(2)$ & $0.545(2)$ & $7.60(8)$ \\
\hline
\end{tabular}
\renewcommand{\arraystretch}{1.0}
\end{center}
\caption{\it Input values for the renormalization constant $Z_P^{\overline{\rm MS}}(2 \gev)$, corresponding to the methods M1 and M2 (see \ref{sec:RCs}), and the chirally extrapolated values of $r_0 / a$ for each value of $\beta$ (see text).}
\label{tab:zpr0a}
\end{table}

Since the renormalization constants $Z_P$ and the values of $r_0/a$ have been evaluated using different ensembles of gauge configurations, their uncertainties have been taken into account in the fitting procedures as follows.
First we generated randomly a set of values of $\left( r_0 /a \right)_i$ and $\left( Z_P \right)_i$ for the bootstrap event $i$ assuming gaussian distributions corresponding to the central values and the standard deviations given in Table \ref{tab:zpr0a}.
Then we added in the definition of the $\chi^2$ the following contribution
 \be
    \sum\nolimits_\beta \frac{\left[ \left( r_0 /a \right)^{fit} _i  - \left( r_0 /a \right)_i \right]^2}{\sigma^2_{r_0 /a}}  + 
    \sum\nolimits_\beta \frac{\left[ \left( Z_P \right)^{fit} _i  - \left( Z_P \right)_i \right]^2}{\sigma^2_{Z_P}} ~ ,
    \label{eq:chi2term}
 \ee
where $\left( r_0 /a \right)^{fit} _i$ and $\left( Z_P \right)^{fit} _i$ are free parameters of the fitting procedure for the bootstrap event $i$. 
The use of Eq.~(\ref{eq:chi2term}) allows the quantities $r_0 / a$ and $Z_P$ to slightly change from their central values (in the given bootstrap event) with a weight in the $\chi^2$ given by their uncertainties.
This procedure corresponds to impose a gaussian prior for $Z_P$ and $r_0 / a$.

Before closing this section we have collected in Table \ref{tab:timeint} the time intervals (conservatively) adopted for the extraction of the PS meson masses (and of the pion decay constant) from the 2-point correlators at each $\beta$ and lattice volume in the light, strange and charm sectors.

\begin{table}[hbt!]
\begin{center}
\renewcommand{\arraystretch}{1.20}
\begin{tabular}{||c|c|c|c||}
\hline
$\beta$ & $V/a^4$ & $[t_{min}, t_{max}]_{(\ell \ell, \ell s)} / a$ & $[t_{min}, t_{max}]_{(\ell c, sc)}/a$ \\
\hline
$1.90$ & $24^3 \times 48$ & $[12,23]$ & $[15,21]$ \\
$1.90$ & $32^3 \times 64$ & $[12,31]$ & $[15,29]$ \\ \hline
$1.95$ & $24^3 \times 48$ & $[13,23]$ & $[16,21]$ \\
$1.95$ & $32^3 \times 64$ & $[13,31]$ & $[16,29]$ \\ \hline
$2.10$ & $48^3 \times 96$ & $[18,40]$ & $[20,40]$ \\
\hline
\end{tabular}
\renewcommand{\arraystretch}{1.0}
\end{center}
\caption{\it Time intervals $[t_{min}, t_{max}] / a$ adopted for the extraction of the PS meson masses (and of the pion decay constant) from the 2-point correlators in the light ($\ell$), strange ($s$) and charm ($c$) sectors.} 
\label{tab:timeint}
\end{table}

\section{Average up and down quark mass}
\label{sec:pionanalysis}

For each ensemble we computed the 2-point PS correlators defined as
 \be
    C(t) = \frac{1}{L^3} \sum\limits_{\vec{x}, \vec{z}} \left\langle 0 \right| P_5 (x) P_5^\dag (z) \left| 0 \right\rangle \delta_{t, (t_x  - t_z )} ~ ,
    \label{eq:P5}
 \ee
where $P_5 (x) = \overline{u}(x) \gamma_5 d(x)$\footnote{We remind that the Wilson parameters of the two valence quarks in any PS meson considered in this work are always chosen to have opposite values.}.
As it is well known at large time distances one has
 \be
    C(t)_{ ~ \overrightarrow{t  \gg a, ~ (T - t) \gg a} ~ } \frac{\mathcal{Z}_\pi}{2M_\pi} \left( e^{ - M_\pi  t}  + e^{ - M_\pi  (T - t)} \right) ~ ,
    \label{eq:larget}
 \ee
so that the pion mass and the matrix element $\mathcal{Z}_\pi = | \langle \pi | \overline{u} \gamma_5 d | 0 \rangle|^2$ can be extracted from the exponential fit given in the r.h.s.~of Eq.~(\ref{eq:larget}). 
The time intervals used for the pion case can be read off from Table \ref{tab:timeint}.
For maximally twisted fermions the value of $\mathcal{Z}_{\pi}$ determines the pion decay constant without the need of the knowledge of any renormalization constant \cite{Frezzotti:2000nk, Frezzotti:2003ni}, namely
 \be
    af_{\pi} = 2a \mu_\ell \frac{\sqrt{a^4 \mathcal{Z}_\pi}}{aM_\pi \mbox{sinh}(aM_\pi)} ~ .
    \label{eq:decaypi}
 \ee
Then we have studied the dependence of the pion mass and decay constant on the renormalized light quark mass
 \be
    m_\ell = (a \mu_\ell) \frac{1}{a Z_P}
    \label{eq:mlR}
 \ee
through simultaneous fits based either on ChPT at next-to-leading order (NLO) or on a polynomial expansion in $m_\ell$.
This was done following two procedures that differ for the choice of the scaling variable.
In the first one we used $r_0 / a$, while in the second one the fictitious meson mass $aM_{s^\prime s^\prime}$ is adopted in order to reduce the impact of discretization effects of the PS meson masses.

\subsection{Analyses in units of $r_0$ (analyses A and B)}
\label{sec:r0}

Since the chiral extrapolation is an important source of uncertainty in our analysis, we have fitted the dependence of both $M_\pi^2$ and $f_\pi$ on the renormalized light quark mass $m_\ell$ using two different fitting functions: the one predicted by ChPT at NLO and a polynomial expansion. 
These two choices correspond to expanding the squared pion mass and decay constant either around the chiral point $m_\ell = 0$ up to higher masses including the effects of chiral logarithms, or around a non-vanishing mass $m_\ell = m_\ell^*$ down to the physical pion point without reaching the chiral limit, where non-analytic terms arise in the expansion.
The ChPT approach at NLO is expected to be more accurate in the region of low $m_\ell$, but to suffer from possible higher order corrections at large values of $m_\ell$, where the polynomial expansion is expected to be more accurate.

Both solutions are in principle legitimate to perform the chiral extrapolation. 
Since both fits turn out to describe our lattice data nicely, the spread between the results obtained using NLO ChPT and those corresponding to the polynomial expansion represents our uncertainty on the chiral extrapolation and it will be used to estimate the corresponding systematics.
This is reasonable also because the polynomial ansatz might underestimate the curvatures of $f_\pi$ and $M_\pi^2 / m_\ell$ at small values of $m_\ell$ (as it does not contain any chiral logarithm), while the NLO ChPT fit applied to the range of our pion data (see Table \ref{tab:MpiL}) might overestimate the curvatures in the small $m_\ell$ region, as suggested by the results of NNLO fits (see later Section \ref{sec:results}) and indicated also by the findings of Refs.~\cite{Frezzotti:2008dr, Baron:2009wt} at $N_f = 2$ and of Refs.~\cite{Borsanyi:2012zv, Durr:2013goa} at $N_f = 2 + 1$.

Let us consider the SU(2) ChPT approach in units of $r_0$ which hereafter will be referred to as analysis A.
The ChPT predictions at NLO can be written in the following way
 \bea
     \label{eq:cptmpi2Ch}
    (M_\pi  r_0 )^2  & = & 2(B r_0 )(m_\ell r_0 )\left[ 1 + \xi_\ell \log \xi_\ell  + P_1 \xi_\ell  + \frac{a^2}{r_0^2} \left( P_2 + \frac{4c_2}{(4\pi f)^2} 
                                       \log \xi_\ell \right) \right] K_{M^2}^{FSE} ~ , \qquad \\
    \label{eq:cptfpiCh}
    (f_\pi  r_0 ) & = & (f r_0 )\left[ 1 - 2\xi_\ell \log \xi_\ell  + P_3 \xi_\ell  + \frac{a^2}{r_0^2} \left( P_4 - \frac{4c_2}{(4\pi f)^2} 
                               \log \xi_\ell \right) \right] K_f^{FSE} ~ ,
 \eea
where $P_1$ - $P_4$ are free parameters and
 \be
     \xi_\ell = \frac{2B m_\ell}{16\pi^2 f^2} ~ ,
     \label{eq:xi}
 \ee
with $B$ and $f$ being the SU(2) low-energy constants (LECs) entering the LO chiral Lagrangian, which have been left free to vary in our fits. 

In Eqs.~(\ref{eq:cptmpi2Ch}-\ref{eq:cptfpiCh}) the parameters $P_1$ and $P_3$ are related to the NLO LECs $\overline{\ell}_3$ and $\overline{\ell}_4$ by
\be
     P_1 = -\overline{\ell}_3 - \log \left( \frac{M_\pi^{\rm phys}}{4 \pi f} \right)^2 ~ , \qquad 
     P_3 = 2 \overline{\ell}_4 + 2 \log \left( \frac{M_\pi^{\rm phys}}{4 \pi f} \right)^2 
     \label{eq:P1P3_def}
\ee
with $M_\pi^{\rm phys}$ being the value of the pion mass at the physical point, while the quantities $K_{M^2}^{FSE}$ and $K_f^{FSE}$ represent the finite size effects (FSE) for the squared pion mass and the pion decay constant, respectively. 
They will be discussed in a while.

For the moment notice the presence of the terms proportional to $ a^2 \log \xi_\ell$ in Eqs.~(\ref{eq:cptmpi2Ch}-\ref{eq:cptfpiCh}). 
These terms originate from the mass splitting between the charged and the neutral pions, which is a discretization effect appearing within the twisted mass formulation.
Its impact on the ChPT expansion of $M_\pi^2$ and $f_\pi$ (see Ref.~\cite{Dimopoulos:2009qv} and references therein) has been worked out in Ref.~\cite{Bar:2010jk}, where a power counting scheme  was adopted in which $a^2 \Lambda_{QCD}^4 \approx 2 B m_\ell$. 
We have expanded the resulting formulae up to ${\cal{O}}(a^2)$, leading to Eqs.~(\ref{eq:cptmpi2Ch}-\ref{eq:cptfpiCh}) with the presence of the parameter $c_2$ which is directly related to the neutral and charged pion mass splitting at LO by
 \be
    \left( M_{\pi^0}^2  - M_{\pi^\pm}^2 \right)_{LO} = 4 a^2 c_2 ~ .
    \label{eq:pi+pi0splitting}
 \ee

In the $\chi^2$-minimization procedure we have given to $c_2$ a prior based on the values found in Ref.~\cite{Herdoiza:2013sla} by analyzing charged and neutral pion data for a set of ETMC ensembles consistent with the one considered in this work\footnote{We treated the prior for $c_2$ in the same way as those for the renormalization constants $Z_P$ and the quantities $r_0 / a$ in Eq.~(\ref{eq:chi2term}).}. 
In Ref.~\cite{Herdoiza:2013sla} two different determinations of $c_2$ are reported, one in which the chiral limit is performed through a constant fit in $M_\pi^2$ and the other one in which the fit was assumed to be linear. 
In the present work we have used an average of the two determinations including the spread in the error, which in units of $r_0$ reads as $c_2 r_0^4 = -1.7 \pm 0.6$.  

On the theoretical side the impact of FSE on $M_\pi$ and $f_\pi$ has been studied within ChPT at NLO in Ref.~\cite{GL87} and using a resummed asymptotic formula in Ref.~\cite{CDH05}, where both leading and sub-leading exponential terms are taken into account and the chiral expansion is applied to the $\pi-\pi$ forward scattering amplitude. 
When the leading chiral representation of the latter is considered, the resummed approach coincides with the NLO result of Ref.~\cite{GL87}.
Viceversa at NNLO the resummation technique includes only part of the two-loop effects as well as of higher-loop effects.
The resummed approach was positively checked against a full NNLO calculation of the pion mass in Ref.~\cite{CH06}, showing that the missing two-loop contributions are actually negligible for $M_\pi L \gsim 2$ and $L \gsim 2$ fm.
Finally, we considered that the cutoff effects, giving rise to the splitting between charged and neutral pions, enter also the determination of FSE, as explicitly worked out within the resummed approach in Ref.~\cite{Colangelo:2010cu}.

Thus, as far as FSE are concerned, we have investigated three different approaches: the NLO ChPT predictions of Ref.~\cite{GL87} (which will be labelled hereafter as GL), the resummed formulae of Ref.~\cite{CDH05} including higher order corrections (labelled as CDH) and the formulae developed in Ref.~\cite{Colangelo:2010cu} which accounts for the $\pi^0 - \pi^+$ mass splitting (labelled as CWW).

The predictions of both CDH and CWW approaches require the knowledge of the LECs $\overline{\ell}_1 - \overline{\ell}_4$ and eventually of the splitting parameter $c_2$. 
The LECs $\overline{\ell}_3$ and $\overline{\ell}_4$, which are related to the $\xi_\ell$-dependent NLO terms in $M_\pi^2$ and $f_\pi$ [see Eqs.~(\ref{eq:cptmpi2Ch}-\ref{eq:P1P3_def})], have been treated as free parameters in our fitting procedures, while for $\overline{\ell}_1$ and $\overline{\ell}_2$ we used the values given in Ref.~\cite{Colangelo:2010cu}.
The CWW corrections depend also on the neutral pion mass $M_{\pi^0}$, which was estimated at LO through Eq.~(\ref{eq:pi+pi0splitting}) using $(M_{\pi^+})_{LO} = 2B m_\ell$. 
We have checked that such values of $M_{\pi^0}$ are consistent with those extracted directly from the neutral PS correlator in Refs.~\cite{Ottnad:2012fv,Herdoiza:2013sla}.

In order to check how well the finite volume corrections predicted by the three chosen approaches are working, we have used the two ensembles $A40.32$ and $A40.24$ (see Table \ref{tab:masses&simudetails}), which correspond to the same quark mass and lattice spacing, but different lattice volumes. 
Notice that the ensemble $A40.24$ has both the lowest value of the quantity $M_\pi L$ (see Table \ref{tab:MpiL}) and the largest pion mass splitting, being $M_{\pi^0} / M_\pi^+ \approx 0.5$ \cite{Ottnad:2012fv,Herdoiza:2013sla}. 
Therefore FSE are expected to be maximal for this ensemble.

The terms $K_{M^2}^{FSE}$ and $K_f^{FSE}$, appearing in the ChPT formulae (\ref{eq:cptmpi2Ch}-\ref{eq:cptfpiCh}), relate the squared pion mass and decay constant calculated at finite volume with their infinite volume counterparts.
For the ensemble $A40.32$ and $A40.24$ we can write
 \bea
    M_{[32]}^2  & = & M_{[\infty ]}^2 K_{M^2, [32]}^{FSE} ~ , \nn \\
    M_{[24]}^2  & = & M_{[\infty ]}^2 K_{M^2, [24]}^{FSE} 
    \label{eq:FSE_Mpi}
 \eea
and in analogous way for $K_{f, [32]}^{FSE}$ and $K_{f, [24]}^{FSE}$ in the case of the decay constant $f_\pi$.
Taking the ratio of the above relations we see that for an ideal correction the ratio of the multiplicative factors $K^{FSE}$ should match the ratio of the uncorrected lattice data independently of the infinite volume values.
The more accurate the correction is, the more the prediction for ($K_{M^2, [32]}^{FSE} / K_{M^2, [24]}^{FSE}$) matches the lattice data $(M_{[32]}^2 / M_{[24]}^2)$.
The corresponding numerical results are reported in Tables \ref{tab:confrFSE_Mpi} and \ref{tab:confrFSE_fpi} for the pion mass and the decay constant, respectively.

\begin{table}[hbt!]
\centering
\renewcommand{\arraystretch}{1.20} 
\begin{tabular}{||c|c|c|c||c||}
\hline
& GL & CDH & CWW & Lattice data $(M_{[32]}^2 / M_{[24]}^2)$\\
\hline
$K_{M^2, [32]}^{FSE} / K_{M^2, [24]}^{FSE}$& $0.988$ & $0.970$&$0.962$&$0.945(25)$  \\
\hline
\end{tabular}
\renewcommand{\arraystretch}{1.0}
\caption{\it Values of the ratio of the FSE correction factor $K_{M^2}^{FSE}$ for the ensembles A40.32 and A40.24, obtained within the approaches $GL$, $CDH$ and $CWW$ (see text), compared with the corresponding ratio of lattice data.}
\label{tab:confrFSE_Mpi}
\end{table}

\begin{table}[hbt!]
\vskip 0.5cm
\centering
\renewcommand{\arraystretch}{1.20} 
\begin{tabular}{||c|c|c|c||c||}
\hline
& GL & CDH & CWW & Lattice data $(f_{[32]} / f_{[24]})$\\
\hline
 $K_{f, [32]}^{FSE} / K_{f, [24]}^{FSE}$ & $1.023$ & $1.040$&$1.054$&$1.050(19)$  \\
\hline
\end{tabular}
\renewcommand{\arraystretch}{1.0}
\caption{\it The same as in Table \ref{tab:confrFSE_Mpi}, but for the decay constant $f_\pi$.} 
\label{tab:confrFSE_fpi}
\end{table}

From these tables one can see that the corrections calculated using the CWW approach are well compatible with the lattice data for both the pion mass and the decay constant.
It is also possible to see how large the relative contribution of the various FSE corrections is.

In table \ref{tab:R_FSE} we collected the values of the coefficients $(K_{M^2, [24]}^{FSE} - 1)$ and $(K_{f, [24]}^{FSE} - 1)$, representing the FSE correction for the ensemble $A40.24$, which, as already noted, is affected by the largest FSE correction in the whole set of ensembles.
By comparing CDH and CWW predictions it can also be seen that the ${\cal{O}}(a^2)$ term related to the pion mass splitting, though not negligible, is not the dominant one and appears to be at the percent level.
In what follows, the pion data will be corrected for FSE using the CWW formulae unless explicitly stated otherwise.

\begin{table}[hbt!]
\centering
\renewcommand{\arraystretch}{1.20} 
\begin{tabular}{||c|c|c|c||}
\hline
& GL & CDH & CWW \\
\hline
$K_{M^2, [24]}^{FSE} - 1$ & $~~~0.0140$ & $~~~0.0377$&$~~~0.0492$ \\
\hline
$K_{f, [24]}^{FSE} - 1$ & $-0.0280$ & $-0.0469$&$-0.0632$ \\
\hline
\end{tabular}
\renewcommand{\arraystretch}{1.0}
\caption{\it Values of $K_{M^2}^{FSE} - 1$ and  $K_f^{FSE} - 1$ for the ensembles A40.24 obtained within the various FSE approaches $GL$, $CDH$ and $CWW$ (see text).} 
\label{tab:R_FSE}
\end{table}

\begin{figure}[hbt!]
\begin{center}
\includegraphics[width=0.8\textwidth]{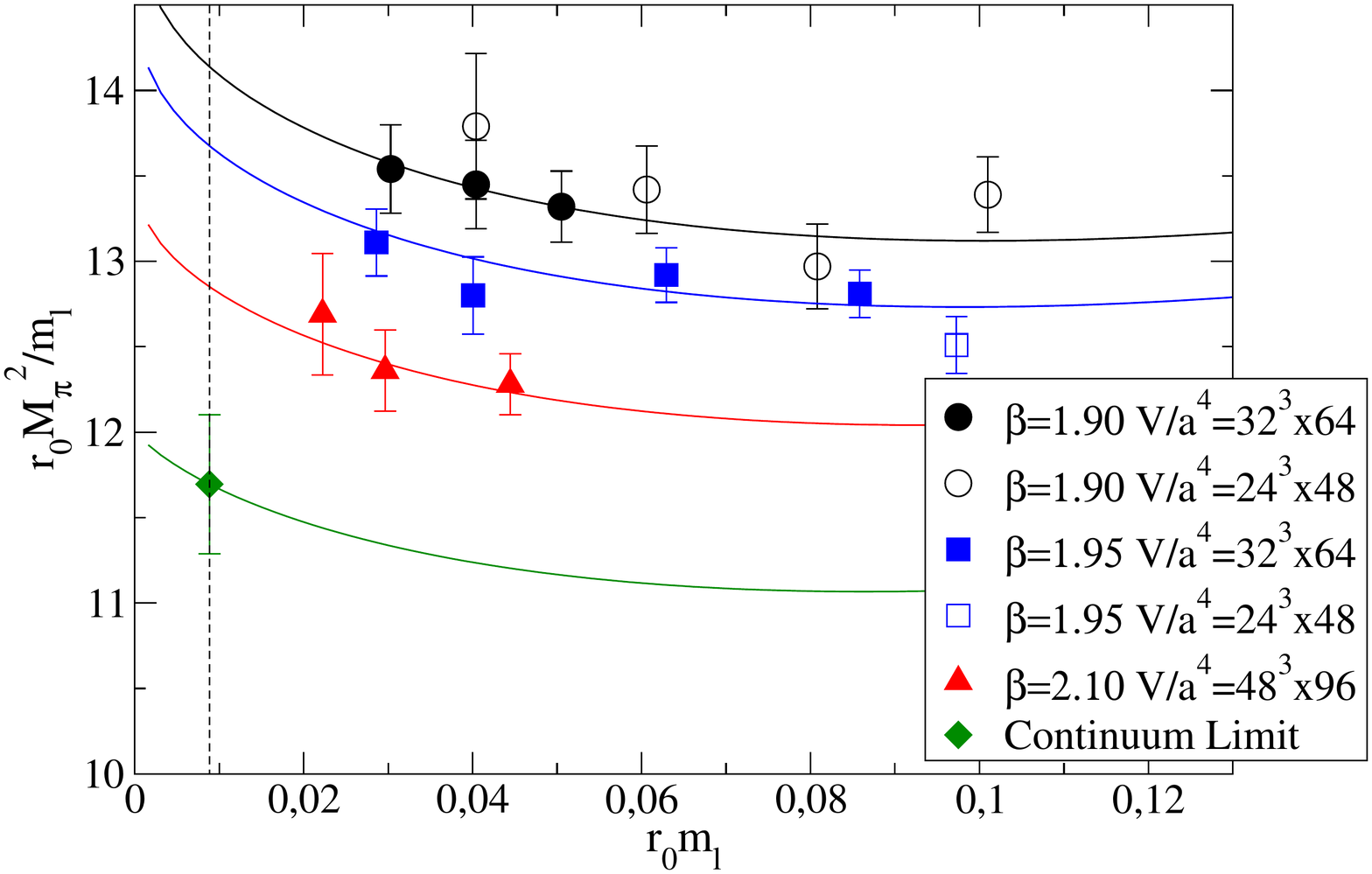}
\end{center}
\vspace*{-1.0cm}
\caption{\it Chiral and continuum extrapolation of $r_0 M_\pi^2 / m_\ell$ based on the NLO ChPT fit given by Eq.~(\ref{eq:cptmpi2Ch}). Lattice data have been corrected for FSE using the CWW approach \cite{Colangelo:2010cu} and correspond to the RCs $Z_P$ calculated with the method M1 (see text).}
\label{fig:mpi2sumlmlCh}

\begin{center}
\includegraphics[width=0.8\textwidth]{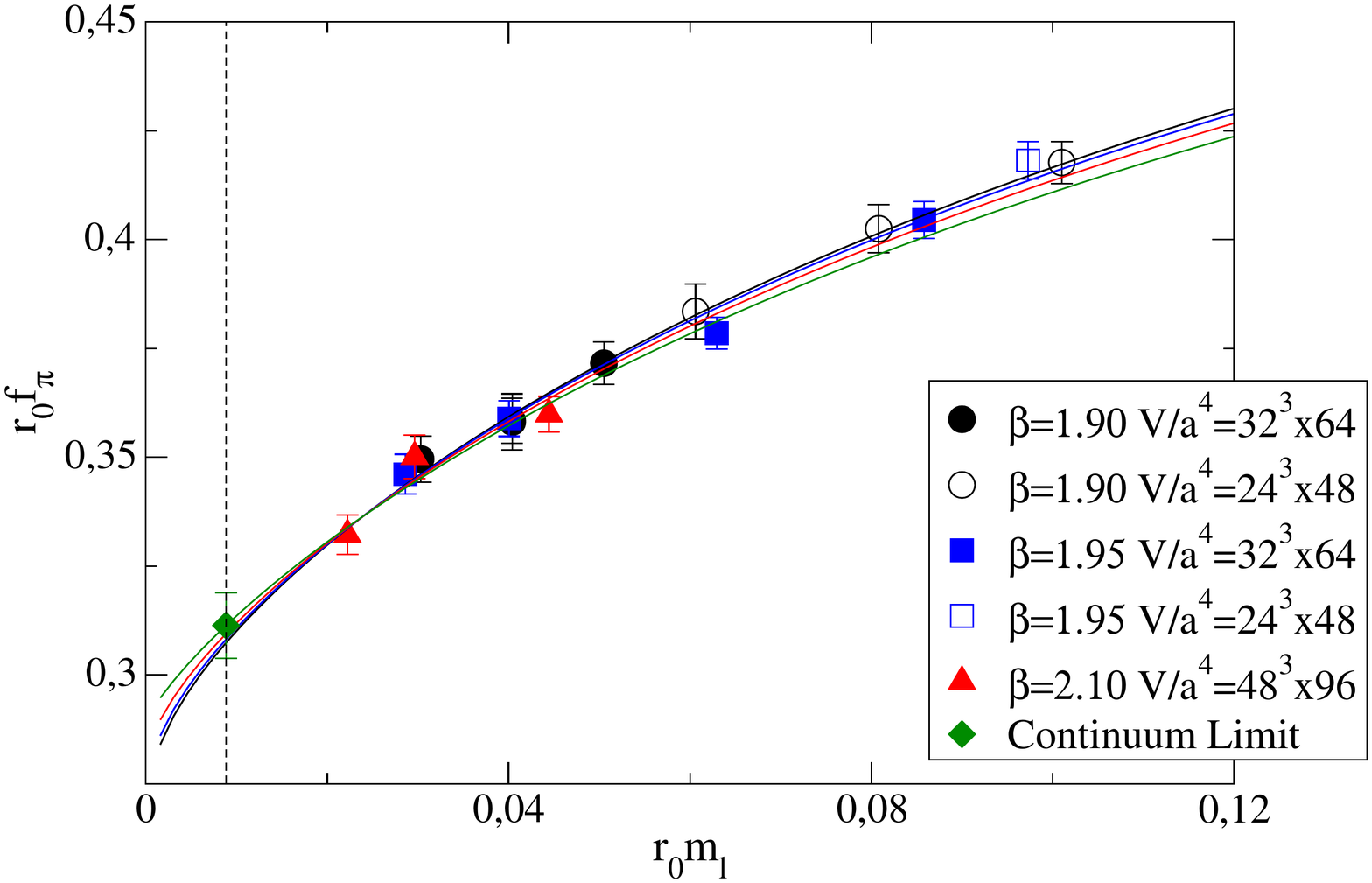}
\end{center}
\vspace*{-1.0cm}
\caption{\it The same as in Fig.~\ref{fig:mpi2sumlmlCh}, but for the decay constant $r_0 f_\pi$.}
\label{fig:fpimlCh}
\end{figure}

The dependence of our lattice data for $r_0 M_\pi^2 /m_\ell$ and $r_0 f_\pi$ on the renormalized quark mass $r_0 m_\ell$ is shown in Figs.~\ref{fig:mpi2sumlmlCh} and \ref{fig:fpimlCh}, respectively. 
The behaviors of the chiral extrapolations for each lattice spacing and in the continuum limit are also presented.
In what follows, unless otherwise stated, the data shown in the figures correspond to the RCs $Z_P$ computed with the method M1.

From Figs.~\ref{fig:mpi2sumlmlCh} and \ref{fig:fpimlCh} it can be seen that the impact of discretization effects using the values of $r_0 / a$ is almost completely negligible in the case of $r_0 f_\pi$, while it is at the level of $\simeq 10 \%$ in the case of $r_0 M_\pi^2 / m_\ell$ (using the difference between the continuum results and the ones at the finest lattice spacing).

The value of the physical average up/down quark mass, $m_{ud}$, can be extracted from the ratio $M_\pi^2 / f_\pi^2$ using as input its experimental value, obtained from the central values of Ref.~\cite{PDG} (see Ref.~\cite{FLAG} for the explanation of the use of the experimental mass of the neutral pion as the pion mass in pure QCD and in the isospin symmetric limit)
 \be
    M_\pi^{exp.} = M_{\pi^0} = 134.98 ~ \mbox{MeV} ~ ,\qquad f_\pi^{exp.} = f_{\pi^+} = 130.41 ~ \mbox{MeV} ~ .
    \label{eq:exp_inputs}
 \ee
The numerical results for $m_{ud}$ as well as those for the lattice spacing and the relevant LECs will be collected and discussed in Section \ref{sec:results}.

As anticipated in Section \ref{sec:intro}, we studied the chiral extrapolation also by replacing the NLO ChPT ansatz with a simple polynomial expansion in the renormalized light quark mass, namely
 \bea
    \label{eq:cptmpi2FP}
     (M_\pi  r_0 )^2  & = & 2 (B r_0) (m_\ell r_0 )\left( {1 + P_1^\prime (m_\ell r_0 ) + P_2^\prime \frac{{a^2 }}{{r_0 ^2 }} + 
                                        P_3^\prime (m_\ell r_0 )^2 } \right) \cdot K_{M^2}^{FSE} ~ , \\
      \label{eq:cptfpiFP}
     (f_\pi  r_0 ) & = & (f r_0) \left( {1 + P_4^\prime (m_\ell r_0 ) + P_5^\prime \frac{{a^2 }}{{r_0 ^2 }} + 
                                 P_6^\prime (m_\ell r_0 )^2 } \right) \cdot K_f^{FSE} ~ ,
 \eea
where $B$, $f$ and $P_1^\prime$ - $P_6^\prime$ are free parameters.
This analysis will be referred to as analysis B.
Since the calculation of $K_{M^2}^{FSE}$ and $K_f^{FSE}$ is based on ChPT, the FSE corrections have been taken from the analysis A and applied directly to the lattice data.

\begin{figure}[hbt!]
\begin{center}
\includegraphics[width=0.8\textwidth]{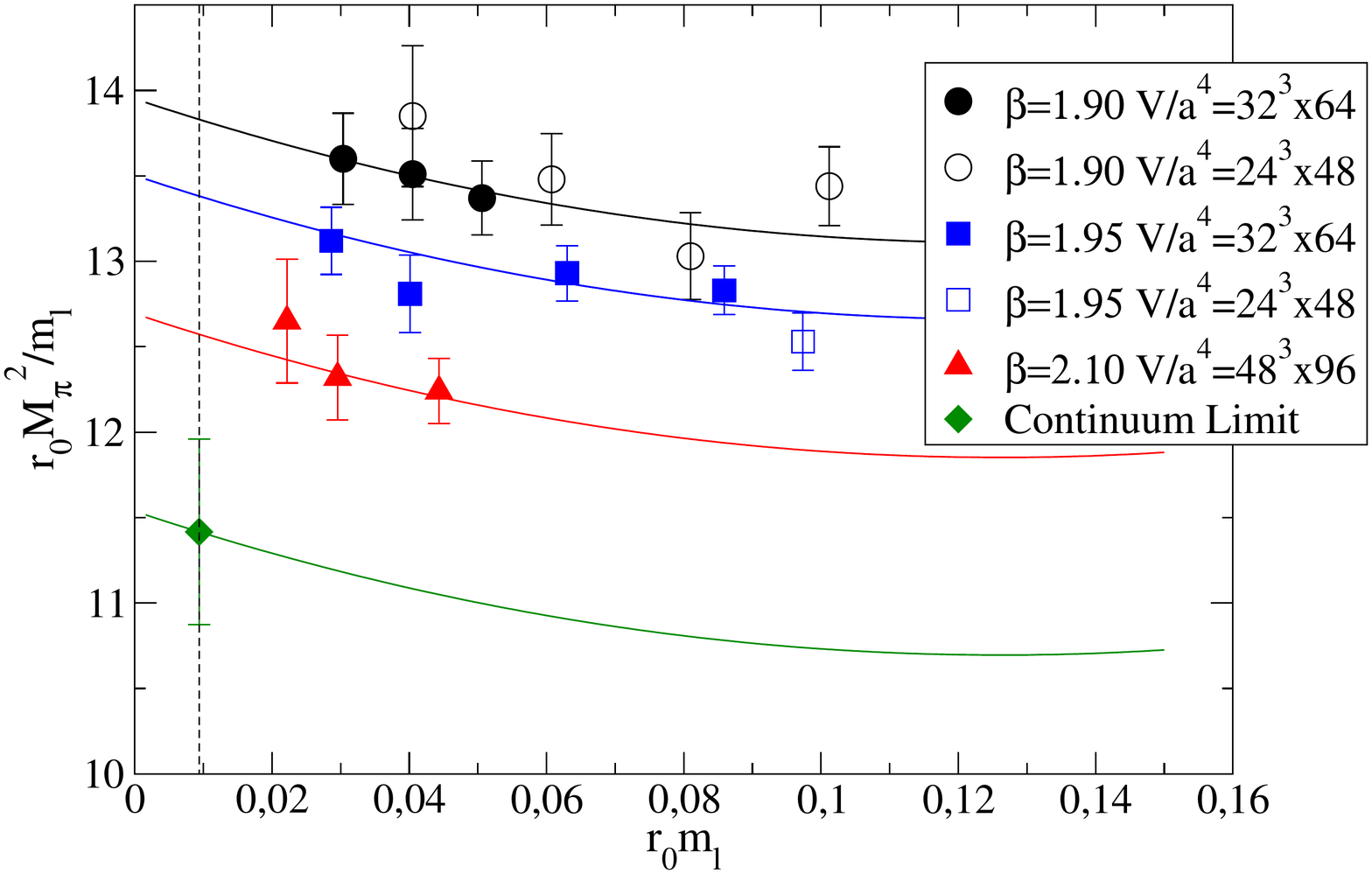}
\end{center}
\vspace*{-1.0cm}
\caption{\it Chiral and continuum extrapolation of $r_0 M_{\pi}^2 / m_\ell$ obtained using the polynomial fit given by Eq.~(\ref{eq:cptmpi2FP}).}
\label{fig:mpi2sumlmlFP}

\begin{center}
\includegraphics[width=0.8\textwidth]{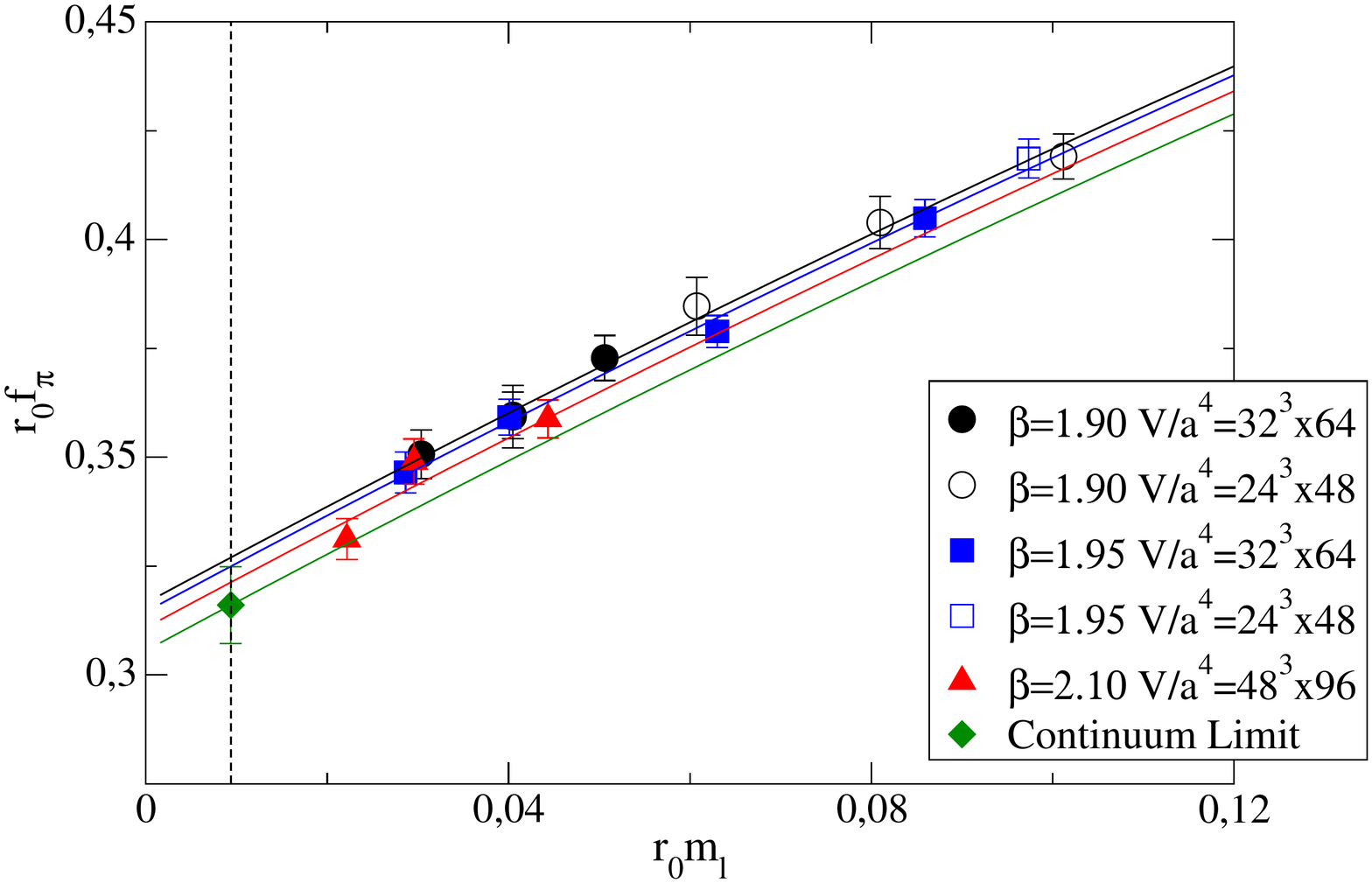}
\end{center}
\vspace*{-1.0cm}
\caption{\it The same as in Fig.~\ref{fig:mpi2sumlmlFP}, but for the decay constant $r_0 f_\pi$.}
\label{fig:fpimlFP}
\end{figure}

The chiral extrapolations of our lattice data for $r_0 M_\pi^2 / m_\ell$ and $r_0 f_\pi$, obtained using the polynomial fits (\ref{eq:cptmpi2FP}-\ref{eq:cptfpiFP}), are shown for each lattice spacing and in the continuum limit in Figs.~\ref{fig:mpi2sumlmlFP} and \ref{fig:fpimlFP}, respectively.

Notice that the impact of discretization effects on $r_0 M_\pi^2 / m_\ell$ obtained using the polynomial fit (see Fig.~\ref{fig:mpi2sumlmlFP}) is very similar to the one found in the case of the NLO ChPT prediction (see Fig.~\ref{fig:mpi2sumlmlCh}), while in the case of $r_0 f_\pi$, at variance with the NLO ChPT fit (see Fig.~\ref{fig:fpimlCh}), the polynomial expansion exhibits visible cutoff effects (see Fig.~\ref{fig:fpimlFP}) though limited at the level of few percent only.
Nevertheless, both the NLO ChPT and the polynomial fits describe quite well the lattice data for the pion mass and the decay constant, yielding only slightly different results, at the percent level, at the physical pion point.

\subsection{Analyses in units of $M_{s^\prime s^\prime}$ (analyses C and D)}
\label{sec:Mss}

The results presented in Figs.~\ref{fig:mpi2sumlmlCh} and \ref{fig:mpi2sumlmlFP} show that the impact of discretization effects using $r_0$ as the scaling variable is at the level of $\simeq 10 \%$ for the squared pion mass.
In order to keep the extrapolation to the continuum limit under better control we repeated the analyses A and B adopting a different choice for the scaling variable, namely instead of $r_0$ we introduced the mass $M_{s^\prime s^\prime}$ of a fictitious PS meson made of two strange-like valence quarks\footnote{To be more precise we consider the fictitious PS meson made of two strange-like quarks $s^\prime$ and $s^{\prime \prime}$ having the same mass, $m_{s^\prime} = m_{s^{\prime \prime}}$, and opposite values of the Wilson r-parameter, $r_{s^\prime} = - r_{s^{\prime \prime}}$. 
For the sake of simplicity we will refer to the mass of such a PS meson as $M_{s^\prime s^\prime}$.}.
The PS mass $M_{s^\prime s^\prime}$ has a very mild dependence on the light-quark mass and is affected by cutoff effects similar to the ones of a K meson. 
Thus, we tried to improve the  continuum extrapolation by considering the ratio $M_\pi^2 / M_{s^\prime s^\prime}^2$ which may exploit a partial cancellation of discretization effects.

To construct the meson mass ratio we first performed a slight interpolation in the strange valence quark mass to get the quantity $a M_{s^\prime s^\prime}$ at a common (but arbitrary) value $r_0 m_{s^\prime} = 0.22$ for each $\beta$ and light quark mass. 
Since, as expected, we found no significant dependence of $a M_{s^\prime s^\prime}$ on the light quark mass, we performed a constant fit in $a \mu_\ell$ to obtain the values of $a M_{s^\prime s^\prime}$ at each $\beta$. 
In this way we find
 \bea
    \left. aM_{s^\prime s^\prime} \right|_{\beta = 1.90, ~ 1.95, ~ 2.10}  & = & \{ 0.3258(2), ~ 0.2896(2), ~ 0.2162(3) \} \quad \mbox{(method M1)} \nn \\
                                                                                                             & = & \{ 0.3391(2), ~ 0.2986(2), ~ 0.2220(3) \} \quad \mbox{(method M2)} ~ .
    \label{eq:aMssvalues}
 \eea
The values of $a M_{s^\prime s^\prime}$ have been used to bring to a common scale all lattice quantities, covering the role that in analysis A and B was played by $r_0 / a$.
The (quite small) errors on $a M_{s^\prime s^\prime}$ are propagated via the bootstrap sampling.  

The new analyses, which will be referred to as analyses C and D, proceed in the same way as in the previous Section, namely in the case of the NLO ChPT fit (analysis C) one employs the ansatz
 \bea
    \label{eq:mpi2suMss2}
    \frac{M_\pi^2}{M_{s^\prime s^\prime}^2} & = & \frac{2 B m_\ell}{M_{s^\prime s^\prime}^2} \left[ 1 + \xi_\ell \log \xi_\ell + P_1 \xi_\ell  + 
                                                                              \left( aM_{s^\prime s^\prime} \right)^2 \left( P_2 + \frac{4c_2}{(4\pi f)^2} 
                                                                              \log \xi_\ell \right) \right] K_{M^2}^{FSE} ~ , \quad \\ 
    \label{eq:fpisuMss}
    \frac{f_\pi}{M_{s^\prime s^\prime}} & = & \frac{f}{M_{s^\prime s^\prime}} \left[ 1 - 2\xi_\ell \log \xi_\ell  + P_3 \xi_\ell  + 
                                                                     \left( aM_{s^\prime s^\prime} \right)^2 \left( P_4 - \frac{4c_2}{(4\pi f)^2} 
                                                                     \log \xi_\ell \right) \right] K_f^{FSE} ~ ,
 \eea
where again the parameters $P_1$ and $P_3$ are related to the NLO LECs $\overline{\ell}_3$ and $\overline{\ell}_4$ through Eq.~(\ref{eq:P1P3_def}).
In the case of the polynomial fit (analysis D) one fits the data with the analogue of Eqs.~(\ref{eq:cptmpi2FP}) and (\ref{eq:cptfpiFP}) expressed in units of $M_{s^\prime s^\prime}$.

\begin{figure}[hbt!]
\begin{center}
\includegraphics[width=0.8\textwidth]{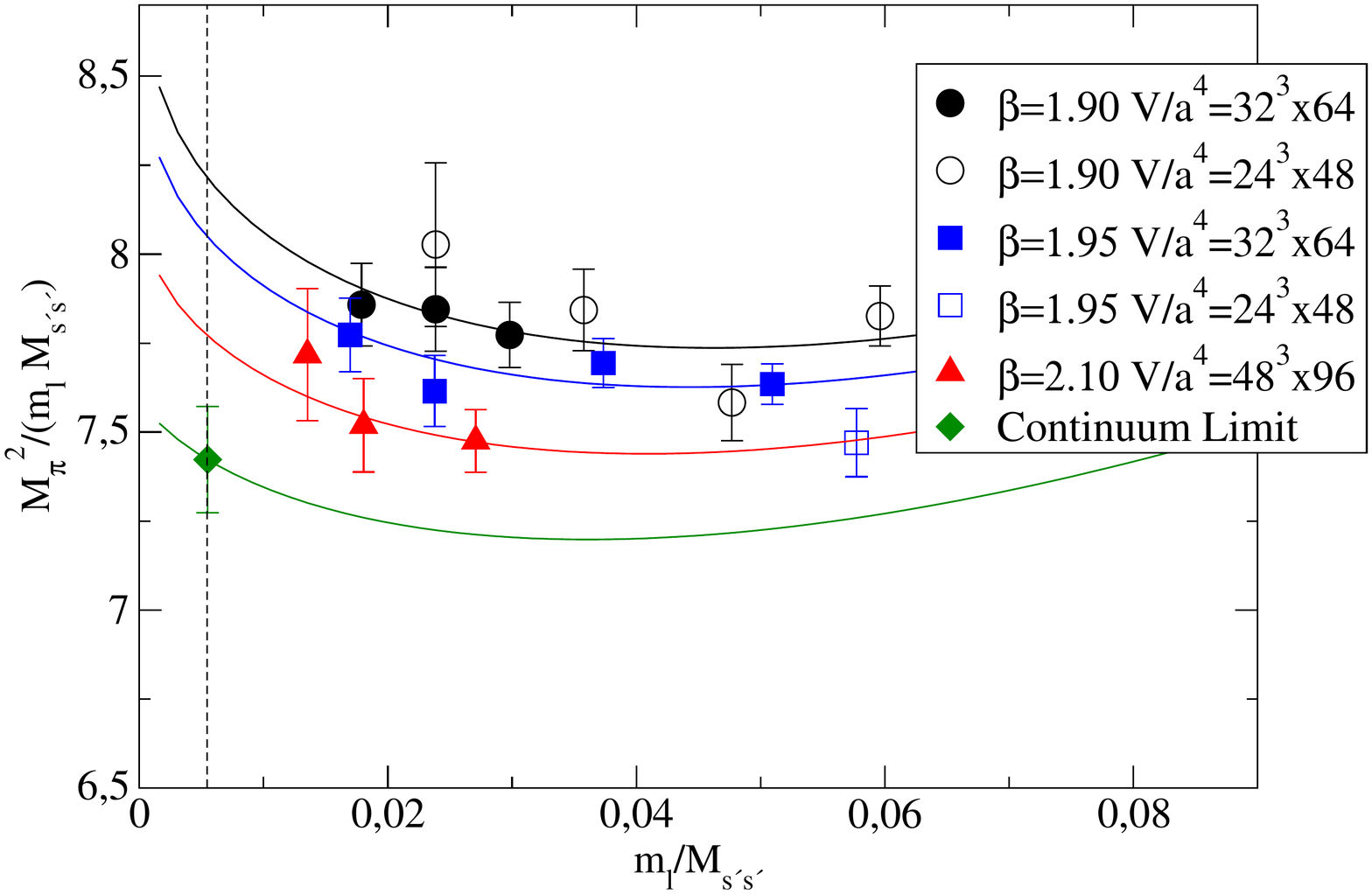}
\end{center}
\vspace*{-1.0cm}
\caption{\it Chiral and continuum extrapolation of $M_\pi^2 / (m_\ell M_{s^\prime s^\prime})$ obtained using the NLO ChPT fit (\ref{eq:mpi2suMss2}).}
\label{fig:M2sumq_mqMssCh}

\begin{center}
\includegraphics[width=0.8\textwidth]{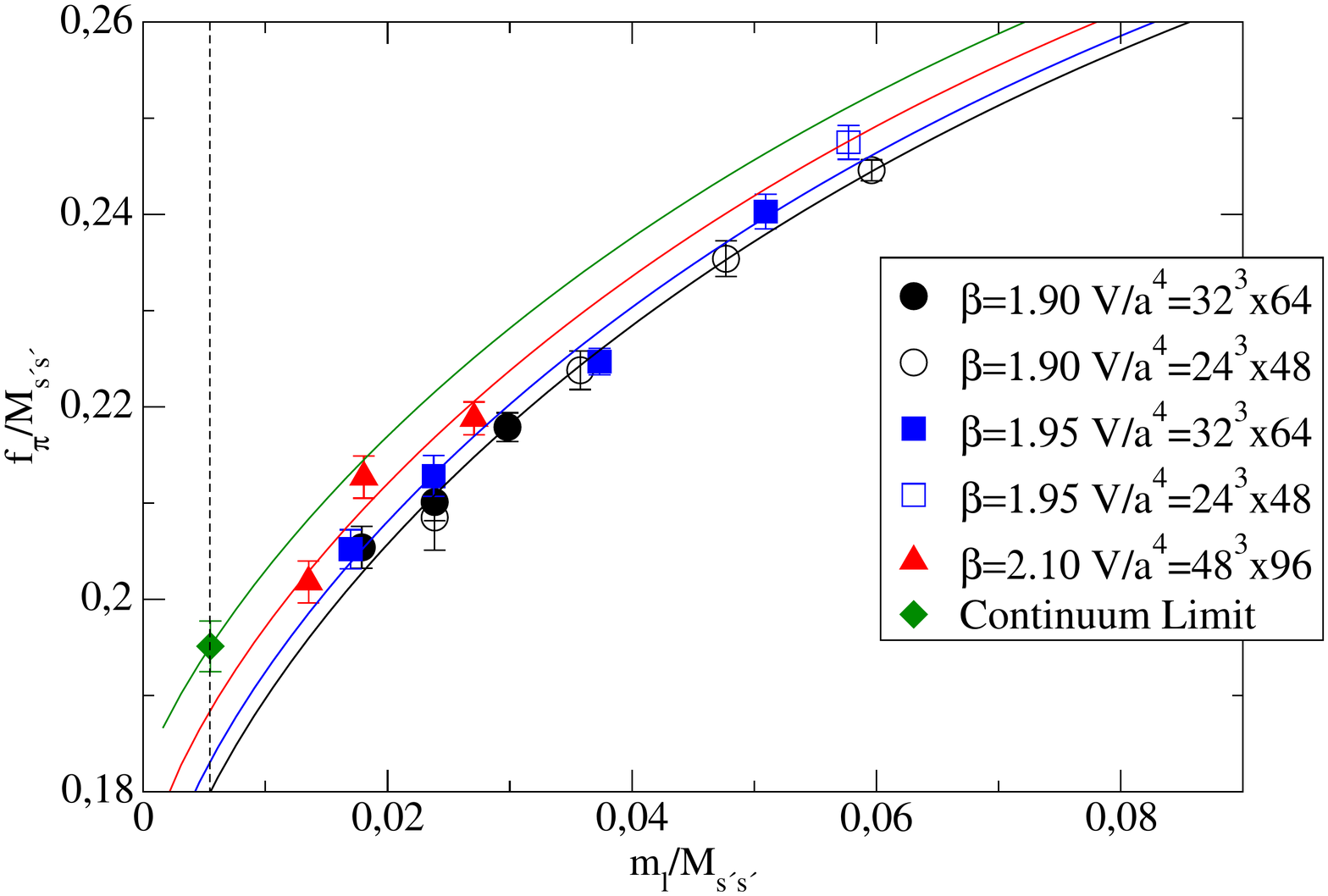}
\end{center}
\vspace*{-1.0cm}
\caption{\it The same as in Fig.~\ref{fig:M2sumq_mqMssCh}, but for the pion decay constant $f_\pi$ in units of $M_{s^\prime s^\prime}$.}
\label{fig:fpi_mqMssCh}
\end{figure}

In Fig.~\ref{fig:M2sumq_mqMssCh} and \ref{fig:fpi_mqMssCh} we show the dependencies of $M_\pi^2 /(m_\ell M_{s^\prime s^\prime})$ and $f_\pi / M_{s^\prime s^\prime}$on  $m_\ell / M_{s^\prime s^\prime}$ at each lattice spacing and in the continuum limit within the analysis C (ChPT fit).
Similar results have been obtained within the analysis D (polynomial fit).

The comparison of Figs.~\ref{fig:mpi2sumlmlCh} and \ref{fig:M2sumq_mqMssCh} clearly shows that, when $M_{s^\prime s^\prime}$ is chosen as the scaling variable, the discretization effects on the squared pion mass are significantly reduced from $\simeq 10 \%$ down to $\simeq 4.5 \%$.
At the same time the discretization effects on the pion decay constant, which are almost negligible in units of $r_0$ (see Fig.~\ref{fig:fpimlCh}), are kept  to be within $\simeq 4 \%$ when $M_{s^\prime s^\prime}$ is used as the scaling variable (see Fig.~\ref{fig:fpi_mqMssCh}).

\subsection{Results for the pion sector}
\label{sec:results}

In this section we present the results of the four analyses (A, B, C, D) carried out in the pion sector.
We have adopted the values of the RCs $Z_P$ corresponding to the methods M1 and M2, so that we end up with eight analyses, which will be referred to as analyses A1, B1, C1, D1 and A2, B2, C2, D2, respectively.

Using the experimental value of the ratio $M_\pi^2 / f_\pi^2$ [see Eq.~(\ref{eq:exp_inputs})], the average up/down quark mass $m_{ud}$ is determined, so that the quantity ($r_0 f_\pi$) is calculated at the physical point within the analyses A1 (A2) and B1 (B2). 
Then, using the experimental value of $f_\pi$ as input,  the Sommer parameter $r_0$ is extracted and this in turn allows to get the values of the lattice spacing at each $\beta$ using the determinations of $r_0 / a$ collected in Table \ref{tab:zpr0a}. 

The analyses C1 (C2) and D1 (D2) proceed in the same way: the average up/down quark mass $m_{ud}$ is determined through the experimental value of the ratio $M_\pi^2 / f_\pi^2$, while the mass $M_{s^\prime s^\prime}$ is obtained by combining the value of $f_\pi / M_{s^\prime s^\prime}$, calculated at the physical point, and the experimental value of $f_\pi$.
However, in order to determine the lattice spacing at each $\beta$ we did not use the quantities $aM_{s^\prime s^\prime}$ given in Eq.~(\ref{eq:aMssvalues}), since they are affected by discretization effects larger than those occurring in the values of $r_0 / a$. 
Thus we proceeded as follows. 
First we converted the results (\ref{eq:aMssvalues}) for $a M_{s^\prime s^\prime}$ to $r_0 M_{s^\prime s^\prime}$ using the values of $r_0 / a$ from Table \ref{tab:zpr0a}, and then we performed a simple fit of the form $r_0 M_{s^\prime s^\prime} = \overline{P}_1 + \overline{P}_2 a^2 / r_0^2$.
Finally, we determined the values of the lattice spacing at each $\beta$ by combining the values of $a / r_0$ with the continuum extrapolation of $r_0 M_{s^\prime s^\prime}$ and the value of $M_{s^\prime s^\prime}$ obtained from the experimental value of $f_\pi$. 

For convenience the results obtained for the quark mass $m_{ud}$, the scaling variables $r_0$ and $M_{s^\prime s^\prime}$, the values of the lattice spacing and the LECs $B$, $f$, $\overline{\ell}_3$ and $\overline{\ell}_4$, are collected in Tables \ref{tab:allpionresult_M1} and \ref{tab:allpionresult_M2}.

\begin{table}[hbt!]
\centering
\small 
                 \renewcommand{\arraystretch}{1.20}
		\begin{tabular}{|c||c|c||c|c|}
		\hline \hline 
		    \multicolumn{1}{|c||}{} & \multicolumn{2}{|c||}{$r_0$ Analysis} & \multicolumn{2}{|c|}{$M_{s^\prime s^\prime}$ Analysis} \\
		    \hline 
		    Quantity & ChPT Fit (A1)& Polyn. Fit (B1)& ChPT Fit (C1)& Polyn. Fit (D1) \\
		    \hline \hline
			$m_{ud}$(MeV) & 3.72(13) & 3.87(17) & 3.66(10) & 3.75(13)  \\ 
			\hline 
			$r_0({\rm GeV}^{-1})$ & 2.39(6) & 2.42(7) & -  & - \\ 
			\hline 
			$r_0$(fm) & 0.470(12) & 0.477(14) & -  & - \\ 
			\hline 
			$M_{s^\prime s^\prime}$(GeV) & - & - & 0.672(9) & 0.654(10) \\ 
			\hline
			a($\beta=1.90$)(fm) & 0.0886(27) & 0.0899(31) & 0.0868(33) & 0.0892(34) \\ 
			\hline 
			a($\beta=1.95$)(fm) & 0.0815(21) & 0.0827(25) & 0.0799(27) & 0.0820(28) \\ 
			\hline 
     		         a($\beta=2.10$)(fm) & 0.0619(11) & 0.0628(13) & 0.0607(14) & 0.0623(15) \\ 
			\hline \hline			
			$B$(MeV) & 2515(90) & 2381(117) & 2551(73) & 2463(95) \\ 
			\hline 
			$f$(MeV) & 121.1(2) & 126.1(7) & 121.3(2) & 125.9(6) \\ 
			\hline
			$\overline{\ell}_3$ & 3.24(25) & - & 2.94(20) & - \\ 
			\hline
			$\overline{\ell}_4$ & 4.69(10) & - & 4.65(8) & - \\ 
			\hline 			
		\end{tabular}
		\renewcommand{\arraystretch}{1.0}
		\caption{\it Summary of the results of the analyses in the pion sector using the set of values of the RCs $Z_P$ from the method M1. }
		\label{tab:allpionresult_M1}

\vskip 1cm
\centering
\small 
                 \renewcommand{\arraystretch}{1.20} 
		\begin{tabular}{|c||c|c||c|c|}
		\hline \hline 
		    \multicolumn{1}{|c||}{} & \multicolumn{2}{|c||}{$r_0$ Analysis} & \multicolumn{2}{|c|}{$M_{s^\prime s^\prime}$ Analysis} \\
		    \hline 
		    Quantity & ChPT Fit (A2)& Polyn. Fit (B2)& ChPT Fit (C2)& Polyn. Fit (D2) \\
		    \hline \hline
			$m_{ud}$(MeV) & 3.63(12) & 3.78(16) & 3.55(9) & 3.63(12)  \\ 
			\hline 
			$r_0({\rm GeV}^{-1})$ & 2.40(6) & 2.42(7) & -  & - \\ 
			\hline 
			$r_0$(fm) & 0.471(11) & 0.477(13) & -  & - \\ 
			\hline 
			$M_{s^\prime s^\prime}$(GeV) & - & - & 0.685(9) & 0.667(10) \\ 
			\hline
			a($\beta=1.90$)(fm) & 0.0887(27) & 0.0898(31) & 0.0865(34) & 0.0888(35) \\ 
			\hline 
			a($\beta=1.95$)(fm) & 0.0816(21) & 0.0826(25) & 0.0796(28) & 0.0817(29) \\ 
			\hline 
     		         a($\beta=2.10$)(fm) & 0.0620(11) & 0.0627(13) & 0.0604(15) & 0.0620(15) \\ 
			\hline \hline			
			$B$(MeV) & 2584(88) & 2438(120) & 2634(67) & 2546(93) \\ 
			\hline 
			$f$(MeV) & 121.1(2) & 126.0(8) & 121.2(2) & 125.9(7) \\ 
			\hline
			$\overline{\ell}_3$ & 3.31(26) & - & 2.93(21) & - \\ 
			\hline
			$\overline{\ell}_4$ & 4.73(10) & - & 4.68(8) & - \\ 
			\hline 			
		\end{tabular}
		\renewcommand{\arraystretch}{1.0}
		\caption{\it The same as in Table \ref{tab:allpionresult_M1}, but using the set of values of the RCs $Z_P$ from the method M2.}
		\label{tab:allpionresult_M2}
\end{table}

It is quite reassuring to find that different ways of handling both the chiral extrapolation and the discretization effects produce consistent results.

Combining the results reported in Tables \ref{tab:allpionresult_M1} and \ref{tab:allpionresult_M2} provides us with the final determinations and the estimates of the various sources of systematic uncertainties. 
For each quantity we have a set of $N$ results (where $N = 4$ or $N = 8$ depending on the specific quantity) coming from the various analyses A1 - D2.
We assign to all analyses the same weight and therefore we assume that the observable $x$ has a distribution $f(x)$ given by $f(x) = (1/N) \sum_{i=1}^N f_i(x)$, where $f_i(x)$ is the distribution provided by the bootstrap sample of the $i$-th analysis and characterized by central value $x_i$ and standard deviation $\sigma_i$.
Thus we estimate the central value and the error for the observable $x$ through the mean value and the standard deviation of the distribution $f(x)$, which are given by
\bea
    \overline{x} & = & \frac{1}{N} \sum_{i=1}^N x_i ~ , \nn \\ 
    \sigma^2 & = & \frac{1}{N} \sum_{i=1}^N \sigma_i^2  + \frac{1}{N} \sum_{i=1}^N (x_i  - \overline{x})^2 ~ .
    \label{eq:combineresults}
 \eea
The second term in the r.h.s.~of Eq.~(\ref{eq:combineresults}), coming from the spread among the results of the different analyses, corresponds to a systematic error which accounts for the uncertainties due to the chiral extrapolation, the cutoff effects and the RCs $Z_P$. 
Finally we add in quadrature to Eq.~(\ref{eq:combineresults}) the systematic uncertainties associated to the calculation of the FSE and to the conversion from the RI$^\prime$-MOM to the $\overline{\rm MS}$ schemes (see \ref{sec:A3}).     
 
Combining all the sources of uncertainties we get the following estimate for the average up/down quark mass in the ${\overline{\rm MS}}$ scheme at a renormalization scale of $2 \gev$:
 \bea
    m_{ud}  & = & 3.70 ~ (13)_{stat + fit} (6)_{Chiral} (5)_{Disc} (5)_{Z_P} (4)_{FSE} (5)_{Pert} \mev \nn \\
                 & = & 3.70 ~ (13)_{stat + fit} (11)_{syst} \mev \nn \\
                 & = & 3.70 ~ (17) \mev  ~ .
    \label{eq:mudresults}
 \eea

The first error includes the statistical one as well as the error associated with the fitting procedure. 
This error is larger than the typical statistical error of the lattice data, being amplified by the chiral and continuum extrapolations. 
For $m_{ud}$ we get a (stat+fit) error equal to $\simeq 3.5 \%$.

In order to separate in Eq.~(\ref{eq:mudresults}) the uncertainties related to the chiral extrapolation, the discretization effects and the choice of the RCs $Z_P$ we split the contribution coming from the second term in the r.h.s.~of Eq.~(\ref{eq:combineresults}) into those related to the differences of the results obtained using $r_0$ or $M_{s^\prime s^\prime}$ (labelled as Disc), chiral or polynomial fits (labelled as Chiral) and the two methods M1 and M2 for the RCs $Z_P$ (labelled as $Z_P$).
In this way we found them to be at the level of $1.6 \%$, $1.6\%$ and $1.4 \%$, respectively.

For the FSE we considered the difference between the result obtained using the most accurate correction, i.e.~the CWW one, and the one corresponding to no FSE correction at all.
This gave rise to an error on $m_{ud}$ equal to $\simeq 1.1 \%$.

The last systematic error appearing in Eq.~(\ref{eq:mudresults}) is the one related to the conversion between the RI$^\prime$-MOM and the $\overline{\rm MS}(2 \gev)$ schemes, estimated to be $\simeq 1.3 \%$ (see \ref{sec:A3}).

Our determination (\ref{eq:mudresults}) for $m_{ud}$ is the first one obtained at $N_f = 2+1+1$. 
The recent lattice averages, provided by FLAG \cite{FLAG} and based on the findings of Refs.~\cite{Blossier:2010cr, Bazavov:2010yq, Arthur:2012opa, Durr:2010vn, Durr:2010aw}, are: $m_{ud} = 3.6(2) \mev$ at $N_f = 2$ and $m_{ud} = 3.42(9) \mev$ at $N_f = 2+1$.
The comparison of these results with our finding (\ref{eq:mudresults}) shows that the partial quenching of the strange and/or charm sea quarks is not yet visible at the (few percent) level of the present total systematic uncertainty.

 For the Sommer scale $r_0$ we get 
 \be
    r_0 = (0.474 \pm 0.014) \rm{fm} ~ ,
    \label{eq:r0result}
 \ee
while the values of the lattice spacing at each $\beta$ are found to be
 \be
    \left. a \right|_{\beta = 1.90, ~ 1.95, ~ 2.10}   = \{ 0.0885(36), ~ 0.0815(30), ~ 0.0619(18)\} \rm{fm} ~ .
    \label{eq:latticespacingsresults}
 \ee
 


As it is known (see the findings of Refs.~\cite{Frezzotti:2008dr, Baron:2009wt} at $N_f = 2$ and of Refs.~\cite{Borsanyi:2012zv, Durr:2013goa} at $N_f = 2 + 1$), a precise determination of the NLO LECs $\overline{\ell}_3$ and $\overline{\ell}_4$ requires refined analyses addressing the impact of the choice of pion mass range used for the chiral extrapolation as well as the effects of NNLO corrections.
Such analyses are beyond the scope of the present work.
Here we mention only that we have performed NNLO fits in the whole mass range covered by our data ($M_\pi < 450 \mev$) as well as NLO fits restricted to pion masses smaller than $300$, $350$ or $400$ MeV.
The results of these fits (see Table \ref{tab:confChPT}) indicate that the curvatures of $M_\pi^2 / m_\ell$ and $f_\pi$ are within the range already selected by the polynomial and the NLO ChPT fits performed in the full range of simulated pion masses.
In particular, for the average up/down quark mass $m_{ud}$, whose determination is one of the main goals of the present work, and for the LECs $B$ and $f$, we have found results always in between those obtained with the polynomial and the NLO ChPT fits.  

\begin{table}[htb!]
\centering
                 \renewcommand{\arraystretch}{1.20} 
		\begin{tabular}{|c||c|c||c|c|}
		\hline\hline 
		    Quantity & NLO Fit (A1) & Polyn. Fit (B1) & NLO Fit & NNLO Fit \\
		                   & $M_\pi < 450 \mev$ & $M_\pi < 450 \mev$ & $M_\pi < 300 \mev$ & $M_\pi < 450 \mev$ \\
		    \hline 
			$m_{ud}$(MeV) & 3.72(13) & 3.87(17) & 3.77(21) & 3.82(16)  \\ 
			\hline 
			$r_0$(fm) & 0.470(12) & 0.477(14) & 0.472(12) & 0.462(10) \\ 
			\hline 
			$B$(MeV) & 2515(90) & 2381(117) & 2474(157) & 2447(107) \\ 
			\hline 		
			$f$(MeV) & 121.1(2) & 126.1(7) & 122.2(8) & 124.0(7) \\ 
			\hline 	
		 	$\overline{\ell}_3$ & 3.24(0.25) & -- & 2.76(1.28) & 3.84(0.88) \\  
			\hline 
			$\overline{\ell}_4$ & 4.69(10) & -- & 4.18(38) & 3.42(37) \\ 
			\hline 		
		\end{tabular}
		\renewcommand{\arraystretch}{1.0}
		\caption{\it Comparison of different chiral extrapolations for various quantities extracted in the pion analyses A1 and B1 (see text). The errors include the (stat + fit) uncertainty.}
		\label{tab:confChPT}
\end{table}

It is interesting to show in detail the impact of the various approaches used to calculate the FSE for the various quantities extracted from the pion analysis. 
The results obtained within the eight analyses A1 - D2 are quite similar to each other. 
In Table \ref{tab:confFSEresults} we have reported the findings corresponding to the analysis A1.

\begin{table}[htb!]
\centering
                 \renewcommand{\arraystretch}{1.20} 
		\begin{tabular}{|c||c|c|c|c|}
		\hline\hline 
		    Quantity & no correction & GL & CDH & CWW   \\
		    \hline 
			$m_{ud}$(MeV) & 3.68(14) & 3.76(14) & 3.73(13) & 3.72(13)  \\ 
			\hline 
			$r_0$(fm) & 0.464(12) & 0.466(12) & 0.468(12) & 0.470(12) \\ 
			\hline 
			$B$(MeV) & 2548(99) & 2497(97) & 2500(93) & 2515(90) \\ 
			\hline 		
			$f$(MeV) & 120.8(1) & 120.9(1) & 120.9(1) & 121.1(2) \\ 
			\hline 	
		 	$\overline{\ell}_3$ & 3.42(20) & 3.35(20) & 3.34(21) & 3.24(25) \\  
			\hline 
			$\overline{\ell}_4$ & 4.83(9) & 4.77(9) & 4.76(9) & 4.69(10) \\ 
			\hline 		
		\end{tabular}
		\renewcommand{\arraystretch}{1.0}
		\caption{\it Comparison of different FSE corrections for various quantities extracted in the pion analysis A1. The errors include the (stat + fit) uncertainty.}
		\label{tab:confFSEresults}
\end{table}

From Table \ref{tab:confFSEresults} it can be seen that, though the FSE corrections in some particular ensemble can be as large as $4.9 \%$ and $6.3 \%$ for the pion mass and decay constant, respectively (see Table \ref{tab:R_FSE} for the ensemble A40.24), the overall final impact on $m_{ud}$, $r_0$ and the LECs $B$, $f$, $\overline{\ell}_3$ and $\overline{\ell}_4$ is limited to be well below the (stat+fit) error.

Before closing this Section, we notice that a set of ETMC data consistent with the ones considered in this work have been analyzed in Ref.~\cite{Baron:2011sf} adopting ChPT at NLO for the chiral extrapolation, but without accounting for the effect of the charged/neutral pion mass splitting and without involving the determinations of the RCs $Z_P$. 
The findings of Ref.~\cite{Baron:2011sf} concerning both the lattice spacings and the LECs $B$, $f$, $\overline{\ell}_3$ and $\overline{\ell}_4$ nicely agree with our results of Tables \ref{tab:allpionresult_M1}-\ref{tab:allpionresult_M2}  within one standard deviation. 
This indicates that the role played in our analyses by the pion mass splitting and by the RCs $Z_P$ is well under control.

\section{Strange quark mass}
\label{sec:strange}

In this Section we present our determination of the strange quark mass $m_s$.
The analysis follows a strategy similar to the one presented for the pion sector. 
As a preliminary step, however, we performed an interpolation of the lattice kaon data to a fixed value of the strange quark mass in order to arrive iteratively at the physical one (see next Section).

As in the pion sector, we handled discretization effects by performing a first analysis which uses $r_0 / a$ as scaling variable, and a second one in which the fictitious PS meson mass $a M_{s^\prime s^\prime}$ is used to build the ratios $M_K / M_{s^\prime s^\prime}$, which are expected to have milder lattice artifacts. 
For both approaches we considered two different chiral extrapolations in the light quark mass $m_\ell$, namely either the predictions of SU(2) ChPT or the polynomial expansion. 
All these analyses are then repeated with the two sets of values of the RCs $Z_P$ obtained within the methods M1 or M2.
In this way, as in the pion sector, there are eight different branches of the analysis.
In all cases the quark masses are converted directly to physical units using the values of the lattice spacing found in the pion sector.

To determine the strange quark mass we made use of several quantities extracted from the pion sector, like the lattice spacing, the LECs $B$ and $f$, the Sommer parameter $r_0$ and the results for the average up/down quark mass $m_{ud}$.
In order to preserve the physical correlations, in each of the eight kaon analyses we adopted the inputs coming from the corresponding pion fit. 
For instance, if SU(2) ChPT is used for the pion, then the same approach is applied to the kaon as well.
The uncertainties on the input quantities are propagated through the bootstrap sampling for each of the branches of the kaon analysis.
Combining the results from all the eight analyses we obtained our final result for $m_s$ and the estimates of the various sources of systematic uncertainty.

\subsection{Chiral extrapolation in units of $r_0$ (analyses A and B)}
\label{sec:strange_r0}

The analysis is performed iteratively.
We start from an initial guess for the physical strange quark mass $m_s$.
Then, adopting a quadratic spline, the lattice data for the kaon masses are interpolated in the strange quark mass to the (guessed) physical value $m_s$ and brought to a common scale using $r_0 / a$. 
A combined fit is performed to extrapolate $M_K^2$ in the light quark mass and in the (squared) lattice spacing to the physical point and to the continuum limit.
Afterwards the value obtained for the kaon mass, converted in physical units using the value of $r_0$ obtained from the pion analyses, is compared with the experimental one.
If the latter is not reproduced, a new guess for $m_s$ is done and the whole process is repeated again.

The experimental value of the kaon mass to be matched is the one in pure QCD corrected for leading strong and electromagnetic isospin breaking effects according to
 \be
      M_K^{exp} = \sqrt{ \frac{M_{K^+}^2  + M_{K^0}^2}{2} - \frac{(1 + \varepsilon + 2 \varepsilon_{K^0} - \varepsilon_m)}{2} 
                            \left( M_{\pi^+}^2 - M_{\pi^0}^2 \right) } \simeq 494.2 (4) ~ \mbox{MeV} ~ ,
    \label{eq:MKexp}
 \ee
where $\varepsilon = 0.7 (3)$, $\varepsilon_{K^0} = 0.3 (3)$ and $\varepsilon_m = 0.04 (2)$ \cite{FLAG}. 

For the analysis A we used the SU(2) ChPT predictions at NLO, which assume the chiral symmetry to be satisfied only by the up and down quarks and read as
 \be
    (r_0 M_K)^2  = P_0 (m_\ell  + m_s ) \left[1 + P_1 m_\ell  + P_3 a^2 \right] K_{M_K^2}^{FSE} ~ . 
    \label{eq:mk2Ch}
 \ee
Alternatively we considered a polynomial fit (analysis B) according to the following expression
 \be
    (r_0 M_K)^2  = P_0^\prime (m_\ell  + m_s ) \left[ 1 + P_1^\prime m_\ell  + P_2^\prime m_\ell^2  + P_3^\prime a^2 \right]  K_{M_K^2}^{FSE} ~ .
    \label{eq:mk2FP}
 \ee
Notice that for the squared kaon mass SU(2) ChPT predicts the absence of chiral logarithms at NLO, so that the expressions (\ref{eq:mk2Ch}) and (\ref{eq:mk2FP}) actually correspond to a linear and a quadratic fit in $m_\ell$, respectively.

The data for the kaon mass have been corrected for FSE using ChPT formulae.
The absence of the chiral log at NLO makes the corresponding FSE correction (GL) vanishing identically, i.e.~$K_{M_K^2}^{FSE} = 1$.
The first non-vanishing correction appears at NNLO and it was calculated in Ref.~\cite{CDH05}.
The pion mass splitting is expected to give a contribution to the FSE also for the kaon mass.
However explicit calculations are not yet available\footnote{A first step in this direction has been done recently in Ref.~\cite{Bar:2014bda}, where however the framework differs by lattice artifacts from the non-unitary setup chosen in this work for valence and sea strange quarks.}.
In Table \ref{tab:confrFSE_MK} the relative size of the FSE correction for the kaon mass is presented, together with a comparison to the lattice data.
It can clearly be seen that: ~ i) FSE on the kaon mass are definitely smaller compared to the pion case (see Table \ref{tab:confrFSE_Mpi}), and ~ ii) even if the contribution from the pion mass splitting is neglected, the CDH predictions appear to work quite well, reproducing the observed ratio of lattice data.  

\begin{table}[hbt!]
\centering
\renewcommand{\arraystretch}{1.20} 
\begin{tabular}{||c|c|c||c||}
         \hline
         & GL & CDH & Lattice data $(M_{K,[32]}^2 / M_{K, [24]}^2)$ \\[1mm]
         \hline
         $K_{M_K^2, [32]}^{FSE} / K_{M_K^2, [24]}^{FSE}$ & $1$ & $0.982$ &$0.980(14)$ \\[1mm]
         \hline
\end{tabular}
\renewcommand{\arraystretch}{1.0}
\caption{\it Values of the ratio of the FSE correction factor $K_{M_K^2}^{FSE}$ in the case of the kaon mass for the ensembles A40.32 and A40.24, obtained within the approaches $GL$ and $CDH$ (see text), compared with the corresponding ratio of lattice data.} 
\label{tab:confrFSE_MK}
\end{table}

\begin{figure}[hbt!]
\begin{center}
\includegraphics[width=0.8\textwidth]{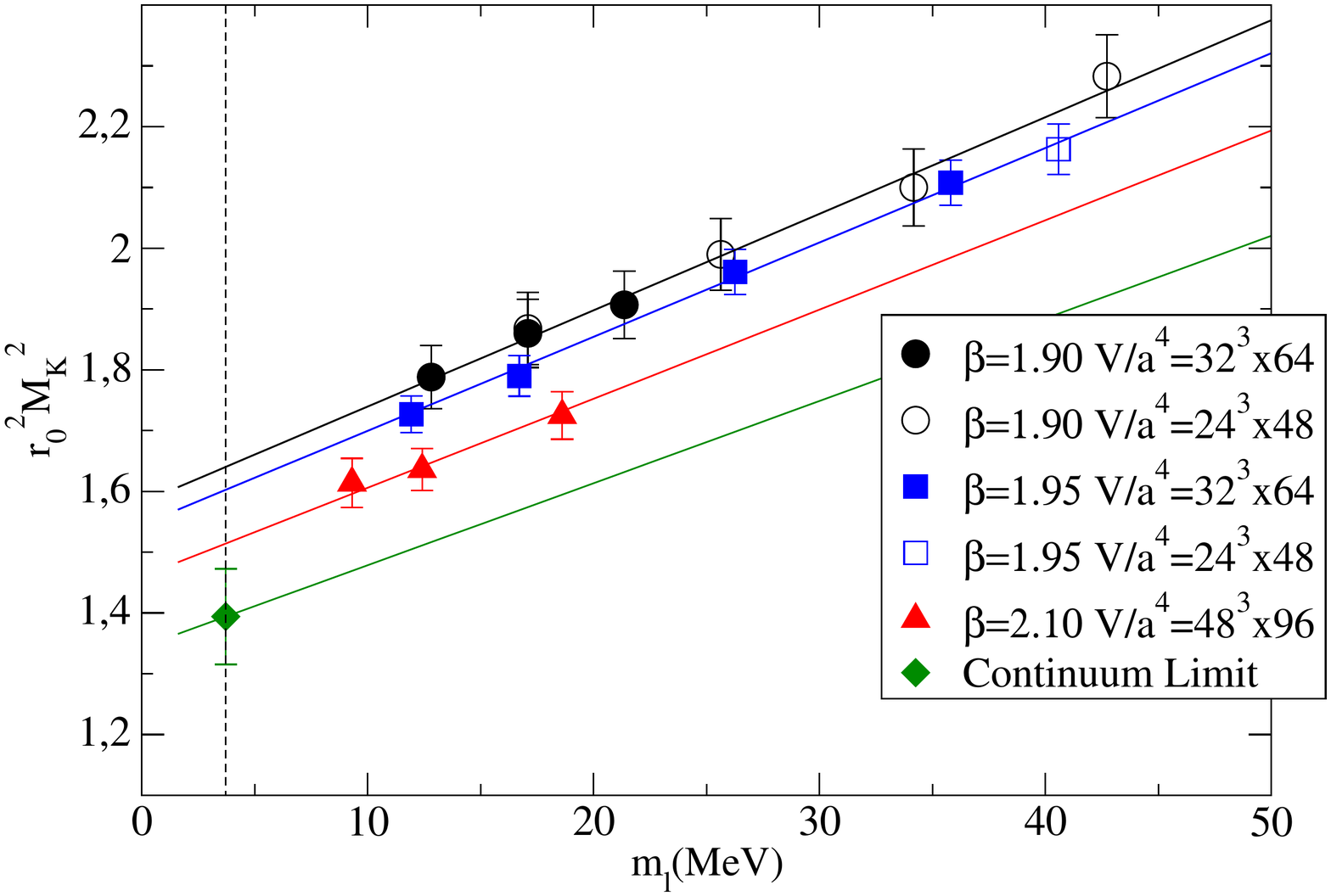}
\end{center}
\vspace*{-1.0cm}
\caption{\it Chiral and continuum extrapolation of $M_K^2$ in units of $r_0$ using the SU(2) ChPT predictions given by Eq.~(\ref{eq:mk2Ch}).}
\label{fig:mK2mlr0Ch}

\begin{center}
\includegraphics[width=0.8\textwidth]{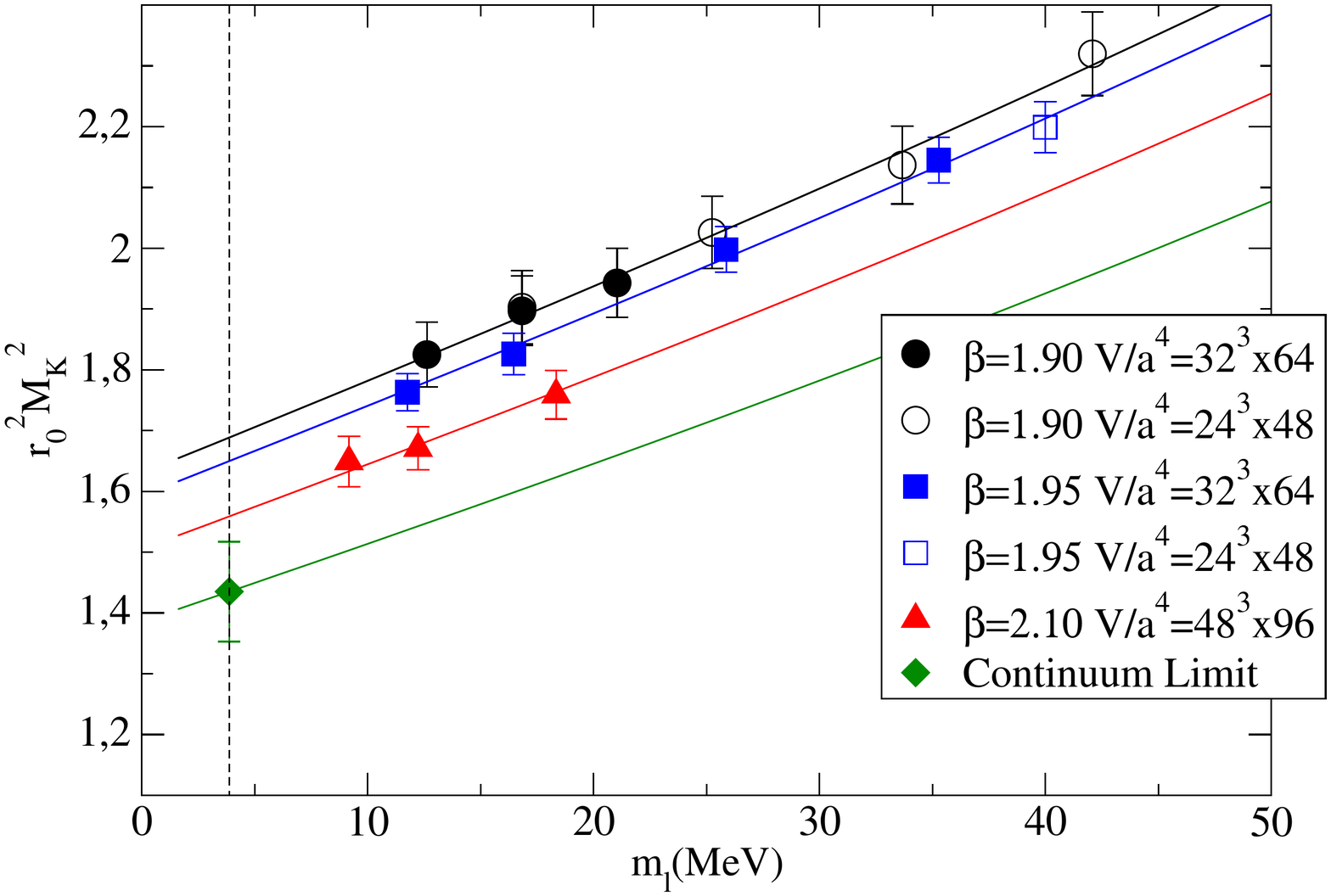}
\end{center}
\vspace*{-1.0cm}
\caption{\it The same as in Fig.~\ref{fig:mK2mlr0Ch}, but in the case of the polynomial fit (\ref{eq:mk2FP}).}
\label{fig:mK2mlr0FP}
\end{figure}

The dependence of $M_K^2$ on the renormalized light quark mass at each lattice spacing as well as its chiral and continuum extrapolation are shown in Figs.~\ref{fig:mK2mlr0Ch} and \ref{fig:mK2mlr0FP} in the cases of the SU(2) ChPT (\ref{eq:mk2Ch}) and polynomial (\ref{eq:mk2FP}) fits, respectively.
In what follows, the kaon data will be corrected for FSE using the CDH formulae \cite{CDH05} unless explicitly stated.

In both cases the lattice data are reproduced quite well by the fitting formulae.
Notice the size of discretization effects, which can be quantified at the level of $\simeq 10\%$ taking the difference between the results at the finest lattice spacing and the ones in the continuum limit.

\subsection{Chiral extrapolations in units of $M_{s^\prime s^\prime}$ (analyses C and D)}
\label{sec:strange_MSS}

Following the same strategy adopted in the pion analyses, the kaon masses simulated at different $\beta$ values can be brought to a common scale by constructing the ratios $M_K^2 / M_{s^\prime s^\prime}^2$, which are expected to suffer only marginally by discretization effects.
The values of $a M_{s^\prime s^\prime}$ for each $\beta$ are given in Eq.~(\ref{eq:aMssvalues}).
The light quark mass $m_\ell$ is expressed directly in physical units by using the values of the lattice spacing found in the corresponding pion analysis.

As for the analyses in units of $r_0$, we considered two different chiral extrapolations, adopting formulae similar to Eqs.~(\ref{eq:mk2Ch}) and (\ref{eq:mk2FP}), but expressed in units of $M_{s^\prime s^\prime}$.
After the chiral extrapolation and the continuum limit are carried out, the result for $M_K / M_{s^\prime s^\prime}$ can be combined with the value of $M_{s^\prime s^\prime}$ obtained in the corresponding pion analysis in order to compare with the experimental kaon mass (\ref{eq:MKexp}). 

\begin{figure}[hbt!] 
\begin{center}
\includegraphics[width=0.8\textwidth]{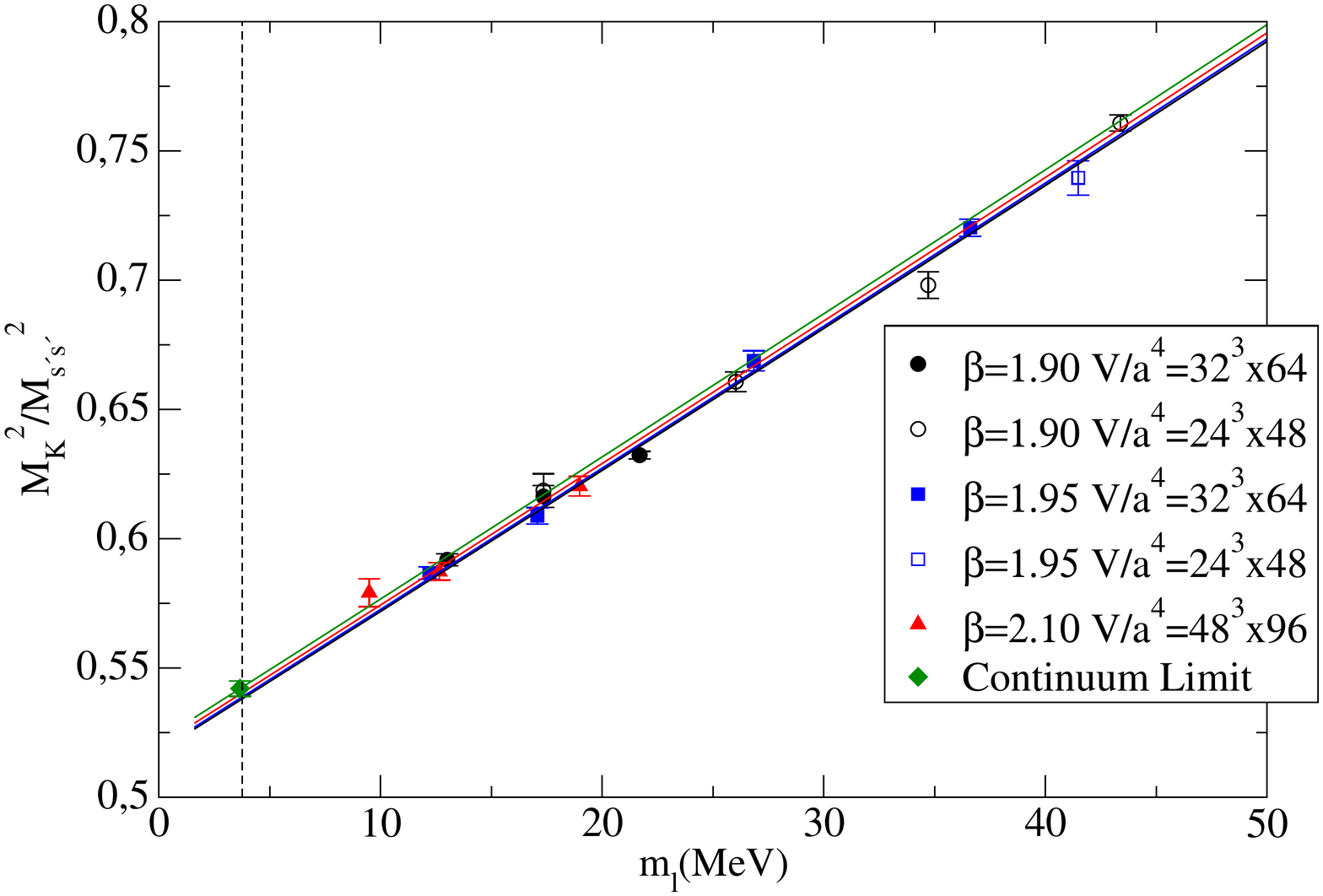}
\end{center}
\vspace*{-1.0cm}
\caption{\it Chiral and continuum extrapolation of $M_K^2$ in units of $M_{s^\prime s^\prime}^2$ using SU(2) ChPT at NLO.}
\label{fig:mK2mlMssCh}
\end{figure}

The dependencies of $M_K^2 / M_{s^\prime s^\prime}^2$ on the renormalized light quark mass at the three values of $\beta$ as well as in the continuum limit are shown in Fig.~\ref{fig:mK2mlMssCh} using the SU(2) ChPT prediction (analysis C).
Results of the same quality are obtained within the analysis D, which makes use of the polynomial fit for the chiral extrapolation.

In the case of the kaon mass the use of the hadron scale $M_{s^\prime s^\prime}$ turns out to be an extremely efficient choice for an almost total cancellation of the discretization effects, namely from $\simeq 10\%$ (see Figs.~\ref{fig:mK2mlr0Ch} and \ref{fig:mK2mlr0FP}) to about $0.4 \%$ (see Fig.~\ref{fig:mK2mlMssCh}).
This allows us to keep the extrapolation to the continuum limit under a very good control in the whole range of values of the renormalized light quark mass. 

\subsection{Results for the kaon sector}
\label{sec:kaon_results}

Our results for the strange quark mass $m_s$ are those reproducing after the chiral and continuum extrapolations the experimental value of the K-meson mass given in Eq.~(\ref{eq:MKexp}).
The results of the eight analyses for the strange quark mass, given in the $\overline{\rm MS}$ scheme at a renormalization scale of $2 \gev$, are shown in Table \ref{tab:msresult}.   

\begin{table}[hbt!]
\begin{center}
\small
\renewcommand{\arraystretch}{1.20}
\begin{tabular}{|c||c|c||c|c|}
  \hline\hline 
  \multicolumn{1}{|c||}{} & \multicolumn{2}{|c||}{$r_0$ Analysis} & \multicolumn{2}{|c|}{$M_{s^\prime s^\prime}$ Analysis} \\
  \hline 
  RCs $Z_P$ & ChPT Fit (A)& Polynomial Fit (B)& ChPT Fit (C)& Polynomial Fit  (D)\\
   \hline 
   method M1 & 101.6(4.4) & 102.5(3.9) & 99.4(2.9) & 100.8(3.2)  \\ 
   \hline 
   method M2 & ~99.0(4.4) & ~99.8(3.9) & 96.3(2.7) & ~97.6(3.0)  \\ 
   \hline 
\end{tabular}
\renewcommand{\arraystretch}{1.0}
\end{center}
\caption{\it Values of the strange quark mass $m_s$ obtained within the eight branches of the analysis. The results are given in {\rm MeV} in the $\overline{\rm MS}(2 \gev)$ scheme.}
\label{tab:msresult}
\end{table}

After combining these results using Eq.~(\ref{eq:combineresults}), we obtain our estimate of the strange quark mass $m_s$ and its systematic uncertainties in the ${\overline{\rm MS}}(2 \gev)$ scheme, namely
 \bea
    m_s & = & 99.6 ~ (3.6)_{stat + fit} (0.6)_{Chiral} (1.1)_{Disc} (1.4)_{Z_P} (0.5)_{FSE} (1.3)_{Pert} \mev \nn \\
            & = & 99.6 ~ (3.6)_{stat + fit} (2.3)_{syst} \mev \nn \\
            & = & 99.6 ~ (4.3) \mev ~ .
    \label{eq:msresults}
 \eea
The chiral extrapolation error has been evaluated from the spread among the results obtained using the chiral and the polynomial fits in units of either $r_0$ or $M_{s^\prime s^\prime}$. 
This corresponds in the error budget to a $0.6 \%$ systematic uncertainty.

The discretization error has been calculated from the spread among the results obtained in units of $r_0$ or $M_{s^\prime s^\prime}$ and represents a $1.1 \%$ uncertainty on $m_s$.

The two different sets of values of $Z_P$, calculated using the methods M1 and M2, introduce an additional uncertainty of $1.4 \%$.

The difference of the results for the strange quark mass obtained without correcting for the FSE and the one obtained using the CDH approach \cite{CDH05} has been conservatively taken as the estimate of the corresponding systematic uncertainty, which turns out to be equal to $0.5 \%$.

The last systematic error appearing in Eq.~(\ref{eq:msresults}) is the one related to the conversion between the RI$^\prime$-MOM and the $\overline{\rm MS}(2 \gev)$ schemes, estimated to be $\simeq 1.3 \%$ (see \ref{sec:A3}).

The largest uncertainty, equal to $3.6 \%$, comes from the statistical error plus the uncertainties due to the fitting procedure. 
The latter is the dominating one and it mainly depends on the distance between the lowest simulated quark mass and the physical point $m_{ud}$ in the chiral extrapolation.

Our determination (\ref{eq:msresults}) for $m_s$ is the first one obtained at $N_f = 2+1+1$. 
The recent lattice averages, provided by FLAG \cite{FLAG} and based on the findings of Refs.~\cite{Blossier:2010cr, Fritzsch:2012wq, Arthur:2012opa, Durr:2010vn, Durr:2010aw, Bazavov:2009fk}, are: $m_s = 101(3) \mev$ at $N_f = 2$ and $m_s = 93.8(2.4) \mev$ at $N_f = 2+1$.
The comparison of these results with our finding (\ref{eq:msresults}) shows that the partial quenching of the strange and/or charm sea quarks is not yet visible at the (few percent) level of the present total systematic uncertainty.

\subsection{The ratio $m_u / m_d$}
\label{sec:mumd}

The light quark mass dependence of the squared kaon mass can be used to calculate the mass difference between the $u$ and $d$ quark masses, leading to an estimate of the ratio $m_u / m_d$. 
In the limit of vanishing electromagnetic interactions the difference between the neutral and charged squared kaon masses can be expanded in terms of the (small) quark mass difference $(m_d - m_u)$ as (see Ref.~\cite{deDivitiis:2011eh} and references therein) 
 \be
    \hat{M}_{K^0}^2 - \hat{M}_{K^+}^2  = (m_d  - m_u) \cdot \left( \frac{\partial M_K^2}{\partial m_\ell} 
    \right)_{m_\ell = m_{ud}} + {\cal{O}}[(m_d - m_u)^2] ~ .
    \label{eq:slopeK}
 \ee
The slope $\left( \partial M_K^2 / \partial m_\ell \right)_{m_\ell = m_{ud}}$ is defined in the isospin symmetric limit and therefore it can be computed directly using our ensembles by taking the derivative of the continuum and infinite volume limits of our fitting formulae, like Eqs.~(\ref{eq:mk2Ch}-\ref{eq:mk2FP}), with respect to $m_\ell$, obtaining
 \bea
     \left( \partial M_K^2 / \partial m_\ell \right)_{m_\ell = m_{ud}} & = & 2.29 ~ (18)_{stat + fit} (17)_{Chiral} (8)_{Disc} (6)_{Z_P} (14)_{FSE} \gev \nn \\
                                                                                                     & = & 2.29 ~ (18)_{stat + fit} (24)_{syst} \gev \nn \\
                                                                                                     & = & 2.29 ~ (30) \gev
      \label{eq:IBslope}
 \eea
We observe that in Ref.~\cite{deDivitiis:2011eh}, using a different method based on the insertion of the isovector scalar density, the slope was found to be equal to $\left( \partial M_K^2 / \partial m_\ell \right)_{m_\ell = m_{ud}} = 2.57 (8) \gev$ at $N_f = 2$.

The charged and neutral kaon masses, $\hat{M}_{K^0}$ and $\hat{M}_{K^+}$, are those defined in pure QCD.
For them we adopt the FLAG estimate $\hat{M}_{K^+}  - \hat{M}_{K^0} = - 6.1 (4) \mev$ \cite{FLAG}, based on the findings of Refs.~\cite{Basak:2008na, Basak:2012zx, Basak:2013iw, Portelli:2010yn, Portelli:2012pn, deDivitiis:2013xla}, and the value $(\hat{M}_{K^+} + \hat{M}_{K^0}) / 2 = 494.2 (4) \mev$ given by Eq.~(\ref{eq:MKexp}).
From Eqs.~(\ref{eq:slopeK}-\ref{eq:IBslope}) we then evaluate $(m_d - m_u)$ and consequently the ratio $m_u  /m _d$ using Eq.~(\ref{eq:mudresults}) for the average value of the up and down quark masses. 
After implementing the above strategy for all the eight branches of the analysis we get the result
 \bea
    \frac{m_u }{m_d} & = & 0.470 ~ (41)_{stat + fit} (26)_{Chiral} (15)_{Disc} (1)_{Z_P} (23)_{FSE} \nn \\
                                & = & 0.470 ~ (41)_{stat + fit} (38)_{syst}\nn \\
                                & = & 0.470 ~ (56) ~ .
    \label{eq:mud}
 \eea
Our $N_f = 2 + 1 + 1$ result is consistent with the FLAG averages $m_u / m_d = 0.50 (4)$ at $N_f = 2$ and $m_u / m_d = 0.46 (3)$ at $N_f = 2+1$ \cite{FLAG}, 
based on the results of Refs.~\cite{deDivitiis:2013xla, Durr:2010vn, Durr:2010aw, Bazavov:2009fk}.

For the up and down quark masses in the ${\overline{\rm MS}}(2 \gev)$ scheme we get
 \bea
        \label{eq:umass}
       m_u & = & 2.36 ~ (20)_{stat + fit} (6)_{Chiral} (8)_{Disc} (3)_{Z_P} (9)_{FSE} (3)_{Pert} \mev \nn \\
               & = & 2.36 ~ (20)_{stat + fit} (14)_{syst} \mev \nn \\
               & = & 2.36 ~ (24) \mev ~ , \\[2mm]
        \label{eq:dmass}
       m_d & = & 5.03 ~ (16)_{stat + fit} (16)_{Chiral} (4)_{Disc} (8)_{Z_P} (7)_{FSE} (7)_{Pert} \mev \nn \\
               & = & 5.03 ~ (16)_{stat + fit} (21)_{syst} \mev \nn \\
               & = & 5.03 ~ (26) \mev ~ .
 \eea

\subsection{Determinations of the strange and charm sea quark masses}
\label{sec:mistuning}

As discussed in Section \ref{sec:simulations}, within the twisted mass formulation adopted in the present work the (renormalized) strange and charm sea quark mass are related to the bare twisted parameters $\mu_\sigma$ and $\mu_\delta$ by
\bea
       \label{eq:mssea}
       m_s^{sea}  = \frac{1}{Z_P} \left( \mu_\sigma - \frac{Z_P}{Z_S} \mu_\delta \right) ~ \\ 
       \label{eq:mcsea}
       m_c^{sea}  = \frac{1}{Z_P} \left( \mu_\sigma + \frac{Z_P}{Z_S} \mu_\delta \right) ~.
 \eea
Using the results found for the RCs $Z_P$ and $Z_P / Z_S$ (see \ref{sec:RCs} for the latter), it turns out that the values of $m_s^{sea}$ obtained from Eq.~(\ref{eq:mssea}) are plagued by large uncertainties that can reach the $20 \%$ level, mainly because of a large cancellation between the two terms in the r.h.s.~of Eq.~(\ref{eq:mssea}).
Moreover, the definition (\ref{eq:mssea}) is affected by the lattice artifacts that unavoidably enter the determination of the RCs.

A more accurate determination of $m_s^{sea}$ can be obtained using the results of Refs.~\cite{Baron:2010bv,Baron:2010th,Baron:2011sf,Ottnad:2012fv, Ottnad13}, where for all the ensembles used in the present work the kaon mass has been determined in the twisted-mass unitary setup, in which the valence quarks are described by the same action (\ref{eq:action_sc}) adopted for the sea quarks.

In terms of the valence ($m_\ell$ and $m_s$) and strange sea ($m_s^{sea}$) quark masses the OS kaon masses, computed in the present study, can be represented as $M_K^{OS} = M_K(m_\ell, m_s; m_s^{sea})$, while the unitary ones correspond to $M_K^{unitary} = M_K(m_\ell, m_s^{sea}; m_s^{sea})$ up to lattice artifacts that may be different in the two setups.
We have then computed for each ensemble the ratio of the unitary over OS values of the combination $2 M_K^2 - M_\pi^2$, namely
 \be
      R_{sea}(m_\ell, m_s, m_s^{sea}) \equiv \frac{ 2 M_K^2(m_\ell, m_s^{sea}; m_s^{sea}) -   M_\pi^2(m_\ell; m_s^{sea})}
                                                                      {2 M_K^2(m_\ell, m_s; m_s^{sea}) - M_\pi^2(m_\ell; m_s^{sea})} ~ . 
      \label{eq:Rmsea}
 \ee
This ratio is equal to the ratio $m_s^{sea} / m_s$ in ChPT at LO and it is equal to unity when $m_s^{sea} = m_s$ up to lattice artifacts corresponding to the difference of the discretization effects in the unitary and OS setups.
Therefore, for each ensemble a smooth local interpolation (carried out with quadratic splines) allows us to find the value of the valence strange quark mass $m_s$ that makes the ratio $R_{sea}(m_\ell, m_s, m_s^{sea})$ equal to unity.

\begin{figure}[hbt!] 
\begin{center}
\includegraphics[width=0.8\textwidth]{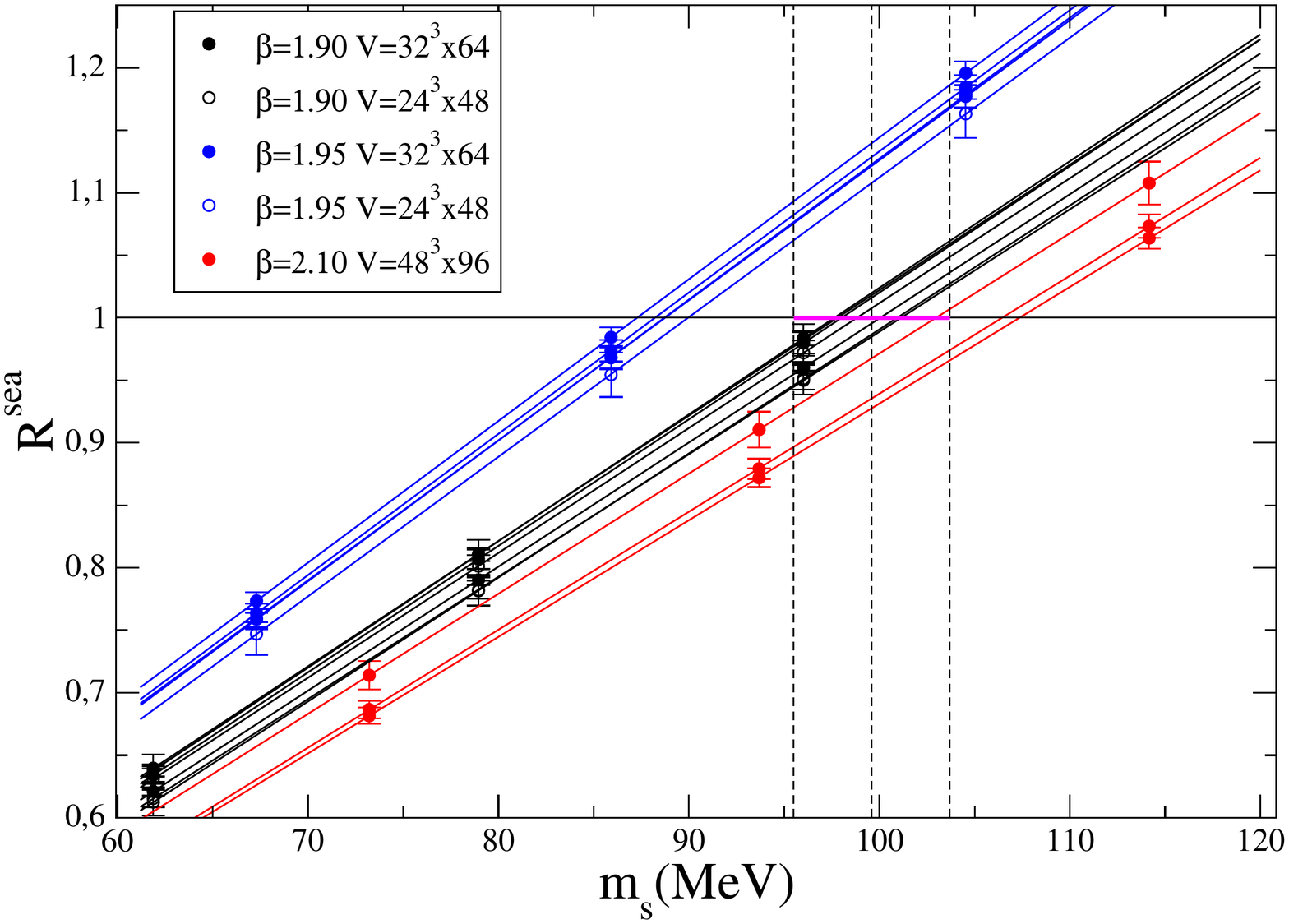}
\end{center}
\vspace*{-0.75cm}
\caption{\it The ratio $R_{sea}(m_\ell, m_s, m_s^{sea})$ for the various ensembles considered in this work versus the valence strange quark mass $m_s$. For each value of $\beta$ and $m_\ell$ the crossing of the interpolation curves of the lattice data with the solid line corresponding to $R_{sea} = 1$ identifies the location of the matching mass $m_s = m_s^{sea}$ up to lattice artifacts (see text). The vertical dashed lines correspond to the determination (\ref{eq:msresults}) of the physical strange quark mass.}
\label{fig:Rsea}
\end{figure}

The results of the above procedure are shown in Fig.~\ref{fig:Rsea}, where it can be seen that the matching mass can be determined with good precision and it is almost independent on the values of the light quark mass for fixed $\beta$.

In this way, using at each $\beta$ the weighted average of the matching masses obtained at the various values of the light-quark mass, we get the results
 \be
      \left. m_s^{sea} \right|_{\beta = 1.90, ~ 1.95, ~ 2.10}  = \{ 99.2 ~ (3.5), ~ 88.3 ~ (3.8),  ~ 106.4 ~ (4.6) \} ~ \mev ~ ,
     \label{eq:msearesults}
 \ee
where each error includes also the spread of the matching mass with respect to the light quark mass (see Fig.~\ref{fig:Rsea}).
The results (\ref{eq:msearesults}) differ from the determination (\ref{eq:msresults}) of the physical strange quark mass by $\approx 10 \%$ at most, with the largest difference at $\beta = 1.95$.

We tried to estimate the effect of the mistuning of the strange sea quark mass using the SU(3) ChPT predictions developed in Refs.~\cite{Bernard:1992mk}-\cite{Bijnens:2006jv} for arbitrary values of sea and valence quark masses.
For the squared pion and kaon masses one gets at NLO
 \bea
     \label{eq:MPi2_msea}
     \Delta M_\pi^2 & \equiv & M_\pi^2(m_\ell; m_s^{sea}) - M_\pi^2(m_\ell; m_s) \nonumber \\
                             & = & \frac{4 B_0 m_\ell}{f_0^2} \left\{ 8 \left[ 2 L_6^r(\mu) - L_4^r(\mu) \right]
                                        \left( \chi_s^{sea} - \chi_s \right) + \frac{1}{6} \overline{A}(\chi_\eta^{sea}) - \frac{1}{6} \overline{A}(\chi_\eta) \right\} ~ , \quad \\
      \label{eq:MK2_msea}
      \Delta M_K^2 & \equiv & M_K^2(m_\ell, m_s; m_s^{sea}) - M_K^2(m_\ell, m_s; m_s) \nonumber \\
                             & = & \frac{2 B_0}{f_0^2} (m_\ell + m_s) \left\{ 8 \left[ 2 L_6^r(\mu) - L_4^r(\mu) \right]
                                       \left( \chi_s^{sea} - \chi_s \right) \right. \nonumber \\ 
                             & - & \left. \frac{1}{3} \overline{A}(\chi_s) \frac{\chi_s - \chi_s^{sea}}{\chi_s - \chi_\eta^{sea}} - 
                                       \frac{1}{3} \overline{A}(\chi_\eta^{sea}) \frac{\chi_\eta^{sea} - \chi_s^{sea}}{\chi_\eta^{sea} - \chi_s} + 
                                       \frac{1}{3} \overline{A}(\chi_\eta) \right\} ~ , \quad
 \eea
where
 \bea
      \chi_\ell & \equiv & 2 B_0 m_\ell ~ , \nonumber \\
      \chi_s & \equiv & 2 B_0 m_s ~ , \qquad \qquad \qquad ~~ \chi_s^{sea} \equiv 2 B_0 m_s^{sea} ~ ,\nonumber \\
      \chi_\eta & \equiv & \frac{1}{3} \left( \chi_\ell + 2 \chi_s \right) ~ , \qquad  \qquad
      \chi_\eta^{sea} \equiv \frac{1}{3} \left( \chi_\ell + 2 \chi_s^{sea} \right) ~  , \nonumber \\
      \overline{A}(\chi) & \equiv & - \frac{\chi}{16 \pi^2} ~ \mbox{log}\left( \frac{\chi}{\mu^2} \right)
      \label{eq:defs}
 \eea
and $B_0$ and $f_0$ are the LO SU(3) LECs, while $L_4^r(\mu)$ and $L_6^r(\mu)$ are the NLO LECs evaluated at the renormalization scale $\mu$.
For the pion decay constant one gets 
 \bea
     \Delta f_\pi & \equiv & f_\pi(m_\ell; m_s^{sea}) - f_\pi(m_\ell; m_s) \nonumber \\
                       & = & \frac{2}{f_0} \left\{ 4 L_4^r(\mu) \left( \chi_s^{sea} - \chi_s \right) +
                                 \frac{1}{2} \overline{A}\left( \frac{\chi_\ell + \chi_s^{sea}}{2} \right) - 
                                 \frac{1}{2} \overline{A}\left( \frac{\chi_\ell + \chi_s}{2} \right) \right\} ~ . \quad 
     \label{eq:fPi_msea}
 \eea
Using from the results quoted in Ref.~\cite{FLAG} the values $B_0 / f_0 = 19 ~ (2) $ and 
 \bea
      2 L_6^r(\mu) - L_4^r(\mu) & = & 0.14 ~ (12) \cdot 10^{-3} ~ , \nn \\
      L_4^r(\mu) & = & 0.09 ~ (34)  \cdot 10^{-3}
      \label{eq:L6L4}
 \eea
at $\mu = M_\rho = 0.770$ GeV, the corrections (\ref{eq:MPi2_msea}), (\ref{eq:MK2_msea}) and (\ref{eq:fPi_msea}) are below the $1 \%$ level at our simulated quark masses and at the physical point.

We have also verified that by including the corrections (\ref{eq:MPi2_msea}), (\ref{eq:MK2_msea}) and (\ref{eq:fPi_msea}) in the lattice data the changes observed in the predictions of our analyses for $m_{ud}$ and $m_s$ are smaller than the other systematic uncertainties.

We close this subsection by presenting the estimate of the charm sea quark mass $m_c^{sea}$.
As in the case of the strange sea quark mass, $m_c^{sea}$ can be estimated either from Eq.~(\ref{eq:mcsea}), which requires the values of the RCs $Z_P$ and $Z_S$, or by investigating the matching between the unitary and OS determinations of the D-meson mass.
In both cases we got consistent results, namely $m_c^{sea}= \{ 1.21 ~ (5), ~ 1.21 ~ (5),  ~ 1.38 ~ (4) \} \gev$ at $\beta = \{1.90, ~ 1.95, ~ 2.10 \}$, which should be compared with the determination of the physical charm quark mass presented in Section \ref{sec:result_charm}.
In the $\overline{MS}(2 \gev)$ scheme the latter reads $m_c = 1.176 ~ (39) \gev$ [see Eq.~(\ref{eq:mcresults_2GeV})].
It follows that, while there is a good agreement within the errors at $\beta = 1.90$ and $1.95$, a $\approx 18 \%$ mistuning is present at $\beta= 2.10$.
Since scaling distortions are not visible in our data, we expect that the mistuning of the charm sea quark mass has a negligible effect with respect to the one of the strange sea quark and, therefore, it does not affect our determination of the quark masses in a significant way.

\subsection{Determination of the ratio $m_s / m_{ud}$}
\label{sec:ratio_sl}

The results for the strange quark mass $m_s$ and for the average up/down quark mass $m_{ud}$ (see Tables \ref{tab:allpionresult_M1}, \ref{tab:allpionresult_M2} and \ref{tab:msresult}) can be used to estimate the ratio $m_s / m_{ud}$.
One gets 
 \bea
    \frac{m_s }{m_{ud}} & = & 26.94 ~ (1.35)_{stat + fit} (0.30)_{Chiral} (0.13)_{Disc} (0.02)_{Z_P} (0.32)_{FSE} \nn \\
                                    & = & 26.94 ~ (1.35)_{stat + fit} (0.46)_{syst} \nn \\
                                    & = & 26.94 ~ (1.43) ~ 
    \label{eq:msmud}
 \eea
with a total uncertainty of $5.3 \%$.

In order to reduce the uncertainty we have investigated an alternative approach, which leads to a more precise determination of the ratio $m_s / m_{ud}$.

Using the kaon and pion lattice data we define the quantity $R(m_s, m_\ell, a^2)$ as
  \be
      R(m_s, m_\ell, a^2) \equiv \frac{m_\ell}{m_s}  \frac{2 M_K^2 - M_\pi^2}{M_\pi^2} ~ ,
      \label{eq:RK}
 \ee
which, by construction, is independent on the values of $Z_P$ as well as of the lattice spacing up to cutoff effects. 

In ChPT at LO the ratio $R(m_s, m_\ell, a^2)$ is equal to unity. 
At the physical point one gets $[(2 M_K^2 - M_\pi^2) / M_\pi^2]^{phys} \simeq 25.8$ and, adopting the estimate (\ref{eq:msmud}) for $(m_s / m_{ud})$, one has $R^{phys} \equiv R(m_s, m_{ud}, 0) \simeq 0.96$.
Therefore, the dependence of $R(m_s, m_\ell, a^2)$ on the strange and light quark masses is expected to give rise to small corrections only.
This is a very useful feature, since the mild dependence on the strange quark mass allow us to interpolate the ratio $R(m_s, m_\ell, a^2)$ at the physical value (\ref{eq:msresults}) with a small sensitivity to the error on $m_s$, while the mild dependence on the light quark mass $m_\ell$ represents a way to reduce the uncertainty introduced by the chiral extrapolation.
In this way a precise determination of the mass ratio $m_s / m_{ud}$ can be obtained as
 \be
      \frac{m_s}{m_{ud}} = \left( \frac{2 M_K^2 - M_\pi^2}{M_\pi^2} \right)^{phys} \frac{1}{R^{phys}} ~ ,
      \label{eq:final}
 \ee
where $R^{phys}$ is computed on the lattice.

\begin{figure}[hbt!] 
\begin{center}
\includegraphics[width=0.8\textwidth]{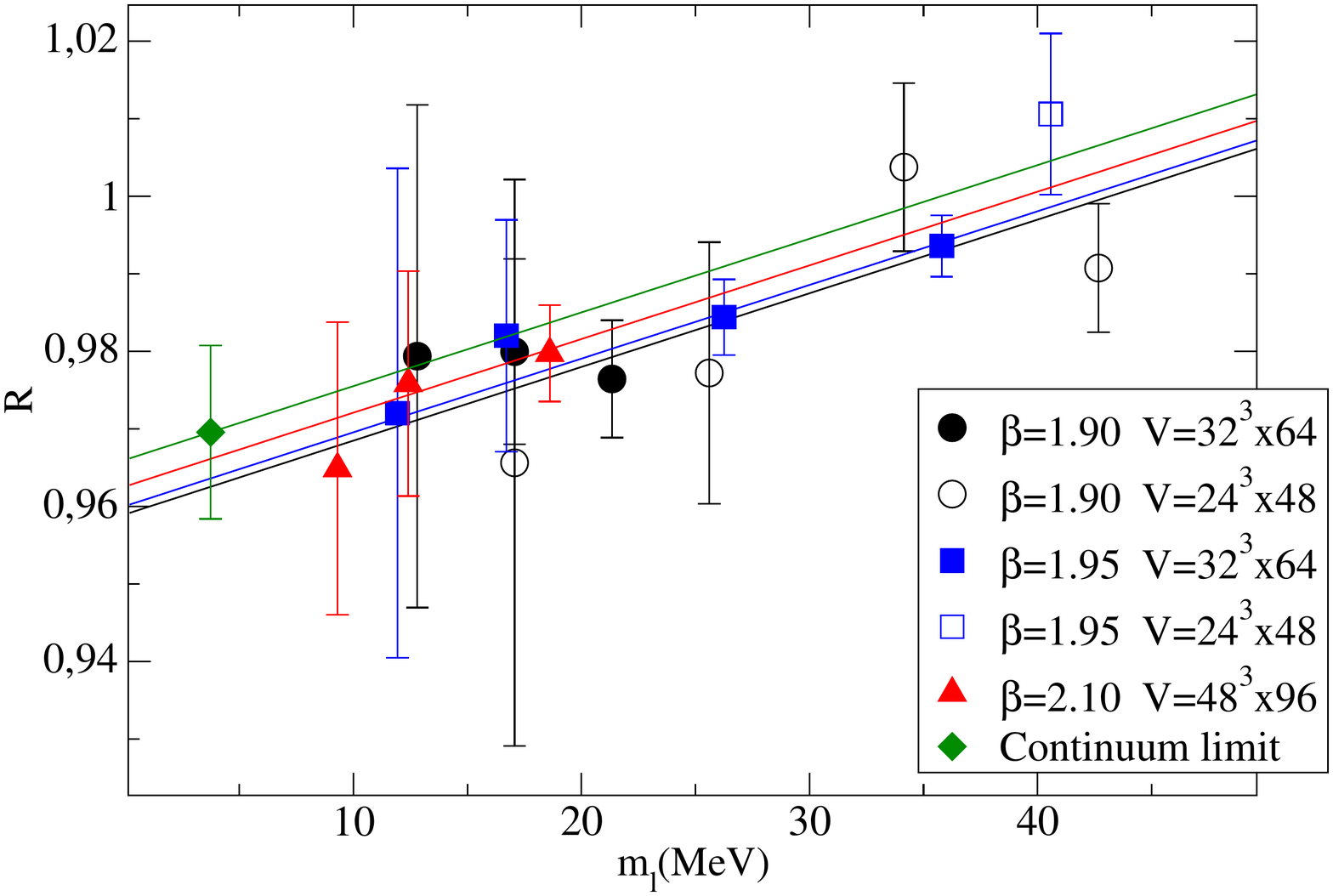}
\end{center}
\vspace*{-0.75cm}
\caption{\it Chiral and continuum extrapolations of the quantity $R(m_s, m_\ell, a^2)$, defined in Eq.~(\ref{eq:RK}), through a linear fit in $m_\ell$. The data are interpolated at the physical strange mass (\ref{eq:msresults}) and corrected for FSE.}
\label{fig:RKml_lin}
\end{figure}

In Fig.~\ref{fig:RKml_lin} the lattice data for $R(m_s, m_\ell, a^2)$, interpolated at the physical strange mass (\ref{eq:msresults}) and corrected for FSE (using the CWW predictions \cite{Colangelo:2010cu} for $M_\pi$ and the CDH ones \cite{CDH05} for $M_K$), are shown versus the light quark mass $m_\ell$ for our ensembles.
As expected, the dependence on the light quark mass is found to be quite mild and the ratio $R(m_s, m_\ell, a^2)$ is close to unity at all the simulated quark masses.

We performed the chiral and continuum extrapolations through a simple fit of the form
 \be
      R(m_s, m_\ell, a^2) = R_0 + R_1 m_\ell + R_3 a^2 ~ .
      \label{eq:RK_lin}
 \ee
The results are presented in Fig.~\ref{fig:RKml_lin} for each $\beta$ value and in the continuum limit.
It can be seen that discretization effects are quite small, being the difference between the result at the finest lattice spacing and the one in the continuum less than $\simeq 1 \%$.
From the result $R^{phys} = 0.9681 ~ (116)_{stat+fit} (7)_{Z_P} (3)_{FSE}$, obtained in the continuum limit and at the physical point, we get from Eq.~(\ref{eq:final}) the result
 \bea
      \frac{m_s}{m_{ud}} & = & 26.66 ~ (32)_{stat+fit} (2)_{Z_P} (1)_{FSE} \nn \\
                                     & = & 26.66 ~ (32) ~ ,
      \label{eq:final_ratio}
 \eea
which has an accuracy at the level of $1.2 \%$. 

For comparison, the updated FLAG averages \cite{FLAG} are $m_s / m_{ud} = 28.1 (1.2)$ at $N_f = 2$ and $m_s / m_{ud} = 27.5 (4)$ at $N_f = 2+1$, based on the findings of Refs.~\cite{Blossier:2010cr, Arthur:2012opa, Durr:2010vn, Durr:2010aw, Bazavov:2009fk}.

\section{Charm quark mass}
\label{sec:charm}

In this Section we present our determination of the mass of the charm quark obtained by analyzing both the $D$- and $D_s$-meson masses, following a strategy similar to the one presented for the $K$-meson.

The lattice data for the $D$- and $D_s$-meson masses are interpolated to the physical strange and charm quark masses using a quadratic spline.
The physical strange quark mass is the one determined in the previous Section, while the physical charm quark mass is defined such that the experimental value of the $D$- or $D_s$-meson mass is reproduced.
Then the dependence of $M_D$ and $M_{D_s}$ on the light quark mass and on the lattice spacing is studied at fixed strange and charm quark masses, and the continuum limit and the chiral extrapolation to the physical point $m_{ud}$ of the light quark mass are performed. 
The analysis based on the $D_s$-meson masses is expected to have smaller systematic uncertainty associated to the chiral extrapolation because of the milder light quark dependence, which occurs only through the sea effects.
Therefore, our final result for the charm quark mass is derived from the $D_s$-meson analysis and the value obtained from fitting the $D$-meson mass is used as a consistency check.

As in the cases of the pion and kaon analyses, the lattice data for the charmed meson masses are converted in units of either the Sommer parameter $r_0$ or the mass $M_{c^\prime s^\prime}$ of a fictitious PS meson, made with one valence strange-like and one valence charm-like quarks (with opposite values of the Wilson $r$-parameter). 
Such a reference mass $M_{c^\prime s^\prime}$, which is expected to have discretization effects close to the ones of $M_D$ or $M_{D_s}$, has been constructed choosing the arbitrary values $r_0 m_{s^\prime} = 0.22$ and $r_0 m_{c^\prime} = 2.4$ at each ensemble. 
As in the case of the mass $M_{s^\prime s^\prime}$, the continuum limit of $M_{c^\prime s^\prime}$ is required and it is calculated by combining the value of $f_\pi / M_{c^\prime s^\prime}$, calculated at the physical point, and the experimental value of $f_\pi$.
 
For the chiral extrapolation in the light quark mass, the Heavy Meson ChPT (HMChPT) predicts no chiral logarithms at NLO for both $D$- and $D_s$-meson masses. 
Therefore, we have adopted either a linear or a quadratic expansion in $m_\ell$ and the latter is used only to estimate the uncertainty related to the chiral extrapolation.


\subsection{Fit in units of $r_0$}
\label{sec:charm_r0}

Our analyses follow closely the strategy already applied to the kaon case. 
We start from an initial guess for the physical charm quark mass $m_c$ and consider the value of the physical strange quark mass $m_s$ given in (\ref{eq:msresults}). 
After a smooth interpolation in the strange and charm quark masses, the $D$- and $D_s$-meson masses, extracted from the corresponding 2-point correlators, are brought to a common scale using $r_0 / a$. 
The light quark mass is directly converted in physical units using the values of the lattice spacing obtained in the pion sector.

As discussed in the previous Section, the dependence of both $r_0 M_D$ and $r_0 M_{D_s}$ on the light quark mass $m_\ell$ is well described by a simple polynomial dependence, namely
 \bea
    \label{eq:Dmass}
    r_0 M_D & = & P_0  + P_1 m_\ell  + P_2 m_\ell^2  + P_3 a^2 ~ , \\
    \label{eq:Dsmass}
    r_0 M_{D_s} & = & P_0^\prime  + P_1^\prime m_\ell  + P_2^\prime m_\ell^2  + P_3^\prime a^2 ~ ,
 \eea
where $P_0$ - $P_3$ and $P_0^\prime$ - $P_3^\prime$ are free parameters.
For both $D$ and $D_s$ mesons we have investigated either a linear (i.e.~with $P_2 = P_2^\prime = 0$ in Eqs.~(\ref{eq:Dmass}-\ref{eq:Dsmass})) or a quadratic fit.

As in the previous analyses, the prior information on $Z_P$ and $r_0/a$ is introduced through the contribution to the $\chi^2$ given in Eq.~(\ref{eq:chi2term}).
Moreover, since the results for the $D$- and $D_s$-meson masses corresponding to the ensembles A40.24 and A40.32 (which differ only for the lattice volume) almost coincide, we did not apply any FSE correction.

\begin{figure}[hbt!]
\begin{center}
\includegraphics[width=0.8\textwidth]{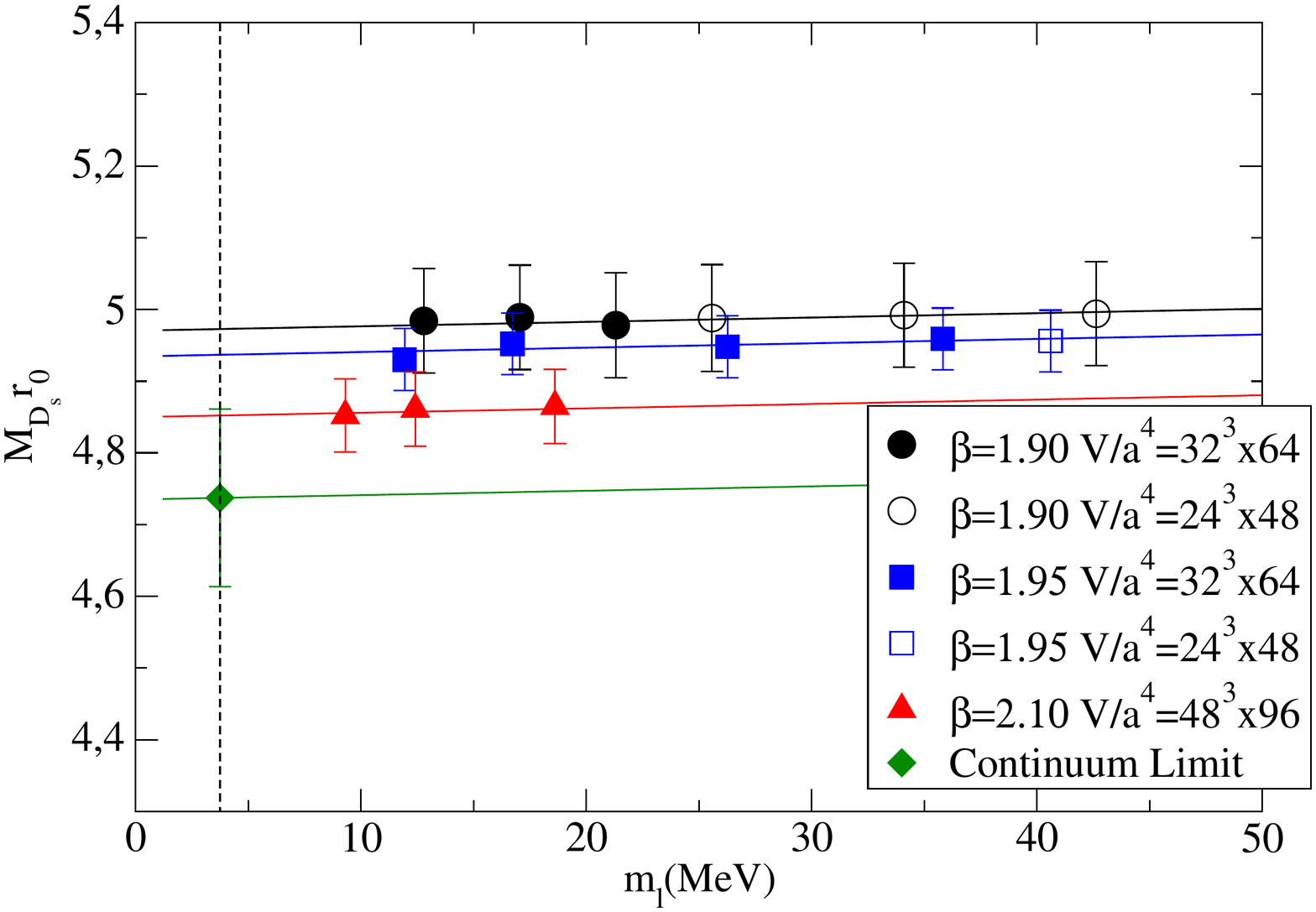}
\end{center}
\vspace*{-1.0cm}
\caption{\it Chiral and continuum extrapolation of $r_0 M_{D_s}$ adopting a linear fit in $m_\ell$, i.e.~with $P_2^\prime = 0$ in Eq.~(\ref{eq:Dsmass}).}
\label{fig:MDs_ml_r0_lin}

\begin{center}
\includegraphics[width=0.8\textwidth]{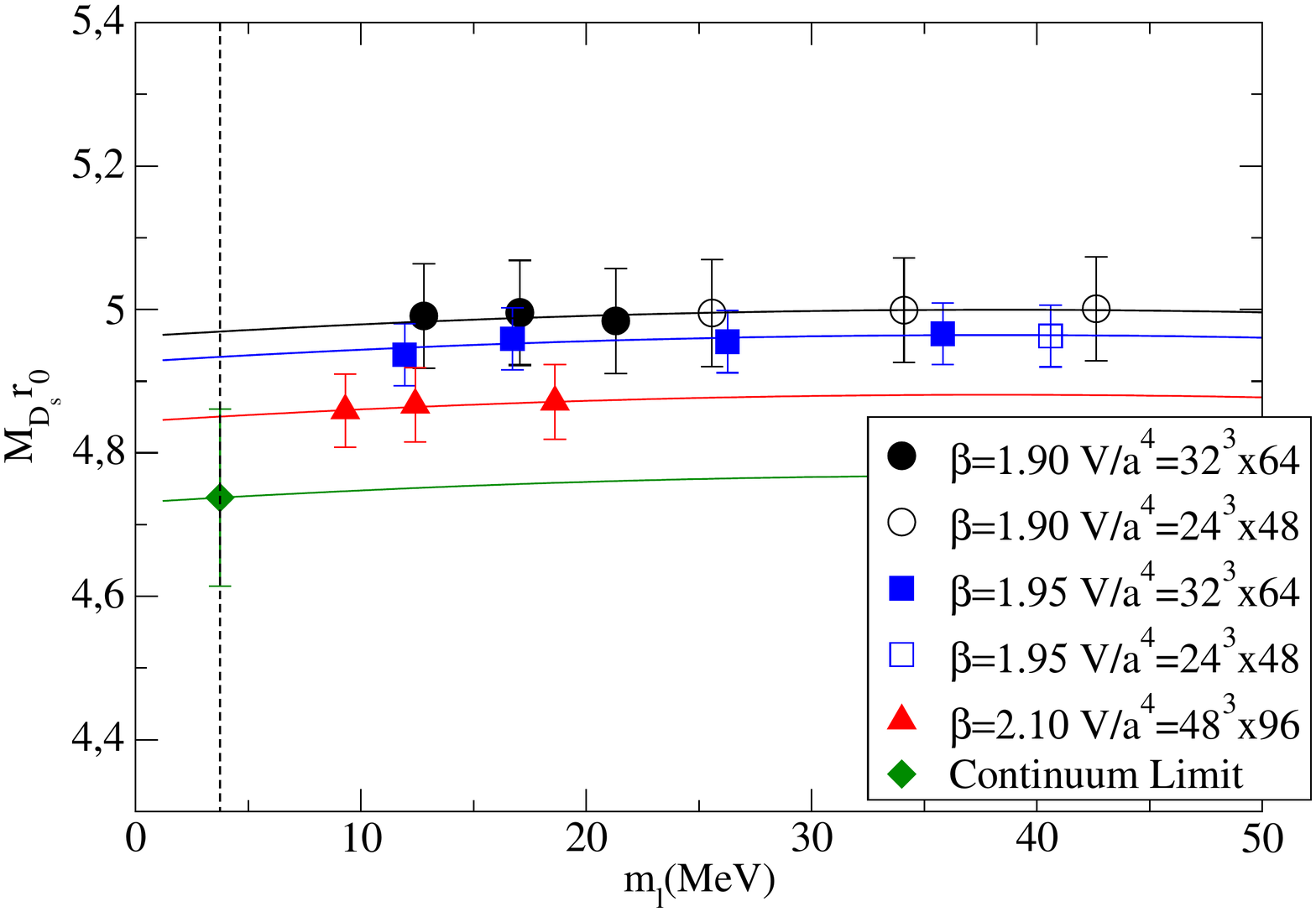}
\end{center}
\vspace*{-1.0cm}
\caption{\it The same as in Fig.~\ref{fig:MDs_ml_r0_lin}, but in case of the quadratic fit of Eq.~(\ref{eq:Dsmass}).}
\label{fig:MDs_ml_r0_quad}
\end{figure}

The dependence of $M_{D_s}$ on the light quark mass $m_\ell$ for each $\beta$ value and in the continuum limit is illustrated in Figs.~\ref{fig:MDs_ml_r0_lin}-\ref{fig:MDs_ml_r0_quad}, adopting a linear or a quadratic fit, respectively.
It can be seen that the discretization effects, which can be quantified by the difference between the results at the finest lattice spacing and those in the continuum limit, are found to be of the order of $3 \%$.

\subsection{Fit in units of $M_{c^\prime s^\prime}$}
\label{sec:charm_Mcs}

The impact of discretization effects can be reduced using the reference meson mass $M_{c^\prime s^\prime}$ as a scaling variable.
Let us divide $a M_{D_s}$ by the mass $a M_{c^\prime s^\prime}$ evaluated for each ensemble choosing the values $r_0 m_{s^\prime} = 0.22$ and $r_0 m_{c^\prime} = 2.4$ for the valence strange-like and charm-like quark masses, respectively.
As in the case of $a M_{s^\prime s^\prime}$ we found no significant dependence of $a M_{c^\prime s^\prime}$ on the light sea quark mass. 
Therefore we performed a constant fit in $a \mu_\ell$ to obtain the following reference values of $a M_{c^\prime s^\prime}$
 \bea
    \left. a M_{c^\prime s^\prime} \right|_{\beta = 1.90, ~ 1.95, ~ 2.10}  & = & \{ 0.8592(3), ~ 0.7681(4), ~ 0.5779(3) \} \quad \mbox{(method M1)} \nn \\
                                                                                                              & = & \{ 0.9009(3), ~ 0.7961(4), ~ 0.5963(3) \} \quad \mbox{(method M2)} ~ .
    \label{eq:aMcsvalues}
 \eea

\begin{figure}[hbt!]
\begin{center}
\includegraphics[width=0.8\textwidth]{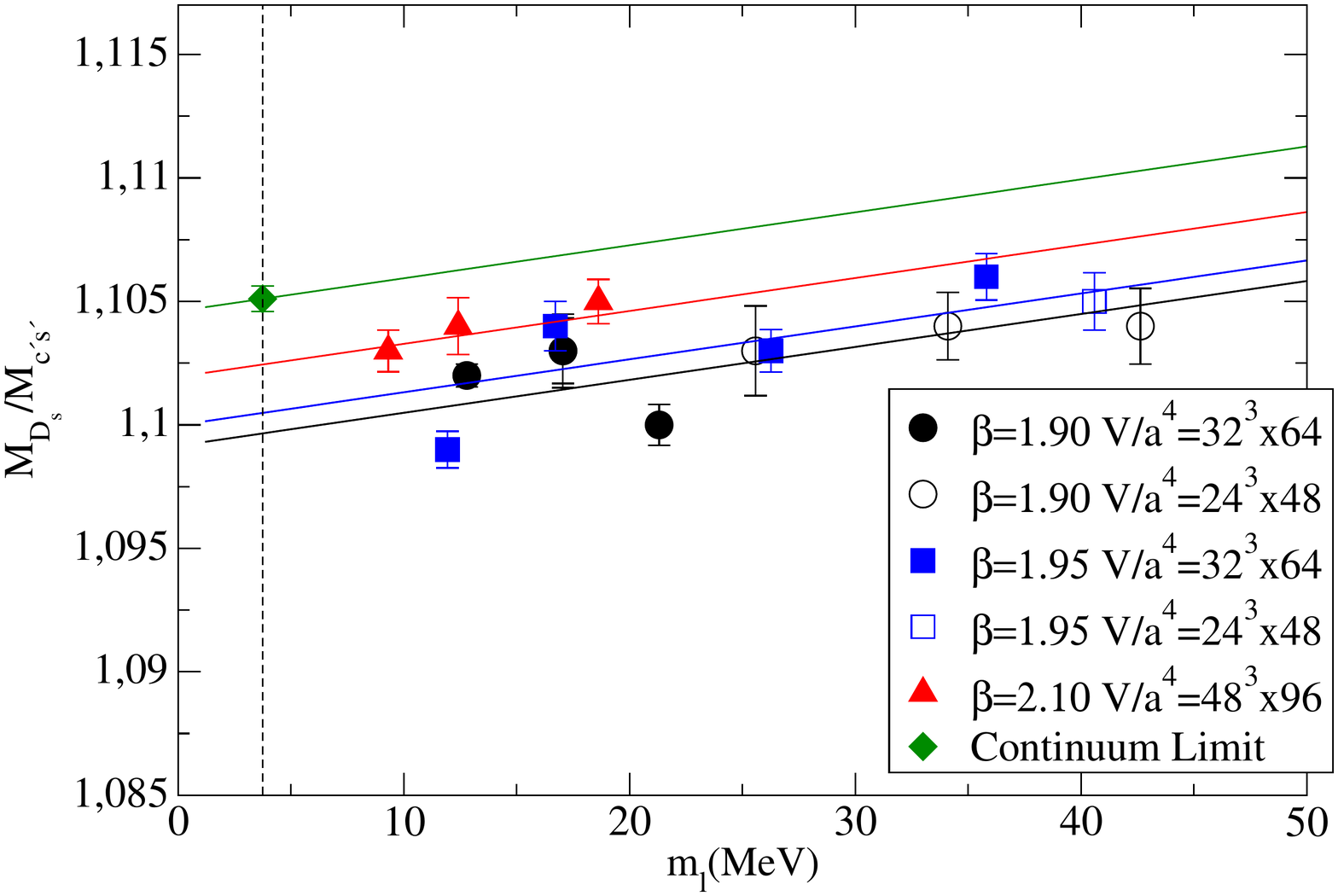}
\end{center}
\vspace*{-1.0cm}
\caption{\it Chiral and continuum extrapolations of $M_{D_s} / M_{c^\prime s^\prime}$ performing a linear fit in $m_\ell$, i.e.~with $\overline{P}_2 = 0$ in Eq.~(\ref{eq:DsmassMcs}).}
\label{fig:MDs_ml_Mcs_lin}

\begin{center}
\includegraphics[width=0.8\textwidth]{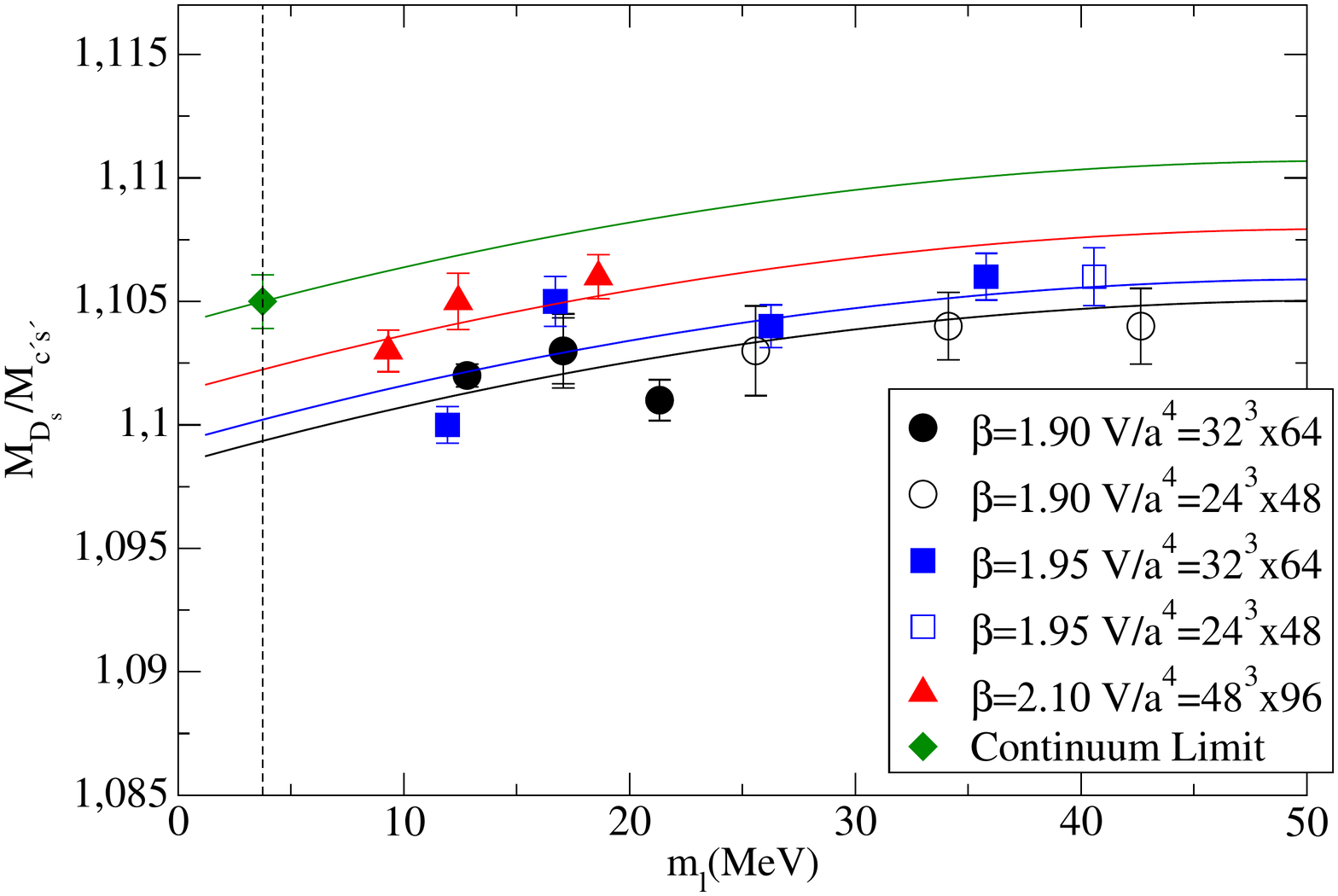}
\end{center}
\vspace*{-1.0cm}
\caption{\it The same as in Fig.~\ref{fig:MDs_ml_Mcs_lin}, but in case of the quadratic fit of Eq.~(\ref{eq:DsmassMcs}).}
\label{fig:MDs_ml_Mcs_quad}
\end{figure}

Then the chiral and continuum extrapolations of $M_{D_s} / M_{c^\prime s^\prime}$ is performed using the fitting formula  
 \be
    \frac{M_{D_s}}{M_{c^\prime s^\prime}} = \overline{P}_0  + \overline{P}_1 m_\ell  + \overline{P}_2 m_\ell^2  + \overline{P}_3 a^2  
    \label{eq:DsmassMcs}
 \ee
and similarly for $M_D / M_{c^\prime s^\prime}$
The dependence of the $D_s$-meson mass on the light quark mass at each $\beta$ and in the continuum limit, corresponding to a linear or a quadratic fit in Eq.~(\ref{eq:DsmassMcs}), are shown in Figs.~\ref{fig:MDs_ml_Mcs_lin} and \ref{fig:MDs_ml_Mcs_quad}, respectively.

The comparison of the results in units of $r_0$ presented in Figs.~\ref{fig:MDs_ml_r0_lin}-\ref{fig:MDs_ml_r0_quad} with those in units of $M_{c^\prime s^\prime}$ shown in Figs.~\ref{fig:MDs_ml_Mcs_lin}-\ref{fig:MDs_ml_Mcs_quad} indicates that discretization effects are strongly reduced in the ratio $M_{D_s} / M_{c^\prime s^\prime}$, as expected.
The gap between the continuum and the finest lattice spacing results decreases from $3 \%$ down to $0.3 \%$ of the continuum result.

\subsection{Results for the charm mass}
\label{sec:result_charm}

After the continuum limit and the extrapolation to the physical light quark mass $m_{ud}$ are performed, the masses of the $D$ and $D_s$ mesons are converted in physical units using the values of either $r_0$ or the continuum extrapolation of $M_{c^\prime s^\prime}$.
Then, by successive iterations the physical charm quark mass $m_c$ is determined by matching the mass of either the $D$- or the $D_s$-meson to the corresponding (isospin averaged) experimental values \cite{PDG}
 \be
    M_D^{exp} = \frac{M_{D^\pm} + M_{D^0}}{2} = 1.867 \gev ~ ,\qquad M_{D_s}^{exp} = M_{D_s^\pm} = 1.969 \gev ~ .
    \label{eq:mdexp}
 \ee

The results for the charm quark mass in the $\overline{\rm MS}(2 \gev)$ scheme, obtained form the $D_s$-meson analysis, are shown in Table \ref{tab:mcresults}.
Each entry in the Table is already the average value evaluated according to Eq.~(\ref{eq:combineresults}) of the results of analyses which differ only for the choice of the set of the input parameters: those coming from pion and kaon analysis A (chiral extrapolation) and B (polynomial extrapolation) when $r_0$ is used as scaling variable, and those coming from the analyses C and D when $M_{c^\prime s^\prime}$ is considered.

It is interesting to compare the results for $m_c$ obtained analyzing the $D_s$-meson mass with those obtained using the $D$-meson mass.
The latter are presented in Table \ref{tab:mcresultfromD}. 
It can be clearly seen that there is indeed a full compatibility, the differences being much smaller than the quoted uncertainties.

\begin{table}[hbt!]
\begin{center}
\renewcommand{\arraystretch}{1.20} 
\begin{tabular}{|c||c|c||c|c|}
  \hline\hline 
  \multicolumn{1}{|c||}{} & \multicolumn{2}{|c||}{Linear~Fit} & \multicolumn{2}{|c|}{Quadratic~Fit} \\
  \hline 
  RCs $Z_P$ & $r_0$ Analysis& $M_{c^\prime s^\prime}$ Analysis& $r_0$ Analysis& $M_{c^\prime s^\prime}$ Analysis\\
                      & (A and B) & (C and D) & (A and B) & (C and D) \\ 
  \hline 
  method M1 & 1.188(32) & 1.198(31) & 1.190(32) & 1.199(31)   \\ 
  \hline 
  method M2 & 1.154(32) & 1.163(31) & 1.157(32) & 1.164(31)   \\ 
  \hline 
\end{tabular}
\renewcommand{\arraystretch}{1.0}
\end{center}
\normalsize
\caption{\it Results for the physical charm quark mass $m_c$ obtained from the various analyses of the $D_s$-meson mass explained in the text. The results are expressed in $\gev$ in the $\overline{\rm MS}(2 \gev)$ scheme.} 
\label{tab:mcresults}

\begin{center}
\renewcommand{\arraystretch}{1.20} 
\begin{tabular}{|c||c|c||c|c|}
  \hline\hline 
  \multicolumn{1}{|c||}{} & \multicolumn{2}{|c||}{Linear~Fit} & \multicolumn{2}{|c|}{Quadratic~Fit} \\
  \hline 
  RCs $Z_P$ & $r_0$ Analysis & $M_{c^\prime s^\prime}$ Analysis & $r_0$ Analysis & $M_{c^\prime s^\prime}$ Analysis \\
                      & (A and B) & (C and D) & (A and B) & (C and D) \\ 
  \hline 
  method M1 & 1.178(35) & 1.179(31) & 1.190(37) & 1.190(32)   \\ 
  \hline 
  method M2 & 1.146(35) & 1.144(31) & 1.158(38) & 1.156(32)   \\ 
  \hline 
\end{tabular}
\renewcommand{\arraystretch}{1.0}
\end{center}
\caption{\it The same as in Table \ref{tab:mcresults}, but using the data for the $D$-meson mass.}
\label{tab:mcresultfromD}
\end{table}

\begin{figure}[hbt!]
\begin{center}
\includegraphics[width=0.8\textwidth]{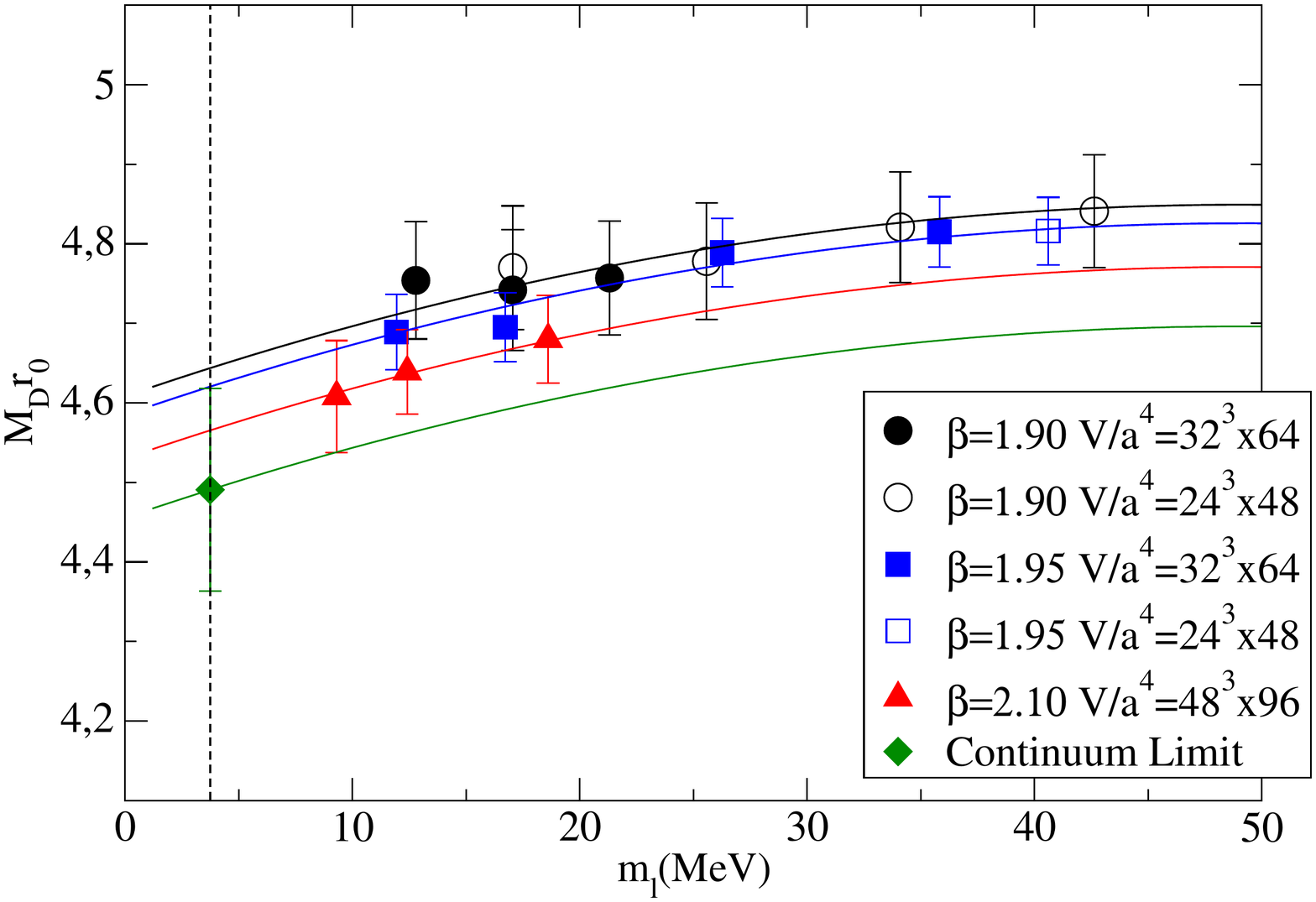}
\end{center}
\vspace*{-1.0cm}
\caption{\it Chiral and continuum extrapolations of $r_0 M_D$ performing a quadratic fit in $m_\ell$.}
\label{fig:MD_ml_r0_quad}
\end{figure}

The quality of the chiral and continuum extrapolation performed on the $D$-meson mass is illustrated in Fig.~\ref{fig:MD_ml_r0_quad} in the case of the quadratic fit in $m_\ell$. 

The results from Table \ref{tab:mcresults} corresponding to the linear fit in the light quark mass have been averaged to get our final result for $m_c$ and its systematic uncertainties in the ${\overline{\rm MS}}(2 \gev)$ scheme, namely
 \bea
      m_c & = & 1.176 ~ (31)_{stat + fit} (2)_{Chiral} (5)_{Disc} (17)_{Z_P} (15)_{Pert} \gev \nn \\
              & = & 1.176 ~ (31)_{stat + fit} (23)_{syst} \gev \nn \\
              & = & 1.176 ~ (39) \gev
     \label{eq:mcresults_2GeV}
 \eea
The results of the quadratic fit in $m_\ell$ have not been included in the average, but they have been considered to estimate the uncertainty related to the chiral extrapolation by taking the difference with the results of the linear fit. 
This error is found to be quite small as expected, since the light quark mass dependence of the $D_s$-meson mass arises only from sea quark effects.

After evolving the perturbative scale from $2 \gev$ to the value of $m_c$ using $\rm N^3LO$ perturbation theory with four quark flavors, one obtains
 \bea
      m_c (m_c ) & = & 1.348 ~ (36)_{stat + fit} (2)_{Chiral} (6)_{Disc} (20)_{Z_P} (19)_{Pert} \gev \nn \\
                         & = & 1.348 ~ (36)_{stat + fit} (28)_{syst} \gev \nn \\
                         & = & 1.348 ~ (46) \gev
     \label{eq:mcresults}
 \eea
with a total uncertainty equal to $3.4 \%$ of the central value.

The strategy followed to separate the various sources of the systematic error is the same as the one used in the pion and kaon cases. 

The first error in Eq.~(\ref{eq:mcresults}) includes the statistical uncertainties combined with the systematic error associated with the fitting procedure, the physical strange quark mass and the scale setting. 
This error is the dominant one and corresponds to a $2.7 \%$ of the central value.

The systematic uncertainty due the chiral extrapolation, estimated from the difference between the results of the linear and quadratic fits, is equal to $0.15 \%$.

The difference among the results obtained using $r_0$ or $M_{c^\prime s^\prime}$ is used to estimate the uncertainty coming from the discretization effects, which results to be of the order of $0.45 \%$.

The effect of choosing the values of $Z_P$ obtained either from the method M1 or M2 gives rise to a systematic uncertainty of $1.5 \%$.

Finally the uncertainty related to the conversion between the RI$^\prime$-MOM and the $\overline{MS}(m_c)$ schemes is estimated to be of the order of $1.4 \%$ (see \ref{sec:A3}).

Our determination (\ref{eq:mcresults}) for $m_c(m_c)$ is the first one obtained at $N_f = 2 + 1 + 1$ and it is consistent with the result $m_c (m_c) = 1.28 (4) \gev$ obtained in Ref.~\cite{Blossier:2010cr} at $N_f = 2$, with the finding $m_c (m_c) = 1.273 (6) \gev$ of Ref.~\cite{McNeile:2010ji} at $N_f = 2 + 1$ as well as with the PDG value $m_c (m_c) = 1.275 (25) \gev$ \cite{PDG}.

\subsection{Determination of the ratio $m_c / m_s$}
\label{sec:ratio_cs}

The results for the strange and charm quark masses given in Tables \ref{tab:msresult} and \ref{tab:mcresults} can be used to evaluate the mass ratio $m_c / m_s$. 
One obtains 
 \bea
    \frac{m_c}{m_s} & = & 11.80 ~ (51)_{stat + fit} (7)_{Chiral} (18)_{Disc} (11)_{Z_P} (6)_{FSE} \nn \\
                              & = & 11.80 ~ (51)_{stat + fit} (23)_{syst} \nn \\
                              & = & 11.80 ~ (56)
    \label{eq:ratiomcms}
 \eea
with a total uncertainty of $4.7 \%$.

In order to improve the precision of this determination we followed an approach similar to the one used in the case of the mass ratio $m_s / m_{ud}$ discussed in Section \ref{sec:ratio_sl}.
Using the lattice data for the masses of the $\eta_c$ and $D_s$ mesons, we define the quantity $\overline{R}(m_c, m_s, m_\ell, a^2)$ as
  \be
      \overline{R}(m_c, m_s, m_\ell, a^2) \equiv \frac{m_s}{m_c}  \frac{(M_{\eta_c} - M_{D_s})(2M_{D_s} - M_{\eta_c})}{2 M_K^2 - M_\pi^2} ~ ,
      \label{eq:Rbar}
 \ee
which by construction is independent of the values of $Z_P$ and of the lattice spacing up to cutoff effects. 
In Eq.~(\ref{eq:Rbar}) the mass $M_{\eta_c}$ of the $\eta_c$ meson corresponds to the (fermionic) connected diagram only, or in other words it is the mass of a fictitious $\overline{c} \tilde{c}$ PS meson with $m_{\tilde{c}} = m_c$ and $r_{\tilde{c}} = - r_c$.

Let us explain the choice of the ratio (\ref{eq:Rbar}).
For a PS meson made of two valence quarks with (renormalized) masses $m_1$ and $m_2$, in which one of the two quarks is around the charm mass, the meson mass $M_{12}$ can be written up to cutoff effects as
 \be
     M_{12} \equiv \overline{A} + \overline{B} (m_1 + m_2) \left[ 1 + \overline{r}(m_1, m_2) \right] ~ ,
     \label{eq:M12}
 \ee
where the function $\overline{r}(m_1, m_2)$ includes higher order contributions in the quark masses.
Therefore, up to cutoff effects, the ratio $\overline{R}(m_c, m_s, m_\ell, a^2)$ has a leading term, which is a constant and receives corrections only from terms in $m_c$ appearing in Eq.~(\ref{eq:M12}) at orders higher than the linear one\footnote{Of course alternative definitions of the ratio $\overline{R}(m_c, m_s, m_\ell, a^2)$ are possible, like for instance $\overline{R}(m_c, m_s, m_\ell, a^2) = (m_s / m_c) (M_{\eta_c} - M_D) / (M_{D_s} - M_D)$. However, the latter definition suffers from much larger statistical errors with respect to the one considered in Eq.~(\ref{eq:Rbar}).} .

As in the case of the ratio $R(m_s, m_\ell, a^2)$ defined in Section \ref{sec:ratio_sl}, the useful features of $\overline{R}(m_c, m_s, m_\ell, a^2)$ are that: ~ i) its interpolation at the physical charm and strange quark masses is only slightly sensitive to the uncertainties on $m_c$ and $m_s$, and ~ ii) its dependence on the light quark mass $m_\ell$ is expected to be mild, so that the uncertainty introduced by the chiral extrapolation is largely reduced.
A determination of the mass ratio $m_c / m_s$ is then obtained from
 \be
      \frac{m_c}{m_s} = \left[ \frac{(M_{\eta_c} - M_{D_s})(2M_{D_s} - M_{\eta_c})}{2M_K^2 - M_\pi^2} \right]^{phys} \frac{1}{\overline{R}^{phys}} ~ ,
      \label{eq:final_cs}
 \ee
where $\overline{R}^{phys} \equiv \overline{R}(m_c, m_s, m_{ud}, 0)$ is computed from lattice data.

\begin{figure}[hbt!] 
\begin{center}
\includegraphics[width=0.8\textwidth]{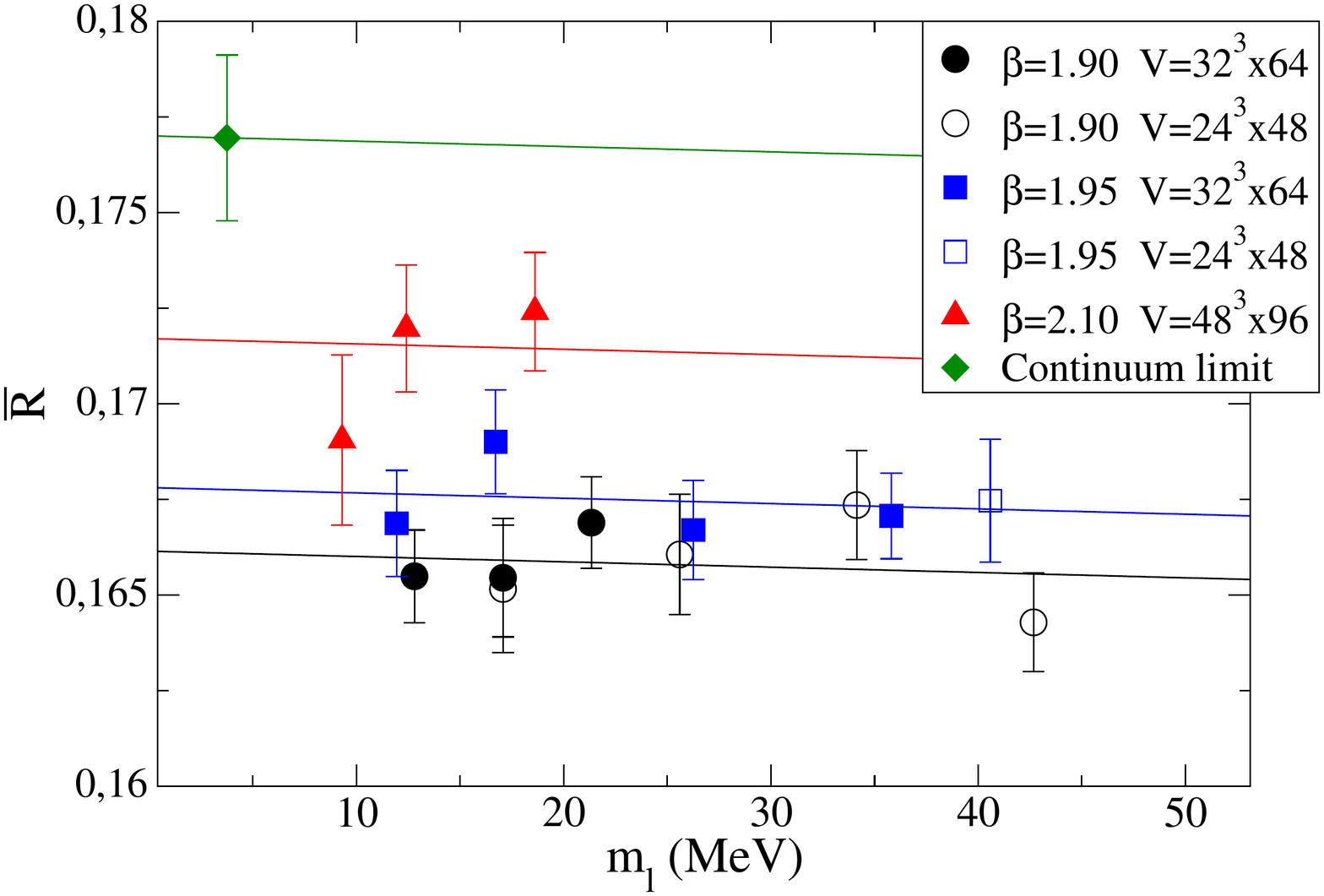}
\end{center}
\vspace*{-0.75cm}
\caption{\it Chiral and continuum extrapolations of $\overline{R}(m_c, m_s, m_\ell, a^2)$, defined in Eq.~(\ref{eq:Rbar}), using a linear fit in $m_\ell$. The data are interpolated at the physical strange and charm quark masses.}
\label{fig:RDml_lin}
\end{figure}

In Fig.~\ref{fig:RDml_lin} the lattice data for $\overline{R}(m_c, m_s, m_\ell, a^2)$, interpolated at the physical strange [Eq.~(\ref{eq:msresults})] and charm [Eq.~(\ref{eq:mcresults})] quark masses, are shown versus the light quark mass $m_\ell$ for all the ensembles.

The chiral and continuum extrapolations are performed through a simple linear fit of the form
 \be
      \overline{R}(m_c, m_s, m_\ell, a^2) = \overline{R}_0 + \overline{R}_1 m_\ell + \overline{R}_3 a^2
      \label{eq:RD_lin}
 \ee
and the results are shown in Fig.~\ref{fig:RDml_lin} as the solid lines at each $\beta$ value and in the continuum.
It can be seen that the dependence on the light quark mass is very mild, allowing to get a precise chiral extrapolation to the physical point, namely $\overline{R}^{phys} = 0.1772 ~ (24)_{stat+fit} (2)_{Z_P}$.

From the PDG \cite{PDG} one gets: $M_{\eta_c} = 2.9837 (7)$ GeV and $M_{D_s^\pm} = 1.9690 (14)$ GeV. 
The disconnected contribution to the physical $\eta_c$ meson, which is neglected in the present calculation, can be estimated from the annihilation rate into gluons, leading to an estimate of $\simeq 2.5$ MeV (see Ref.~\cite{Davies:2009ih} and references therein).
Assuming a $50 \%$ error on the latter, the ``connected" $\eta_c$ mass to be used in Eq.~(\ref{eq:final_cs}) is equal to $2.981 (1)$ GeV.
Thus, for the mass ratio $m_c / m_s$ we obtain the result
 \bea
      \frac{m_c}{m_s} & = & 11.62 ~ (16)_{stat+fit} (1)_{Z_P} \nn \\ 
                                & = & 11.62 ~ (16)
      \label{eq:mcms}
 \eea
with an error of $1.4 \%$. 
Note that the systematic uncertainty related to the FSE has not been reported in Eq.~(\ref{eq:mcms}), since it was found to be much less than all the other uncertainties. 

For comparison recent results for the ratio $m_c / m_s$ are: $m_c / m_s = 12.0 (3)$ \cite{Blossier:2010cr} and $m_c / m_s = 11.27 (40)$ \cite{Durr:2011ed} at $N_f = 2$, and $m_c / m_s = 11.85 (16)$ \cite{Davies:2009ih} at $N_f = 2 + 1$.

\section{Conclusions}
\label{sec:conclusions}

We have presented results for the up, down, strange and charm quark masses, obtained with $N_f = 2 + 1 + 1 $ twisted-mass Wilson fermions.
We have used the gauge configurations produced by the ETMC, which include in the sea, besides two light mass degenerate quarks, also  the strange and the charm quarks with masses close to their physical values. 
Such a setup is the closest one to the real world, adopted till now only by the ETM \cite{Baron:2010bv, Baron:2010th, Baron:2011sf, Ottnad:2012fv} and the MILC \cite{Bazavov:2010ru} Collaborations.

The analysis includes data at three values of the lattice spacing and pion masses in the range $210 \div 450$ MeV, allowing for accurate continuum limit and controlled chiral extrapolation.
In order to estimate the systematic error associated  with the chiral extrapolation we studied the dependence on the light quark mass by using different fitting formulae based either on the predictions of ChPT or on polynomial expressions. 

As for the continuum limit, in order to lower as much as possible the impact of discretization effects and to keep the continuum extrapolation under control we investigated two different procedures, which both use $f_\pi$ to set the scale.
The first one involves the Sommer parameter $r_0$ as the intermediate scaling variable, while in the second one we used the mass of a fictitious pseudoscalar meson made of two strange-like quarks (or a strange-like and a charm-like quark), $M_{s^\prime s^\prime}$ (or $M_{c^\prime s^\prime}$), trying to exploit cancellation of discretization effects in ratios like $M_K / M_{s^\prime s^\prime}$ (or $M_{D_s} / M_{c^\prime s^\prime}$).
For the kaon and $D_s$(D) meson masses these ratios really lead to a significant reduction of discretization effects. 

To account for FSE we used the resummed asymptotic formulae developed in Ref.~\cite{Colangelo:2010cu} for the pion sector, which include the effects due to the neutral and charged pion mass splitting (present in the twisted mass formulation), and the formulae of Ref.~\cite{CDH05} for the kaon sector.
We checked the accuracy of these predictions for FSE on the lattice data obtained at fixed quark masses and lattice spacings, but different lattice sizes.

The quark mass renormalization has been carried out non-perturbatively using the RI$^\prime$-MOM method, adopting dedicated ensembles of gauge configurations produced by ETMC with $N_f = 4$ degenerate flavors of sea quarks.

The main results obtained in this paper for the up, down, strange and charm quark masses and for some important quark mass ratios have been collected in Section \ref{sec:intro}, see Eqs.~(\ref{eq:qmasses})-(\ref{eq:R&Q}).

\section*{Acknowledgements}
We warmly thank our colleagues of the ETM Collaboration for fruitful discussions.

\noindent We acknowledge the CPU time provided by the PRACE Research Infrastructure under the project PRA027 ``QCD Simulations for Flavor Physics in the Standard Model and Beyond'' on the JUGENE BG/P system at the J\"ulich SuperComputing Center (Germany), and by the agreement between INFN and CINECA under the specific initiative INFN-RM123 on the Fermi BG/P - BG/Q system at CINECA (Italy).

\noindent We thank the AuroraScience collaboration for providing us access to the Aurora system in FBK, Trento.

\noindent G.~H.~acknowledges the support by DFG (SFB 1044).

\noindent D.~P.~acknowledges partial support by the Helmholtz International Center for FAIR within the framework of the LOEWE program launched by the State of Hesse.

\noindent S.~R.~thanks the Donald Smits Center for Information Technology of the University of Groningen.

\noindent V.~L., S.~S.~and C.~T.~thank MIUR (Italy) for partial support under Contract No. PRIN 2010-2011.

\noindent R.~F.~and G.C.~R.~thank MIUR (Italy) for partial support under Contract No. PRIN 2009-2010. 

\noindent This work is supported in part by the DFG and the NSFC through funds provided to the Sino-German CRC 110.

\appendix
\section{Renormalization constants}
\label{sec:RCs}

In order to obtain results for the quark mass $m_f$ ($f = u, d, s, c$) in the $\overline{\rm MS}$ scheme at a given renormalization scale, chosen to be $2$ GeV in the present study, a necessary step is the evaluation of the quark mass renormalization constant (RC) in a suitable intermediate lattice renormalization scheme, which here we take to be the RI$^\prime$-MOM scheme~\cite{RIMOM}.

In the lattice framework employed in the present paper, which technically is a mixed action setup based on twisted mass Wilson fermions (see Section \ref{sec:simulations}), what we really need is the renormalization constant of the valence quark mass $\mu_f$ appearing in the valence fermion action (\ref{eq:OS}).
As discussed in Refs.~\cite{Frezzotti:2004wz,Frezzotti:2003ni,Frezzotti:2000nk}, such a renormalization constant, $Z_\mu$, is independent of the flavor $f$ and the sign of $r_f$ in Eq.~(\ref{eq:OS}), as well as of all the sea Wilson parameters. 
The RC $Z_\mu$ can be conveniently chosen and evaluated as
\be
    Z_\mu = \frac{1}{Z_P} ~ ,
\ee
i.e.~as the inverse of the renormalization constant $Z_P$ of the pseudoscalar, flavor non-singlet density $P_{ff^\prime} = \bar q_f \gamma_5 q_{f^\prime}$, where $q_f$ and $q_{f^\prime}$ are two distinct valence flavors of maximally twisted Wilson fermions with action as in Eq.~(\ref{eq:OS}) and $r_{f^\prime}=-r_f$. 

Since RI$^\prime$-MOM is a mass-independent scheme \cite{Weinberg73}, the RCs of operators with non-vanishing anomalous dimension must be defined in the massless limit of the UV-regulated theory, i.e.~QCD  with $N_f = 4$ massless quark flavors. 
For this purpose the ensembles with fixed (non small) strange and charm sea quark masses, summarized in Table \ref{tab:masses&simudetails} and employed to compute physical observables (the so-called ``production ensembles''), are not well suited. 
Rather one needs to produce dedicated ensembles with $N_f = 4$ ``moderately light'' and, for simplicity, degenerate dynamical quarks in a lattice setup whose chiral limit coincides with the one of the lattice formulation chosen for the ``production ensembles''. 
Doing so for a sequence of progressively smaller dynamical quark mass values allows for a controlled extrapolation of massive RC-estimators to the desired chiral limit.

With an eye to Section \ref{sec:simulations}, a moment of thought reveals that in the chiral limit the relevant lattice regulated theory is unique (up to a choice of sign for the Wilson parameters $r_u$, $r_d$, $r_s$, $r_c$) and corresponds to the Iwasaki action in the pure gauge sector and the standard Wilson action for $N_f = 4$ massless fermions in the quark sector\footnote{Taking the chiral limit of the lattice action in Eq.~(\ref{eq:OS}) one obtains the massless standard Wilson action written in a quark basis where the critical Wilson term appears multiplied by $- i \gamma_5 r_f$.}. 

Since RC-dedicated simulations have eventually to be performed outside the chiral limit, different numerical strategies are conceivable. The simplest and most attractive one for a RI$^\prime$-MOM scheme computation of RCs probably amounts to working with two degenerate maximally twisted doublets with twisted masses $\mu_{u, d, s, c} = \mu$, which is obtained by setting $m_0 = m_{cr}$, $r_d = -r_u$ and $r_c = -r_s$.
In such a setup RC-estimators are free of ${\cal{O}}(a)$ lattice artifacts at arbitrary values of the twisted mass $\mu$ and momenta $p$ \cite{ETM-RC-Nf2}.  

However, for at least two ($\beta=1.90$ and $\beta=1.95$) of the three $\beta$-values considered in this paper, the strategy outlined above could not be carried out due to numerical difficulties in implementing the maximal twist setup. In fact, in the region of small PCAC quark masses, which must be accessed when $m_0$ approaches $m_{cr}$, Monte Carlo simulation instabilities were observed leading to very large autocorrelation times~\cite{Dimopoulos:2011wz}. 
Hence we opted for an alternative strategy to achieve ${\cal{O}}(a)$ improvement, already proposed in Ref.~\cite{Frezzotti:2003ni}, that does not require to work at maximal twist.

The method is based on averaging results obtained at opposite values of the PCAC quark mass and thus requires a doubling of the (in any case reasonably low) CPU time cost for producing ensembles at non-zero standard and twisted quark mass. 
Naturally, as it is customary in RI$^\prime$-MOM scheme studies of RCs, we have to consider several values of the valence mass parameters for each given choice of sea mass parameters~\cite{Dimopoulos:2011wz,ETM:2011aa} in order to have stable and reliable valence quark mass chiral extrapolations. 

The corresponding RC computational setup, which can be viewed as a partially (un)quenched setting for Wilson tmLQCD with $N_f = 4$ mass degenerate quark flavors at generic twist angle(s), is outlined in \ref{sec:A1}, where also the choice of the relevant quark mass parameters is discussed.

In \ref{sec:A2} we recall why ${\cal{O}}(a)$ artifacts get canceled in correlation functions of parity-even (multi-)local operators upon averaging results obtained at opposite values of the PCAC quark mass. 
This is sufficient to prove the ${\cal{O}}(a)$ improvement of $Z_q$ and, with little more effort, of $Z_P$ and the other RCs of bilinear quark operators.
In \ref{sec:A3} we report on the numerical parameters of our RC-dedicated simulations and the analysis procedure we followed. 
The latter is illustrated in its key aspects for a few typical examples. 
Our final results for the RCs in the RI$^\prime$-MOM scheme for the three $\beta$ values considered here are given in Tables \ref{tab:final1} and 
\ref{tab:final2} together with few remarks on the conversion to the $\overline{\rm MS}$ scheme at the 2~GeV scale.

\subsection{RC computational setup} 
\label{sec:A1}

The lattice setup for the computation of the RCs can be summarized as follows.
In the so-called twisted basis, which is the one adopted for the definition and the determination of the RCs, the full $N_f = 4$ (possibly partially quenched) local action is of the form
\be
    S^{(N_f = 4)} = S_{YM}[U] + S_{tm}^{\rm sea}[\chi_f^{\rm sea}, U] + S_{tm}^{\rm val} [\chi_f,\phi_f,U] ~ , \qquad \mbox{f = u, d, s, c} ~ ,
    \label{eq:S-MA-full}
\ee
where $S_{YM}[U]$ stands for the Iwasaki gluon action.
The sea quark sector action reads 
\be
   S_{tm}^{\rm sea} = a^4 \sum_{x, f}  \bar\chi_f^{\rm sea} \Big{[}\gamma \cdot \tilde\nabla + W_{cr}  +
                                  (m_{0, f}^{\rm sea} - m_{cr}) + ir_f^{\rm sea} \mu_f^{\rm sea} \gamma_5 \Big{]}\chi_f^{\rm sea} ~ , \\
    \label{eq:SEA-LTB}
\ee
with $\gamma \cdot \tilde\nabla = \frac{\gamma_\mu}{2} (\nabla_\mu + \nabla_\mu^*)$, $W_{cr} = -\frac{a}{2} \nabla_\mu^*\nabla_\mu + m_{cr}$ and $r_d^{\rm sea} = -r_u^{\rm sea}$, $r_c^{\rm sea} = -r_s^{\rm sea}$. 
This choice guarantees positivity of the fermion determinant for the case of fully degenerate quark flavors of interest here, where we set
\be
    \mu_u^{\rm sea} = \mu_d^{\rm sea} = \mu_s^{\rm sea} = \mu_c^{\rm sea} \equiv \mu^{\rm sea} ~ .
    \label{eq:mu-sea} 
\ee
In the valence fermion sector we have $S_{tm}^{\rm val} = S^{\rm val} + S^{\rm ghost}$, where
\be
    S^{\rm val} = a^4 \sum_{x, f} \bar\chi_f^{\rm val} \Big{[} \gamma \cdot \tilde\nabla - 
                          \frac{a}{2} \nabla_\mu^*\nabla_\mu + m_{0, f}^{\rm val} + 
                          ir_f^{\rm val} \mu_f^{\rm val} \gamma_5 \Big{]} \chi_f^{\rm val} ~ , 
    \label{eq:VAL-LTB}
\ee
while the ghost sector term $S^{\rm ghost}$, that must appear to cancel the valence determinant and ensure locality, will be immaterial in what follows. 
As usual, the possible values of the parameters $r_f^{\rm val,sea}$ are restricted to $\pm 1$, while the twisted mass parameters $a\mu_f^{\rm val,sea}$ are assumed to be non-negative.
 
In the partially quenched situation of interest, with all flavors mass-degenerate, a convenient and chiral covariant choice of renormalized quark mass parameters is given by \cite{Frezzotti:2004wz}
\bea
    {\cal{M}}^{\rm sea, val} & = & Z_P^{-1} {\cal{M}}_0^{\rm sea, val} = Z_P^{-1} \sqrt{ (Z_A m_{PCAC}^{\rm sea, val})^2 + 
                                                   (\mu^{\rm sea, val})^2 } ~ , \nonumber \\
    {\mbox{tg}} ~ \theta_f^{\rm sea, val} & = & \frac{Z_A m_{PCAC}^{\rm sea, val}}{r_{f}^{\rm sea, val} \mu^{\rm sea, val}} ~ .
    \label{eq:SEAM+VALMpar} 
\eea
Here $Z_A$ stands for the RC of the (flavor non-singlet) axial current for untwisted Wilson fermions and $m_{PCAC}^{\rm sea}$ denotes the standard PCAC quark mass computed in the unitary setup, while $m_{PCAC}^{\rm val}$ is the analogous quantity that is obtained from correlators defined in terms of valence quark fields, with valence mass parameters possibly different from their sea counterparts.
More precisely, the angles $\theta_f^{\rm sea}$ and $\theta_f^{\rm val}$ are fully determined via the formulae ${\cal{M}}^{\rm sea/val} \cos(\theta_f^{\rm sea/val}) = r_f^{\rm sea/val} \mu^{\rm sea/val}$ and ${\cal{M}}^{\rm sea/val} \sin(\theta_f^{\rm sea/val}) = Z_A m_{PCAC}^{\rm sea/val}$.

We mention in passing that, out of maximal twist, in the partially quenched framework the valence PCAC mass vanishes at a value of $m_0^{\rm val}$ different from the $m_{cr}$ defined in the unitary setup. 
This known fact is properly taken into account if the mass parameters of Eq.~(\ref{eq:SEAM+VALMpar}) are employed in the analysis\footnote{This feature represents a slight numerical complication for the determination of the valence critical mass with respect to the case of maximally twisted LQCD, where the linearly UV-divergent standard valence mass counterterm, being constrained by symmetry~\cite{Frezzotti:2004wz} to depend only on even powers of $\mu^{\rm sea}$ and $\mu^{\rm val}$, can receive no O($a^0 (\mu^{\rm val}-\mu^{\rm sea})$) contribution.}. 

The parameter choice of Eq.~(\ref{eq:SEAM+VALMpar}) is convenient because the renormalized correlators and all the derived quantities in the target continuum limit theory are expected to depend only on ${\cal{M}}^{\rm val}$ and ${\cal{M}}^{\rm sea}$ and not on the twist angles. 
This property is most transparent (see next subsection) in the so-called physical quark basis appropriate for generic twist angles
\be
    \omega_f^{\rm sea, val} = \frac{\pi}{2} - \theta_f^{\rm sea, val} ~ ,
    \label{eq:OMEGA-def}
\ee
where the quark/antiquark fields are defined by the chiral transformation 
\bea
    q_f^{\rm sea, val} & = & \exp\Big(\frac{i}{2}\omega_f^{\rm sea, val}\gamma_5\Big) \chi_f^{\rm sea, val} ~ , \nonumber \\
    \bar q_f^{\rm sea, val} & = & \bar\chi_f^{\rm sea, val} \exp\Big(\frac{i}{2}\omega_f^{\rm sea, val}\gamma_5\Big) ~ .
    \label{eq:PHYS-BASIS-def}
\eea
We recall that the multiplicatively renormalizable quark masses ${\cal{M}}_0^{\rm sea, val}$ in Eq.~(\ref{eq:SEAM+VALMpar}) differ from their classical level analogs\footnote{We define $m_{cr,f}^{\rm val}$ as the value of $m_{0,f}^{\rm val}$ where $m_{PCAC}^{\rm val} = 0$ at given sea quark masses.}
\bea
   {\cal{M}}_{0,class}^{\rm sea} & = & \sqrt{ (m_{0,f}^{\rm sea} - m_{cr})^2 \!+\! (\mu^{\rm sea})^2 } ~ , \nonumber \\
   {\cal{M}}_{0,class}^{\rm val}  & = & \sqrt{ (m_{0,f}^{\rm val} - m_{cr,f}^{\rm val})^2 \!+\! (\mu^{\rm val})^2 } ~ ,
\eea
which are trivially defined in terms of bare parameters of the Lagrangians (\ref{eq:SEA-LTB}) and (\ref{eq:VAL-LTB}). 
This is due to loop effects induced by the chiral-breaking Wilson terms. 
For the same reason the twist angles $\omega_f^{\rm sea, val}$ differ from their tree-level counterparts
\bea
    \omega_{f,class}^{\rm sea} & = & \frac{\pi}{2} - {\rm atan}\Big( \frac{m_{0,f}^{\rm sea} - m_{cr}}{r_f^{\rm sea} \mu^{\rm sea}} \Big) ~ , \nonumber \\
    \omega_{f,class}^{\rm val} & = & \frac{\pi}{2} - {\rm atan}\Big( \frac{m_{0,f}^{\rm val} - m_{cr,f}^{\rm val}}{r_f^{\rm val} \mu^{\rm val}} \Big) ~ . 
    \label{eq:CLASS-ANGLE} 
\eea

\subsection{${\cal{O}}(a)$ improvement via $\theta$-average} 
\label{sec:A2}

In the physical quark basis the parity $P$ entails the standard fermion field transformations\footnote{In this Section to lighten notation we shall omit the flavor labels.} ($x_P \equiv (x_0, -\vec{x}) $)
\bea
    q^{\rm sea, val}(x) & \to & \gamma_0 q^{\rm sea,val}(x_P) ~ , \nonumber \\
    \overline{q}^{\rm sea, val}(x) & \to & \overline{q}^{\rm sea,val}(x_P) \gamma_0 ~ ,
\eea
(besides the obvious ones necessary for gauge fields) making immediate to build and/or identify $P$-even ($P$-odd) operators. 

The lattice vacuum expectation value (vev) of a (multi-)local operator $O$ of definite parity admits a Symanzik local effective Lagrangian (SLEL) description of the form
\be
    \langle O(y, z, ...) \rangle^{\rm latt}_{m,\omega} = \langle O(y, z, ...) \rangle^{L_4} - a \int d^4 x
    \langle O(y, z, ...) L_5(x) \rangle^{L_4} + {\rm O}(a^2) 
    \label{eq:SLELexp} 
\ee
where $L_4$ is the formal Lagrangian of the partially quenched (Euclidean) continuum QCD,
\bea
    L_4 & = & \frac{1}{4} F \cdot F + \ell_4^{\rm sea} + \ell_4^{\rm val} ~ , \nonumber \\
    \ell_4^{\rm sea, val} & = & \overline{q}^{\rm sea, val} (\Dslash + {\cal{M}}^{\rm sea, val}) q^{\rm sea, val} 
\eea
with four degenerate quark flavors of renormalized mass ${\cal{M}}^{\rm sea}$ and ${\cal{M}}^{\rm val}$, while the dimension-five Symanzik operator, $L_5$, takes the form\footnote{In $L_5$ the occurrence of terms like $\bar q^{\rm sea,val} (\gamma_5 \gamma \cdot D) q^{\rm sea,val}$ is forbidden by charge conjugation invariance, while $F \cdot \tilde{F}$ terms are ruled out by the $P \times (u \leftrightarrow d) \times (c \leftrightarrow s)$ symmetry (thanks to $r_{u,c}^{\rm sea} = -r_{d,s}^{\rm sea}$).}
\bea
    L_5 & = & \ell_5^{\rm sea} + \ell_5^{\rm val} + {\cal{M}}^{\rm sea}\cos(\omega^{\rm sea}) [c_g F\cdot F + c_q^{\rm sea} 
                     \ell_4^{\rm sea}] + {\cal{M}}^{\rm val}\cos(\omega^{\rm val}) c_q^{\rm val} \ell_4^{\rm val} ~ , \nonumber \\[2mm]
    \ell_5^{\rm sea, val} & = & c_{Pauli}^{\rm sea, val} ~ \overline{q}^{\rm sea, val} \exp(-i\omega^{\rm sea, val}\gamma_5) 
                                              i \sigma \cdot F q^{\rm sea, val} + \nonumber \\
           & + & c_{Kin}^{\rm sea, val} ~ \overline{q}^{\rm sea, val} \exp(-i\omega^{\rm sea, val}\gamma_5) (- D^2) 
                     q^{\rm sea, val} + \nonumber \\ 
           & + & c_{M2}^{\rm sea, val} ({\cal{M}}^{\rm sea, val})^2 ~ \overline{q}^{\rm sea, val} \exp(-i\omega^{\rm sea, val} 
                     \gamma_5) q^{\rm sea, val} ~ , 
   \label{eq:L5form}
\eea 
with the various $c_{...}^{sea} = c_{...}^{sea}[g_0^2, (\theta^{\rm sea})^2]$ and $c_{...}^{val} = c_{...}^{val}[g_0^2, (\theta^{\rm val})^2,(\theta^{\rm sea})^2]$ being appropriate O(1) coefficient functions.

In the physically interesting case where $O$ is $P$-even, one can check by inserting the expression (\ref{eq:L5form}) of $L_5$ into Eq.~(\ref{eq:SLELexp}) that terms linear in $a$ appear either as vev's of $P$-odd operators (in the target continuum $L_4$-theory) times a factor $\sin(\omega^{\rm sea})$ or $\sin(\omega^{\rm val})$, or as vev's of $P$-even operators multiplied by a factor $\cos(\omega^{\rm sea})$ or $\cos(\omega^{\rm val})$. 
The former terms vanish by parity, which is a symmetry of the target $L_4$-theory, while the latter are in general non-zero. 
They however get canceled if the lattice correlator $\langle O \rangle^{\rm latt}_{{\cal{M}},\omega}$ of Eq.~(\ref{eq:SLELexp}) is averaged with its counterpart $\langle O \rangle^{\rm latt}_{{\cal{M}}, \pi-\omega}$, as $\cos(\pi-\omega^{\rm sea, val}) = - \cos(\omega^{\rm sea, val})$. 
Notice that, in view of the twist angle definition (\ref{eq:OMEGA-def}), the average over $\omega^{\rm sea, val}$ and $\pi-\omega^{\rm sea, val}$ corresponds to averaging over $\theta^{\rm sea, val}$ and $-\theta^{\rm sea, val}$. 

This ${\cal{O}}(a)$ improvement property can be viewed~\cite{Dimopoulos:2011wz,ETM:2011aa} as a consequence of the formal invariance of the lattice mixed action~(\ref{eq:S-MA-full}) rewritten in the physical quark basis (\ref{eq:PHYS-BASIS-def}) under the spurionic transformation ${\cal D}_d \times ({\cal{M}}_{0,class}^{\rm sea,val} \to - {\cal{M}}_{0,class}^{\rm sea,val}) \times P \times ( \theta_{0,class}^{\rm sea,val} \to - \theta_{0,class}^{\rm sea,val}) $.
Since ${\cal D}_d \times ({\cal{M}}_{0,class}^{\rm sea,val} \to - {\cal{M}}_{0,class}^{\rm sea,val})$ just counts the parity of the dimension of Lagrangian terms, the above spurionic invariance implies that in the SLEL of the vev's of multiplicatively renormalizable $P$-even (multi-)local operators all the lattice artifact contributions with odd powers of $a$ appear with a coefficient also odd in $\theta^{\rm sea}$ and $\theta^{\rm val}$. 
Hence they get canceled upon averaging vev's taken at opposite values of $\theta^{\rm sea}$ and $\theta^{\rm val}$. 
We remark that by definition (see Eqs.~(\ref{eq:SEAM+VALMpar}) and (\ref{eq:CLASS-ANGLE})) a sign change in $\theta^{\rm sea,val}$ is equivalent to a sign change in $\theta_{class}^{\rm sea,val}$. 
In the following this way of removing the cutoff effects of first order (as a matter of fact of all odd integer orders) in $a$ will be referred to as $\theta$-average.

\subsubsection{${\cal{O}}(a)$ improvement of $Z_q$}

In the RI$^\prime$-MOM scheme the quark wave function renormalization constant, $Z_q$, at the scale $p^2$ is defined by the condition
\be
    Z_q^{-1} \frac{-i}{12} {\rm Tr}\Big[ \frac{\pslash S_\chi^{-1}(p)}{p^2} \Big] = 
    Z_q^{-1} \frac{-i}{12} {\rm Tr}\Big[ \frac{\pslash S_q^{-1}(p)}{p^2} \Big] =1
    \label{eq:Zq-def} 
\ee
where 
\bea
    \label{eq:Sq-def}
    S_q(p) & = & a^4 \sum_x e^{-ipx} \langle q_f^{\rm val}(x) \bar q_f^{\rm val}(0) \rangle^{\rm latt}_{{\cal{M}}, \omega} ~ , \\
    \label{eq:Schi-def}
    S_\chi(p) & = & a^4 \sum_x e^{-ipx} \langle \chi_f^{\rm val}(x) \bar\chi_f^{\rm val}(0) \rangle^{\rm latt}_{{\cal{M}},\omega} 
                             \nonumber \\
                   & = & e^{-i\omega^{\rm val}\gamma_5/2} S_q(p) e^{-i\omega^{\rm val}\gamma_5/2} \; ,
\eea
are the (momentum space) lattice propagators of the valence quark field of flavor $f$ in the chiral limit expressed in the physical (q) and twisted ($\chi$) basis, respectively.
 
In practice one imposes the condition (\ref{eq:Zq-def}) at non-zero quark mass obtaining $Z_q$-estimators that must be subsequently extrapolated to the chiral limit.
Applying to the massive quark propagator $S_q(p)$ the arguments on leading cutoff effects developed in the introductory part of \ref{sec:A2} and noting the $P$-invariance of $S_q^{-1}(p)$, it follows that in the lattice expression~(\ref{eq:Zq-def}) the cutoff effects linear in $a$ get canceled if $S_q^{-1}(p)$ is replaced by its $\theta$-average, i.e.~by the average of $S_q^{-1}(p)$ evaluated at $({\cal{M}}, \omega)$ and its analog evaluated at $({\cal{M}}, \pi-\omega)$. 
The $\theta$-average procedure guarantees ${\cal{O}}(a)$ improvement already at the level of the RC-estimators in the massive theory.

\subsubsection{${\cal{O}}(a)$ improvement of $Z_P$, $Z_S$ and $Z_T$ }

With the usual notation, according to which RCs are denoted by the name they would have in the twisted basis, where the fermionic sector of the Lagrangian is given by Eqs.~(\ref{eq:SEA-LTB}) and (\ref{eq:VAL-LTB}), the formulae that define in the chiral limit the RCs of quark bilinear operators in the RI$^\prime$-MOM scheme read
\be
    \frac{Z_q}{Z_\Gamma} =  {\rm Tr} \Big[ S_{\chi_1}^{-1}(p) \Big( a^8 \sum_{x,y} e^{-ip(x-y)} \langle \chi_1^{\rm val}(x) 
                                             (\overline{\chi}_1^{\rm val}\Gamma\chi_2^{\rm val})(0) \overline{\chi}_2^{\rm val}(y) 
                                             \rangle^{\rm latt}_{{\cal{M}},\omega} \Big) S_{\chi_2}^{-1}(p) P_\Gamma \Big] ~ ,
    \label{eq:RC-GammaPP}
\ee
where $\Gamma = 1\!\!1, \gamma_5, \gamma_\mu, \gamma_\mu \gamma_5, \sigma_{\mu \nu}$, $P_\Gamma$ is a Dirac projector satisfying ${\rm Tr}(\Gamma P_\Gamma) = 1$, while $\chi_1^{\rm val}$ and $\chi_2^{\rm val}$ are valence quark fields with flavor indices $f = 1$ and $f =2 $, and parameters $r_1^{\rm val}$ and $r_2^{\rm val}$, respectively.

For the case of $r_2^{\rm val} = -r_1^{\rm val}$ in the valence fermion Lagrangian (\ref{eq:VAL-LTB}), passing to the physical quark basis we have $\omega_2^{\rm val}= - \omega_1^{\rm val}$. 
If $\Gamma = \gamma_5$ in Eq.~(\ref{eq:RC-GammaPP}) we thus find the identity 
\bea
    && {\rm Tr}\Big{[} S_{\chi_1}^{-1}(p) \Big( \sum_{x,y} e^{-ip(x-y)} \langle \chi_1^{\rm val}(x) (\bar \chi_1^{\rm val} 
          \gamma_5 \chi_2^{\rm val})(0) \bar\chi_2^{\rm val}(y) \rangle^{\rm latt}_{{\cal{M}},\omega} \Big) S_{\chi_2}^{-1}(p) 
          P_{\gamma_5} \Big{]} \nonumber \\
    && = {\rm Tr}\Big{[} S_{q_1}^{-1}(p) \Big( \sum_{x,y} e^{-ip(x-y)} \langle q_1^{\rm val}(x) (\bar q_1^{\rm val} 
             \gamma_5  q_2^{\rm val})(0) \bar q_2^{\rm val}(y) \rangle^{\rm latt}_{{\cal{M}},\omega} \Big) S_{q_2}^{-1}(p) 
             P_{\gamma_5} \Big{]} ~ .
    \label{eq:ID} 
\eea
From this identity, applying the arguments developed in the introductory part of \ref{sec:A2} to $a^8 \sum_{x,y} e^{-ip(x-y)}$ $\langle q_1^{\rm val}(x) (\bar q_1^{\rm val}\gamma_5  q_2^{\rm val})(0) \bar q_2^{\rm val}(y) \rangle^{\rm latt}_{{\cal{M}},\omega}$, as well as to $S_{q_1}^{-1}(p)$ and $S_{q_2}^{-1}(p)$, and noting that the Dirac trace of their combination in the r.h.s.~is a parity invariant form factor, we conclude that taking the $\theta$-average of the lattice expression in Eq.~(\ref{eq:ID}), improved estimators of $Z_q/Z_P$ for all values of ${\cal{M}}^{\rm val}$ and ${\cal{M}}^{\rm sea}$ are obtained. 
Once an ${\cal{O}}(a)$ improved determination of $Z_q$ is available, $Z_P$ can be extracted with only O($a^2$) artifacts by appropriate chiral extrapolations.  

The argument for the ${\cal{O}}(a)$ improvement via $\theta$-average of the lattice estimators of $Z_S$ and $Z_T$ is identical to the one given above for $Z_P$, because for $\Gamma = 1$ or $\Gamma = \sigma_{\mu\nu}$ and $r_2^{\rm val} = -r_1^{\rm val}$ we find identities completely analogous to Eq.~(\ref{eq:ID}) -- of course with $\gamma_5$ and $P_{\gamma_5}$ replaced by the relevant Dirac matrix $\Gamma$ and associated projector $P_\Gamma$.

\subsubsection{${\cal{O}}(a)$ improvement of $Z_V$ and $Z_A$ }

As the massive lattice estimators for $Z_q / Z_{V,A}$ in the RI$^\prime$-MOM approach  are provided by Eq.~(\ref{eq:RC-GammaPP}) with $\Gamma = \gamma_\mu$ or $\Gamma = \gamma_\mu \gamma_5$, passing from the twisted to the physical quark basis, in the case of $r_2^{\rm val} = -r_1^{\rm val}$, identities analogous to Eq.~(\ref{eq:ID}) are obtained, but (owing to anti-commutation of $\Gamma$ with the $\gamma_5$ occurring in the equation relating $\chi^{\rm val}_f$ and $q^{\rm val}_f$) with a more complicated r.h.s. 

If, for instance, $\Gamma=\gamma_\mu$, upon setting $-\omega_2^{\rm val}= \omega_1^{\rm val} \equiv \omega^{\rm val}$, we find
\bea
    \hspace*{-1.1cm} 
    && {\rm Tr}\Big{[}S_{\chi_1}^{-1}(p) \Big( \sum_{x,y} e^{-ip(x-y)} \langle \chi_1^{\rm val}(x) (\bar \chi_1^{\rm val} 
          \gamma_\mu \chi_2^{\rm val})(0) \bar\chi_2^{\rm val}(y) \rangle^{\rm latt}_{{\cal{M}},\omega} \Big) 
          S_{\chi_2}^{-1}(p) P_{\gamma_\mu} \Big{]} \nonumber \\
    \hspace*{-1.1cm} 
    && = C^2 ~ {\rm Tr}\Big{[} S_{q_1}^{-1}(p) \Big( \sum_{x,y} e^{-ip(x-y)} \langle q_1^{\rm val}(x)  V_{12,\mu}(0)
             \bar q_2^{\rm val}(y) \rangle^{\rm latt}_{{\cal{M}},\omega} \Big) S_{q_2}^{-1}(p) P_{\gamma_\mu} \Big{]} + \nonumber \\
    \hspace*{-1.1cm} 
    && + S^2 ~ {\rm Tr}\Big{[} S_{q_1}^{-1}(p) \Big( \sum_{x,y} e^{-ip(x-y)} \langle q_1^{\rm val}(x) A_{12,\mu}(0)
          \bar q_2^{\rm val}(y) \rangle^{\rm latt}_{{\cal{M}},\omega} \Big) S_{q_2}^{-1}(p) P_{\gamma_\mu\gamma_5} \Big{]} +
          \nonumber \\
    \hspace*{-1.1cm} 
    && -i C S ~ {\rm Tr}\Big{[} S_{q_1}^{-1}(p) \Big( \sum_{x,y} e^{-ip(x-y)} \langle q_1^{\rm val}(x) V_{12,\mu}(0)
          \bar q_2^{\rm val}(y) \rangle^{\rm latt}_{{\cal{M}},\omega} \Big) S_{q_2}^{-1}(p) P_{\gamma_\mu\gamma_5} \Big{]} + 
          \nonumber \\
    \hspace*{-1.1cm} 
    && +i C S ~ {\rm Tr}\Big{[} S_{q_1}^{-1}(p) \Big( \sum_{x,y} e^{-ip(x-y)} \langle q_1^{\rm val}(x) A_{12,\mu}(0)
          \bar q_2^{\rm val}(y) \rangle^{\rm latt}_{{\cal{M}},\omega} \Big) S_{q_2}^{-1}(p) P_{\gamma_\mu} \Big{]} 
    \label{eq:ID2}
\eea
with $C = \cos\omega$, $S = \sin\omega$, $V_{12,\mu} = \bar q_1^{\rm val}\gamma_\mu  q_2^{\rm val}$ and $A_{12, \mu} = \bar q_1^{\rm val}\gamma_\mu\gamma_5  q_2^{\rm val}$.
Looking at the r.h.s.~of this identity, we note that the expressions with pre-factors $C^2$ and $S^2$ ($\pm i CS$) are parity-even (parity-odd) form factors. 
Then applying the Symanzik analysis arguments developed in the introductory part of \ref{sec:A2} to $\sum_{x, y} e^{-ip(x-y)}$ $\langle q_1^{\rm val}(x) V_{12, \mu}(0)$ $[A_{12, \mu}(0)] \bar q_2^{\rm val}(y) \rangle^{\rm latt}_{{\cal{M}},\omega}$, as well as to $S_{q_1}^{-1}(p)$ and $S_{q_2}^{-1}(p)$, we see that
\begin{itemize}
\item to order zero in $a$, the terms with pre-factors $\pm i CS$ vanish by parity, while those with pre-factors $C^2$ and $S^2$ are non-zero (and coinciding in the limit of unbroken chiral symmetry);
\item to first order in $a$ the contributions that do not vanish by parity are those obtained either by inserting the $P$-even piece of $L_5$ in the terms on the the r.h.s.~of Eq.~(\ref{eq:ID2}) with pre-factors $C^2$ and $S^2$ or by inserting the $P$-odd piece of $L_5$ in the terms with pre-factors $\pm i CS$.
\end{itemize}
Taking also into account the $\omega$- and $\omega^{\rm sea}$-dependence of  $L_5$ (see Eq.~(\ref{eq:L5form})), one checks that, while the contributions of zero (actually even integer) order in $a$ are even under $\omega \to \pi - \omega$, all the contributions of first (actually odd integer) order in $a$ are odd under $\omega \to \pi - \omega$. 
Hence by taking the $\theta$ average of the lattice expression (\ref{eq:ID2}), the lattice artifacts of odd order in $a$ get canceled, leaving out the contributions of order $a^{2n}$.

This proves the ${\cal{O}}(a)$ improvement by $\theta$ average of the massive lattice estimators of $Z_V$, from which the RC is extracted after chiral extrapolations.
Identical arguments clearly hold as well for $Z_A$, because for $\Gamma = \gamma_\mu\gamma_5$ we find an identity completely analogous to Eq.~(\ref{eq:ID2}) with the axial and vector operators and the associated projectors properly interchanged.

We have focused here on the choice $r_2^{\rm val} = -r_1^{\rm val}$ for the valence parameters of quark bilinear operators, because this is the case with smallest statistical fluctuations in the numerical evaluation of RCs and which the results quoted in the following refer to. 
The discussion of the alternative (and computationally more noisy) choice $r_2^{\rm val} = r_1^{\rm val}$ could be carried out along similar lines\footnote{If $r_2^{\rm val} = r_1^{\rm val}$, however, the identities obtained when passing from the twisted to the physical quark basis have a simple r.h.s.~for the case of $Z_{V, A}$ and a more complicated one (like the one of Eq.~(\ref{eq:ID2})) for $Z_{P, S, T}$.}, finding again that upon $\theta$ average the estimators of the RCs of all quark bilinear operators are ${\cal{O}}(a)$ improved.

\subsection{Numerical details and results}
\label{sec:A3}

In Table~\ref{tab:results} we report the information on the relevant simulation parameters for the three ensembles we have considered in this paper. 
Except for the $\theta$-average, which is implemented in order to achieve the ${\cal{O}}(a)$ improvement out of the maximal twist, the other parts of the analysis follow closely the procedure described in Ref.~\cite{ETM-RC-Nf2}.

\begin{table}[!htb]
\begin{center}
\begin{tabular}{|l|c|c|c|c|c|c|}
\hline 
& {\scriptsize $a\mu^{\textrm{sea}}$} & {\scriptsize $am_{\textrm{PCAC}}^{\textrm{sea}}$} & {\scriptsize $am_{0}^{\textrm{sea}}$} 
& {\scriptsize $\theta^{\textrm{sea}}$} & {\scriptsize $a\mu^{\textrm{val}}$} & {\scriptsize $am_{\textrm{PCAC}}^{\textrm{val}}$}
\tabularnewline
\hline 
\hline 
\multicolumn{7}{|c}{{\scriptsize $\beta=1.90$ ($L=24$, $T=48$)}}\tabularnewline
\hline 
{\scriptsize A4m} & {\scriptsize 0.0080} & {\scriptsize -0.0390(01)} & {\scriptsize 0.0285(01)} & {\scriptsize -1.286(01)} & 
{\scriptsize \{0.0060, 0.0080, 0.0120, } & {\scriptsize -0.0142(02)}\tabularnewline
{\scriptsize A4p} &  & {\scriptsize 0.0398(01)} & {\scriptsize 0.0290(01)} & {\scriptsize +1.291(01)} & {\scriptsize 0.0170, 0.0210
,0.0260\}} & {\scriptsize +0.0147(02)}\tabularnewline
\hline 
{\scriptsize A3m} & {\scriptsize 0.0080} & {\scriptsize -0.0358(02)} & {\scriptsize 0.0263(01)} & {\scriptsize -1.262(02)} & 
{\scriptsize \{0.0060, 0.0080, 0.0120, } & {\scriptsize -0.0152(02)}\tabularnewline
{\scriptsize A3p} &  & {\scriptsize 0.0356(02)} & {\scriptsize 0.0262(01)} & {\scriptsize +1.260(02)} & {\scriptsize 0.0170, 0.0210
,0.0260\}} & {\scriptsize +0.0147(03)}\tabularnewline
\hline 
{\scriptsize A2m} & {\scriptsize 0.0080} & {\scriptsize -0.0318(01)} & {\scriptsize 0.0237(01)} & {\scriptsize -1.226(02)} & 
{\scriptsize \{0.0060, 0.0080, 0.0120, } & {\scriptsize -0.0155(02)}\tabularnewline
{\scriptsize A2p} &  & {\scriptsize +0.0310(02)} & {\scriptsize 0.0231(01)} & {\scriptsize +1.218(02)} & {\scriptsize 0.0170, 0.021
0,0.0260\}} & {\scriptsize +0.0154(02)}\tabularnewline
\hline 
{\scriptsize A1m} & {\scriptsize 0.0080} & {\scriptsize -0.0273(02)} & {\scriptsize 0.0207(01)} & {\scriptsize -1.174(03)} & 
{\scriptsize \{0.0060, 0.0080, 0.0120, } & {\scriptsize -0.0163(02)}\tabularnewline
{\scriptsize A1p} &  & {\scriptsize +0.0275(04)} & {\scriptsize 0.0209(01)} & {\scriptsize +1.177(05)} & {\scriptsize 0.0170, 0.021
0,0.0260\}} & {\scriptsize +0.0159(02)}\tabularnewline
\hline 
\multicolumn{7}{|c}{{\scriptsize $\beta=1.95$ ($L=24$, $T=48$)}}\tabularnewline
\hline 
{\scriptsize B1m} & {\scriptsize 0.0085} & {\scriptsize -0.0413(02)} & {\scriptsize 0.0329(01)} & {\scriptsize -1.309(01)} & 
{\scriptsize \{0.0085, 0.0150, 0.0203,} & {\scriptsize -0.0216(02)}\tabularnewline
{\scriptsize B1p} &  & {\scriptsize +0.0425(02)} & {\scriptsize 0.0338(01)} & {\scriptsize +1.317(01)} & {\scriptsize{} 0.0252, 0.02
98\}} & {\scriptsize +0.0195(02)}\tabularnewline
\hline 
{\scriptsize B7m} & {\scriptsize 0.0085} & {\scriptsize -0.0353(01)} & {\scriptsize 0.0285(01)} & {\scriptsize -1.268(01)} & 
{\scriptsize \{0.0085, 0.0150, 0.0203,} & {\scriptsize -0.0180(02)}\tabularnewline
{\scriptsize B7p} &  & {\scriptsize +0.0361(01)} & {\scriptsize 0.0285(01)} & {\scriptsize +1.268(01)} & {\scriptsize{} 0.0252, 0.02
98\}} & {\scriptsize +0.0181(01)}\tabularnewline
\hline 
{\scriptsize B8m} & {\scriptsize 0.0020} & {\scriptsize -0.0363(01)} & {\scriptsize 0.0280(01)} & {\scriptsize -1.499(01)} & 
{\scriptsize \{0.0085, 0.0150, 0.0203,} & {\scriptsize -0.0194(01)}\tabularnewline
{\scriptsize B8p} &  & {\scriptsize +0.0363(01)} & {\scriptsize 0.0274(01)} & {\scriptsize +1.498(01)} & {\scriptsize{} 0.0252, 0.02
98\}} & {\scriptsize +0.0183(02)}\tabularnewline
\hline 
{\scriptsize B3m} & {\scriptsize 0.0180} & {\scriptsize -0.0160(02)} & {\scriptsize 0.0218(01)} & {\scriptsize -0.601(06)} & 
{\scriptsize \{0.0060,0.0085,0.0120,0.0150,} & {\scriptsize -0.0160(02)}\tabularnewline
{\scriptsize B3p} &  & {\scriptsize +0.0163(02)} & {\scriptsize 0.0219(01)} & {\scriptsize +0.610(06)} & {\scriptsize 0.0180,0.0203,
0.0252,0.0298\}} & {\scriptsize +0.0162(02)}\tabularnewline
\hline 
{\scriptsize B2m} & {\scriptsize 0.0085} & {\scriptsize -0.0209(02)} & {\scriptsize 0.0182(01)} & {\scriptsize -1.085(03)} & 
{\scriptsize \{0.0085, 0.0150, 0.0203,} & {\scriptsize -0.0213(02)}\tabularnewline
{\scriptsize B2p} &  & {\scriptsize +0.0191(02)} & {\scriptsize 0.0170(02)} & {\scriptsize +1.046(06)} & {\scriptsize{} 0.0252, 0.02
98\}} & {\scriptsize +0.0191(02)}\tabularnewline
\hline 
{\scriptsize B4m} & {\scriptsize 0.0085} & {\scriptsize -0.0146(02)} & {\scriptsize 0.0141(01)} & {\scriptsize -0.923(04)} & 
{\scriptsize \{0.0060,0.0085,0.0120,0.0150,} & {\scriptsize -0.0146(02)}\tabularnewline
{\scriptsize B4p} &  & {\scriptsize +0.0151(02)} & {\scriptsize 0.0144(01)} & {\scriptsize +0.940(07)} & {\scriptsize 0.0180,0.0203,
0.0252,0.0298\}} & {\scriptsize +0.0151(02)}\tabularnewline
\hline 
\multicolumn{7}{|c}{{\scriptsize $\beta=2.10$ ($L=32$, $T=64$)}}\tabularnewline
\hline 
{\scriptsize C5m} & {\scriptsize 0.0078} & {\scriptsize -0.00821(11)} & {\scriptsize 0.0102(01)} & {\scriptsize -0.700(07)} & 
{\scriptsize \{0.0048,0.0078,0.0119,} & {\scriptsize -0.0082(01)}\tabularnewline
{\scriptsize C5p} &  & {\scriptsize +0.00823(08)} & {\scriptsize 0.0102(01)} & {\scriptsize +0.701(05)} & {\scriptsize 0.0190,0.0242
,0.0293\}} & {\scriptsize +0.0082(01)}\tabularnewline
\hline 
{\scriptsize C4m} & {\scriptsize 0.0064} & {\scriptsize -0.00682(13)} & {\scriptsize 0.0084(01)} & {\scriptsize -0.706(09)} & 
{\scriptsize \{0.0039,0.0078,0.0119,} & {\scriptsize -0.0068(01)}\tabularnewline
{\scriptsize C4p} &  & {\scriptsize +0.00685(12)} & {\scriptsize 0.0084(01)} & {\scriptsize +0.708(09)} & {\scriptsize 0.0190,0.0242
,0.0293\}} & {\scriptsize +0.0069(01)}\tabularnewline
\hline 
{\scriptsize C3m} & {\scriptsize 0.0046} & {\scriptsize -0.00585(08)} & {\scriptsize 0.0066(01)} & {\scriptsize -0.794(07)} & 
{\scriptsize \{0.0025,0.0046,0.0090,0.0152,} & {\scriptsize -0.0059(01)}\tabularnewline
{\scriptsize C3p} &  & {\scriptsize +0.00559(14)} & {\scriptsize 0.0064(01)} & {\scriptsize +0.771(13)} & {\scriptsize 0.0201,0.0249
,0.0297\}} & {\scriptsize +0.0056(01)}\tabularnewline
\hline 
{\scriptsize C2m} & {\scriptsize 0.0030} & {\scriptsize -0.00403(14)} & {\scriptsize 0.0044(01)} & {\scriptsize -0.821(17)} & 
{\scriptsize \{0.0013,0.0030,0.0080,0.0143,} & {\scriptsize -0.0040(01)}\tabularnewline
{\scriptsize C2p} &  & {\scriptsize +0.00421(13)} & {\scriptsize 0.0045(01)} & {\scriptsize +0.843(15)} & {\scriptsize 0.0195,0.0247
,0.0298\}} & {\scriptsize +0.0042(01)}\tabularnewline
\hline 
\end{tabular}
\caption{\it Simulation and correlator analysis details. Here $L_4 \equiv T$ and $L_{1,2,3} \equiv L$.}
\label{tab:results}
\end{center}
\end{table}

For each ensemble in the table we compute the RC-estimators at values of momenta, $p_{\mu}=\left(2\pi/L_{\mu}\right)n_{\mu}$, with components lying in the following intervals 
\bea
    n_{\mu} & = & \left(\left[0,2\right],\left[0,2\right],\left[0,2\right],\left[0,3\right]\right) \nonumber \\
                  &    & \left(\left[2,3\right],\left[2,3\right],\left[2,3\right],\left[4,7\right]\right) , ~~~ \mbox{for }\beta=1.95,  \nonumber \\[2mm] 
    n_{\mu} & = & \left(\left[0,2\right],\left[0,2\right],\left[0,2\right],\left[0,3\right]\right) \nonumber \\
                  &    & \left(\left[2,5\right],\left[2,5\right],\left[2,5\right],\left[4,9\right]\right) , ~~~ \mbox{for }\beta=1.90 ~ \mbox{~ and ~} 2.10
\eea
and $L_{\mu}$ denoting the lattice size in the direction $\mu$. 
Anti-periodic boundary conditions on the quark fields in the time direction are implemented by a shift of the time component of the four-momentum by the constant $\Delta p_4 = \pi / L_4$. 
The final analysis of the RC estimators has been performed at four-momenta that pass the  ``democratic'' momentum cut defined by 
\be
    \Delta_4(p) \equiv \frac{\sum_{\mu} \tilde p_{\mu}^4}{(\sum_{\mu} \tilde p_{\mu}^2)^2} < 0.29, 
\ee
where 
\be
     \tilde p_{\mu} \equiv \frac{1}{a} \sin (ap_{\mu}) ~ .
\ee

\begin{figure}[!htb]
\begin{center}
{\includegraphics[width=0.475\textwidth]{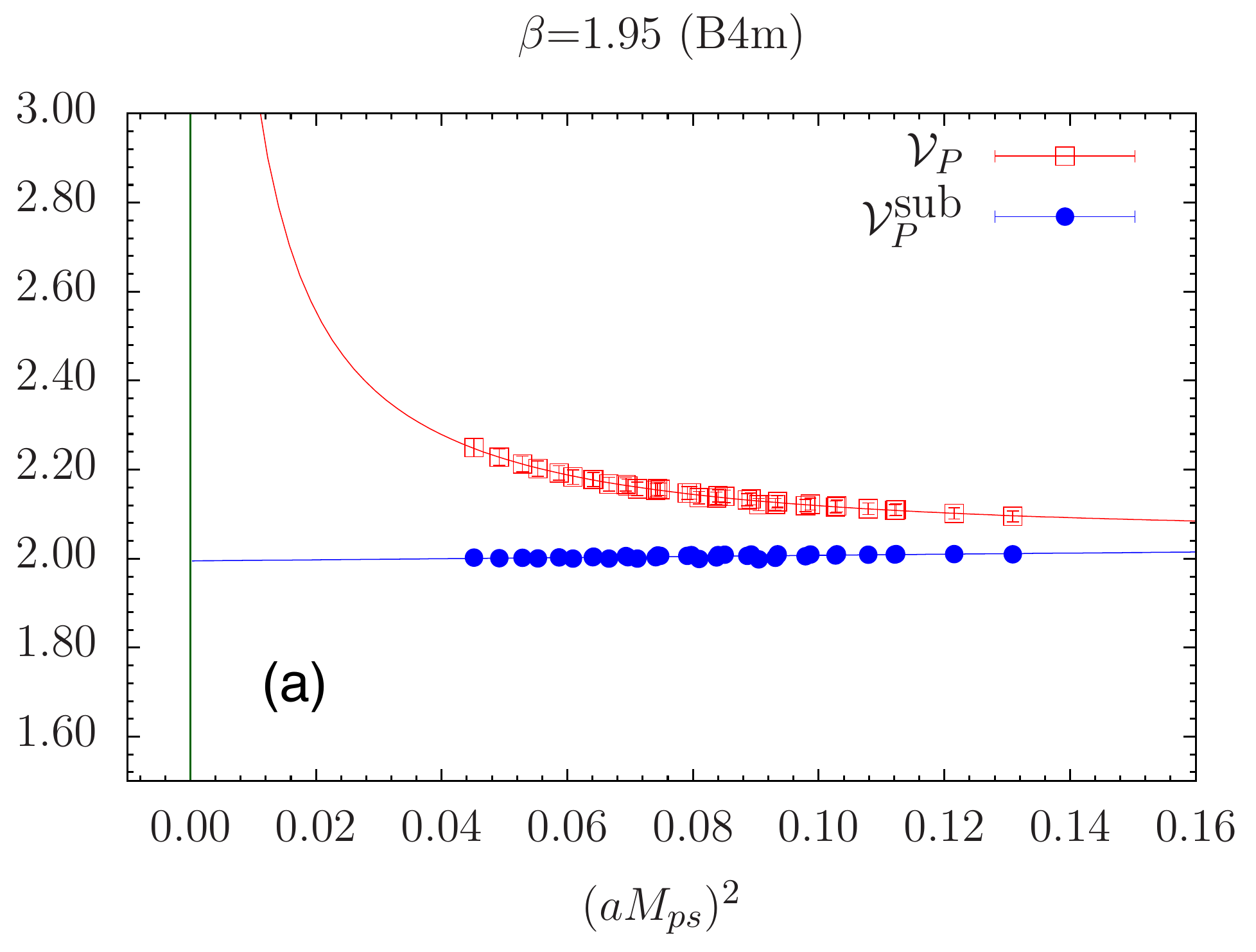}
\hspace{.5cm}
\includegraphics[width=0.475\textwidth]{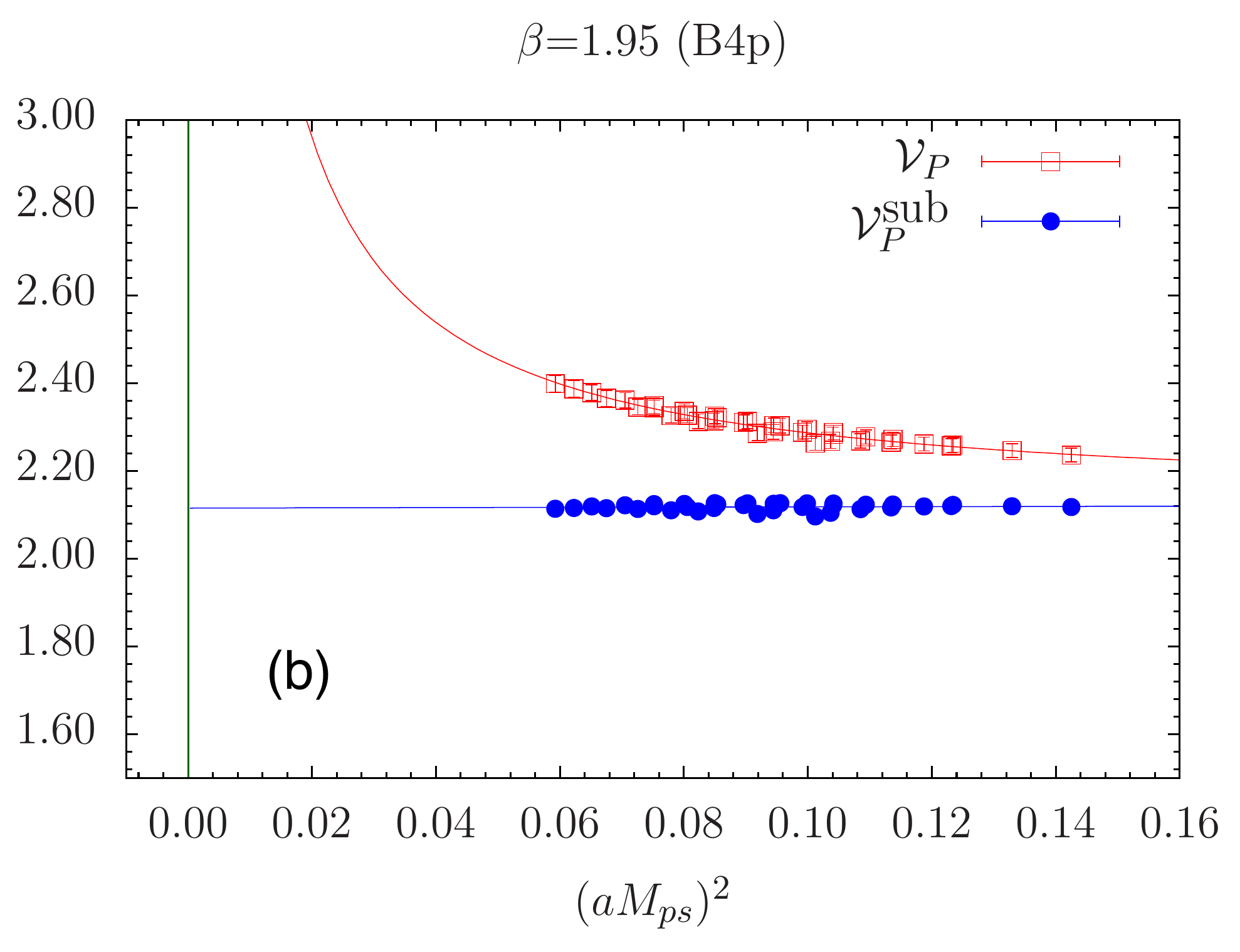}}
\end{center}
\vspace*{-0.5cm}
\caption{\it Amputated pseudoscalar density two-point correlators before (${\cal V}_P$, red squares) and after (${\cal V}_P^{\rm{sub}}$, blue dots) the Goldstone pole subtraction, at $\beta=1.95$ and $(a\tilde p)^2\approx 1.5$. Panels (a) and (b) correspond to data from ensembles B4m and B4p, respectively.}
\label{fig:ZP_GPS}

\begin{center}
{\includegraphics[width=0.475\textwidth]{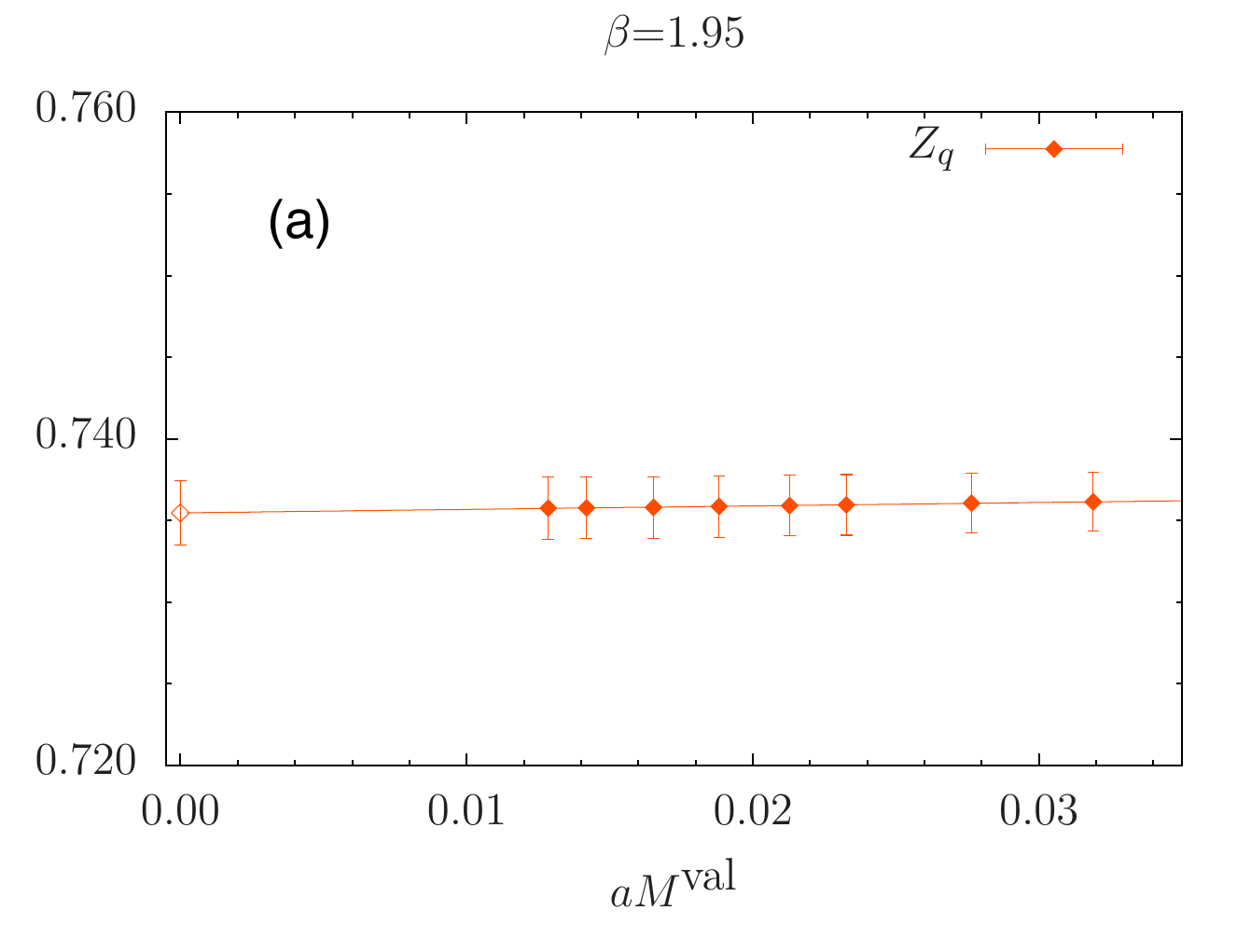}
\hspace{.5cm}
\includegraphics[width=0.475\textwidth]{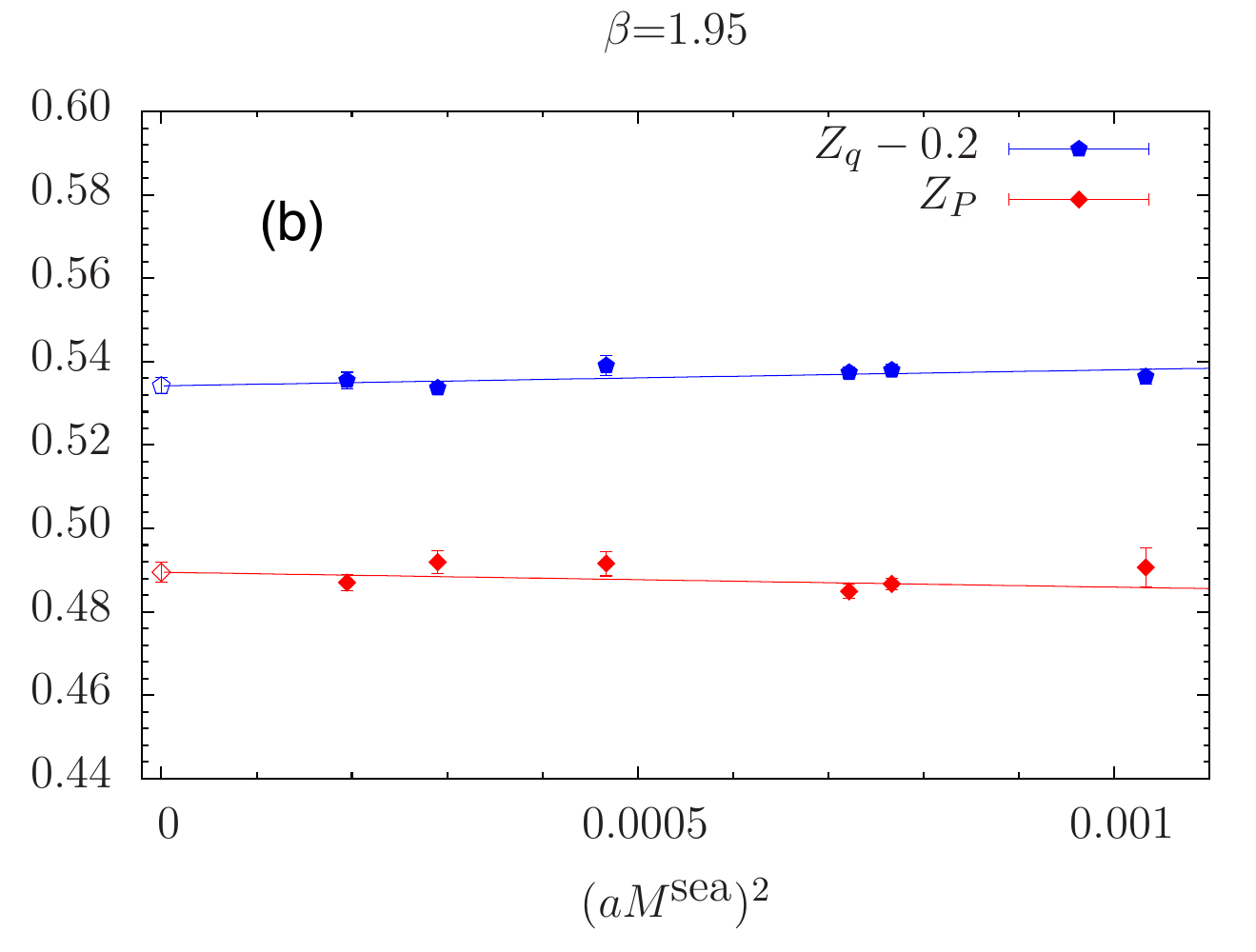}}
\end{center}
\vspace*{-0.5cm}
\caption{\it Chiral extrapolations of $\theta$-averaged B4m and B4p data at $\beta = 1.95$ and $(a\tilde p)^2 \approx 1.5$. 
Panel (a): valence quark mass extrapolation of $Z_q$ data. Panel (b): sea quark mass extrapolation of $Z_q$ and $Z_P$ data. The blue dots have been displaced by $0.2$ from the red squares for better visibility.}
\label{fig:val_sea_chi_lim}
\end{figure}

As a typical example, in Fig.~\ref{fig:ZP_GPS} we show the effect of the subtraction of the Goldstone pole in the amputated two-point correlators for the ensembles B4m and B4p (the most critical ones at $\beta=1.95$). 
The quantities ${\cal V}_P$ and ${\cal V}_P^{\rm{sub}}$ are defined according to Eqs.~(3.4) and~(3.12)-(3.13) of Ref.~\cite{ETM-RC-Nf2}. 

In Fig.~\ref{fig:val_sea_chi_lim}(a) we plot the valence mass extrapolation of $Z_q$, while Fig.~\ref{fig:val_sea_chi_lim}(b) shows the sea quark mass extrapolation in the chiral limit of $\theta$-averaged B4m and B4p data for the cases of $Z_q$ and $Z_P$ at $\beta=1.95$ and $(a \tilde{p})^2 \approx 1.5$.  

Our final estimates of the RCs of all the quark bilinear operators in the RI$^\prime$-MOM(3 GeV), $\overline{\rm MS}(3 \gev)$ and $\overline{\rm MS}(2 \gev)$ schemes are collected in Tables \ref{tab:final1}-\ref{tab:final4}.
They have been obtained after subtracting the perturbative cutoff effects up to O($a^2 g_{\rm boost}^2$), where as usual $g_{\rm boost}^2 = 6/(\beta \langle P \rangle)$, with $\langle P \rangle$ the average plaquette value. 
Perturbative estimates of the discretization effects on the Green function of the operators of interest can be found in Ref.~\cite{Constantinou:2009tr}. 
The RCs $Z_q$, $Z_P$ and $Z_S$, obtained in the RI$^\prime$-MOM(3 GeV) scheme, have been converted to the $\overline{\rm MS}(3 \gev)$ and $\overline{\rm MS}(2 \gev)$ schemes using the appropriate N$^3$LO formulae from Ref.~\cite{Chetyrkin:1999pq}, while for the RC $Z_T$ the N$^2$LO formula of Ref.~\cite{Gracey_2003} has been employed.
The uncertainty due to the matching between the RI$^\prime$-MOM and the $\overline{\rm MS}$ schemes, performed at the scale of $3 \gev$, cannot be neglected.
From the convergence of the matching at LO, NLO, N$^2$LO and N$^3$LO orders we estimate an uncertainty of $\simeq 1.3 \%$ due to higher perturbative orders\footnote{The matching for the quark mass between the RI$^\prime$-MOM and the $\overline{\rm MS}$ schemes is given by \cite{Chetyrkin:1999pq}: $m^{\overline{\rm MS}} / m^{\rm RI^\prime-MOM} = 1 - 0.4244 \alpha_s - 0.7102 \alpha_s^2 - 1.4782 \alpha_s^3$. Using $\alpha_s(3 \gev) = 0. 256$, corresponding to $N_f = 4$ and $\Lambda_{QCD} = 296 \mev$ \cite{PDG}, one gets $m^{\overline{\rm MS}} / m^{\rm RI^\prime-MOM} = 1 - 0.1085 - 0.04654 - 0.02480$. At least an approximate factor $1/2$ relates each term of the series with its next one and therefore we take $1/2$ of the last term as our estimate of the uncertainty in the perturbative matching.}. 
In the same way the evolution in the $\overline{\rm MS}$ scheme from $3$ to $2 \gev$ (or to $m_c$) has an uncertainty of the order of $\simeq 0.1 \%$ ($0.5 \%$). By adding in quadrature such an uncertainty to the one due to the matching, the perturbative error in the conversion from  the RI$^\prime$-MOM($3 \gev$) to $\overline{\rm MS}(2 \gev)$ (or $\overline{\rm MS}(m_c)$) schemes is found to be equal to $\simeq 1.3 \%$ ($1.4 \%$). This error, being not related to a genuine lattice uncertainty, is added directly as a further systematic error to our determinations of the quark masses, separately from the one due to the choice of the RCs from the methods M1 and M2 [see Eqs.~(\ref{eq:mudresults}, \ref{eq:msresults}, \ref{eq:umass}, \ref{eq:dmass}, \ref{eq:mcresults_2GeV}, \ref{eq:mcresults})], and it is not reported in Tables \ref{tab:MS3GeV} and \ref{tab:final3}. 

Recently, the RCs $Z_V$, $Z_A$, $Z_P$ and $Z_S$ have been computed perturbatively up to three loops in Ref.~\cite{Brambilla:2014bja} at $\beta = 1.95$ and $2.10$.
The comparison with our non-perturbative results of Tables \ref{tab:final1} and \ref{tab:final4} shows a remarkable, fair agreement within the quoted errors.

\begin{table}[!h]
\begin{center}
\renewcommand{\arraystretch}{1.20}
\begin{tabular}{||c|c|c|c|c|c||}  \hline
$\beta$ & Method  & $Z_q$ & $Z_P$ & $Z_S$ & $Z_T$ \\ \hline \hline
1.90 &  M1 &     0.721(5)   & 0.423(6) & 0.598(10) & 0.714(5) \\
        &  M2 &     0.737(2)   & 0.459(4) & 0.702(3) &  0.702(3) \\
        \hline
1.95 &  M1 &    0.733(4)    & 0.424(4) & 0.608(7) &   0.721(4) \\
        &  M2 &    0.741(1)    & 0.454(2) & 0.684(2) &   0.707(2)\\
        \hline
2.10 &  M1 &    0.766(4)    & 0.478(2) & 0.649(5)  &  0.749(4)\\
        &  M2 &    0.767(2)    & 0.505(2) & 0.695(3)  &  0.742(2)\\
       \hline
\end{tabular}
\renewcommand{\arraystretch}{1.0}
\end{center}
\caption{RCs $Z_q$, $Z_P$, $Z_S$ and $Z_T$ obtained in the RI$^\prime$-MOM scheme at the renormalization scale $1 / a(\beta)$ (see  Eq.~(\ref{eq:latticespacingsresults})), using the methods M1 and M2.}
\label{tab:final1}


\begin{center}
\renewcommand{\arraystretch}{1.20}
\begin{tabular}{||c|c|c|c|c|c||}  \hline
$\beta$ & Method & $Z_q$ & $Z_P$ & $Z_S$ & $Z_T$ \\ \hline \hline
1.90 &  M1 & 0.713(05) & 0.480(07) & 0.678(11) & 0.692(05)\\
        &  M2 & 0.729(02) & 0.521(04) & 0.796(03) & 0.680(03)\\
      \hline
1.95 &  M1 & 0.727(04) & 0.462(04) & 0.663(08) & 0.705(04)\\
        &  M2 & 0.736(01) & 0.495(02) & 0.746(02) & 0.691(02) \\
      \hline
2.10 &  M1 & 0.767(04) & 0.468(02) & 0.635(05) & 0.753(04)\\
        &  M2 & 0.768(02) & 0.494(02) & 0.680(03) & 0.746(02)\\
     \hline
\end{tabular}
\renewcommand{\arraystretch}{1.0}
\end{center}
\caption{\it The same as in Table \ref{tab:final1}, but at the renormalization scale of 3 GeV.}
\label{tab:final2}
\end{table}

\begin{table}[!htb]
\begin{center}
\renewcommand{\arraystretch}{1.20}
\begin{tabular}{||c|c|c|c|c|c||}  \hline
$\beta$ & Method  & $Z_q$ & $Z_P$ & $Z_S$ & $Z_T$ \\ \hline \hline
1.90 &  M1 &          0.705(05)   & 0.587(08)  & 0.830(14)  &  0.684(05)\\
        &  M2 &          0.720(02)   & 0.637(06)  &  0.974(04)   &  0.672(03)\\
         \hline
1.95 &  M1 &          0.719(04)   &  0.566(05) &  0.812(09) &  0.697(04)\\
        &  M2 &          0.727(01)   &  0.606(03) &  0.913(03) &  0.683(02)\\
         \hline
2.10 &  M1 &          0.759(04)   & 0.572(02) & 0.777(06)  &  0.744(04)\\
        &  M2 &          0.760(02)   & 0.605(02) & 0.832(04)  &  0.737(02)\\
         \hline
\end{tabular}
\renewcommand{\arraystretch}{1.0}
\end{center}
\caption{RCs $Z_q$, $Z_P$, $Z_S$ and $Z_T$ obtained in the $\overline{\rm MS}(3 \gev)$ scheme using the methods M1 and M2.}
\label{tab:MS3GeV}


\begin{center}
\renewcommand{\arraystretch}{1.20}
\begin{tabular}{||c|c|c|c|c|c||}  \hline
$\beta$ & Method & $Z_q$ & $Z_P$ & $Z_S$ & $Z_T$ \\ \hline \hline
1.90 &  M1 & 0.712(05) & 0.529(07) & 0.747(12) & 0.711(05)\\
        &  M2 & 0.728(02) & 0.574(04) & 0.877(03) & 0.700(03)\\
      \hline
1.95 &  M1 & 0.726(04) & 0.509(04) & 0.713(09) & 0.724(04)\\
        &  M2 & 0.735(01) & 0.546(02) & 0.822(02) & 0.711(02) \\
      \hline
2.10 &  M1 & 0.766(04) & 0.516(02) & 0.700(06) & 0.774(04)\\
        &  M2 & 0.767(02) & 0.545(02) & 0.749(03) & 0.767(02)\\
     \hline
\end{tabular}
\renewcommand{\arraystretch}{1.0}
\end{center}
\caption{\it The same as in Table \ref{tab:MS3GeV}, but at the renormalization scale of 2 GeV.}
\label{tab:final3}
\end{table}


\begin{table}[!htb]
\begin{center}
\renewcommand{\arraystretch}{1.20}
\begin{tabular}{||c|c|c|c|c||}  \hline
$\beta$ & Method & $Z_V$ & $Z_A$ & $Z_P/Z_S$ \\ \hline \hline
                     & M1  & 0.587(04) & 0.731(08) & 0.699(13) \\
1.90              & M2  & 0.608(03) & 0.703(02) & 0.651(06) \\
                     & WTI & 0.5920(04) & - & -\\
 \hline
                     & M1  & 0.603(03) & 0.737(05) & 0.697(07) \\
1.95              & M2  & 0.614(02) & 0.714(02) & 0.666(04) \\
                     & WTI & 0.6095(03) & - & -\\
\hline
                     & M1  & 0.655(03) & 0.762(04) & 0.740(05) \\
2.10              & M2  & 0.657(02) & 0.752(02) & 0.727(03) \\
                     & WTI & 0.6531(02) & - & - \\ 
 \hline
\end{tabular}
\renewcommand{\arraystretch}{1.0}
\end{center}
\caption{\it RCs $Z_V$, $Z_A$ and $Z_P/Z_S$ obtained with the methods M1 and M2. We also present the accurate results for $Z_V$ obtained using the Ward-Takahashi identity (WTI) (for more details see Section 2.3 of Ref.~\cite{ETM-RC-Nf2}).} 
\label{tab:final4}
 \end{table}

\begin{figure}[!htb]
\begin{center}
\includegraphics[width=0.8\textwidth]{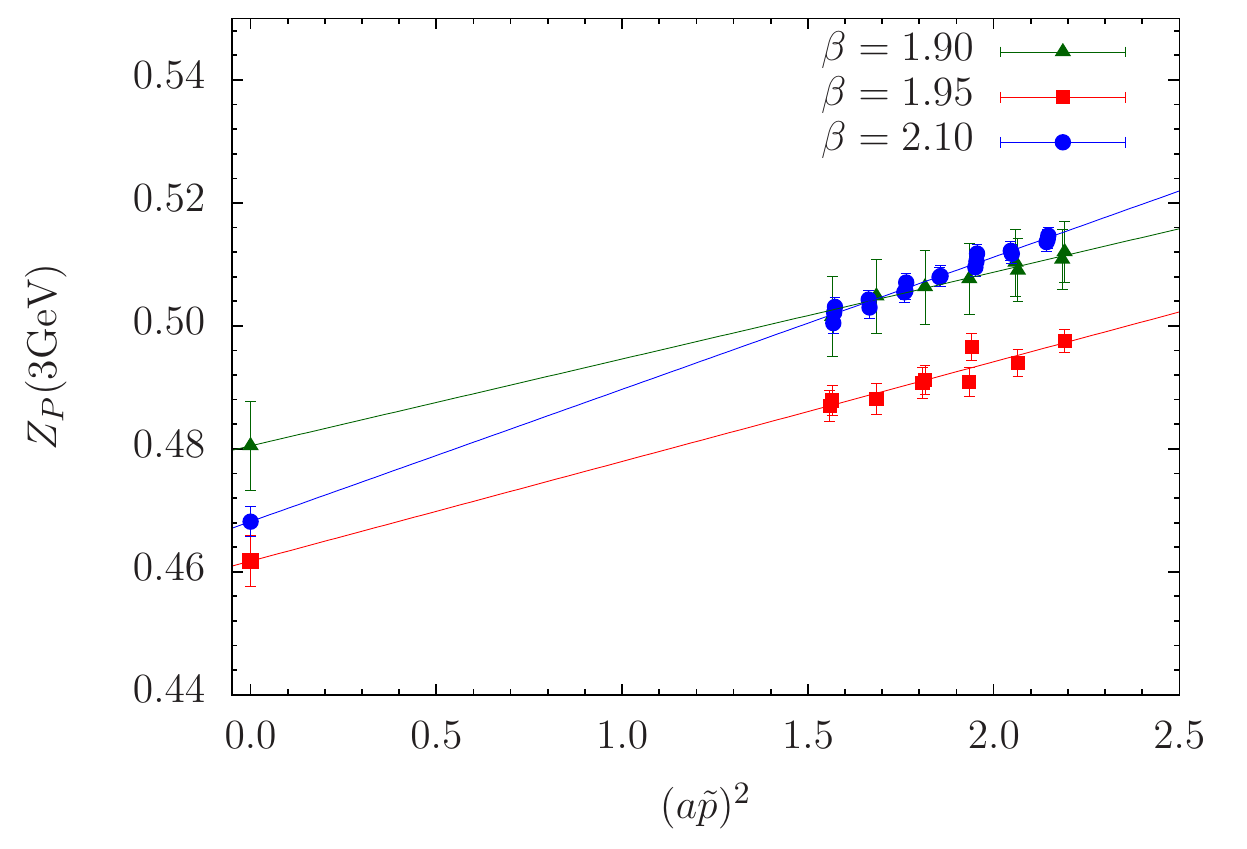}
\end{center}
\vspace*{-0.5cm}
\caption{\it Extrapolation of $Z_P$, obtained in the $\rm{RI}$-$\rm{MOM}(3 \gev)$ scheme, as a function of $(a \tilde{p})^2$ at $\beta = 1.90$ (green triangles), $\beta = 1.95$ (red squares) and $\beta = 2.10$ (blue dots).}
\label{fig:ZP_vs_ap2_allbeta}
\end{figure}

In Tables \ref{tab:final1}-\ref{tab:final3} we have given our results for the RCs derived from two different methods, M1 and M2, that differ for the way one deals with the residual $(a\tilde p)^2$ discretization effects \cite{ETM-RC-Nf2}. 
The method M1 consists in extrapolating the RCs linearly to $(a\tilde p)^2 \rightarrow 0$, after fitting the functions $Z_{q, \Gamma}(\mu = 3 \gev; (a\tilde p)^2)$ in the wide momentum interval $(a \tilde{p})^2 \in [1.5,2.2]$. 
The slopes of the fits, $\lambda_{q, \Gamma} = dZ_{q, \Gamma}(\mu = 3 \gev; (a\tilde p)^2)/d(a\tilde p)^2$ at each value of $\beta$ exhibit only a very mild dependence on the coupling constant. 
Following the discussion of Ref.~\cite{ETM-RC-Nf2} (see Section 3.2.2 and in particular the arguments leading to Eq.~(3.24) of that reference), we assume a simple linear dependence of $\lambda_{q, \Gamma}$ on $\beta$, and perform a simultaneous extrapolation of $Z_{q, \Gamma}(\mu = 3 \gev; (a\tilde p)^2)$ towards $(a\tilde p)^2 \rightarrow 0$ for the three values of $\beta$ (see for instance Fig.~\ref{fig:ZP_vs_ap2_allbeta}). 

\begin{figure}[!htb]
\begin{center}
\includegraphics[width=0.8\textwidth]{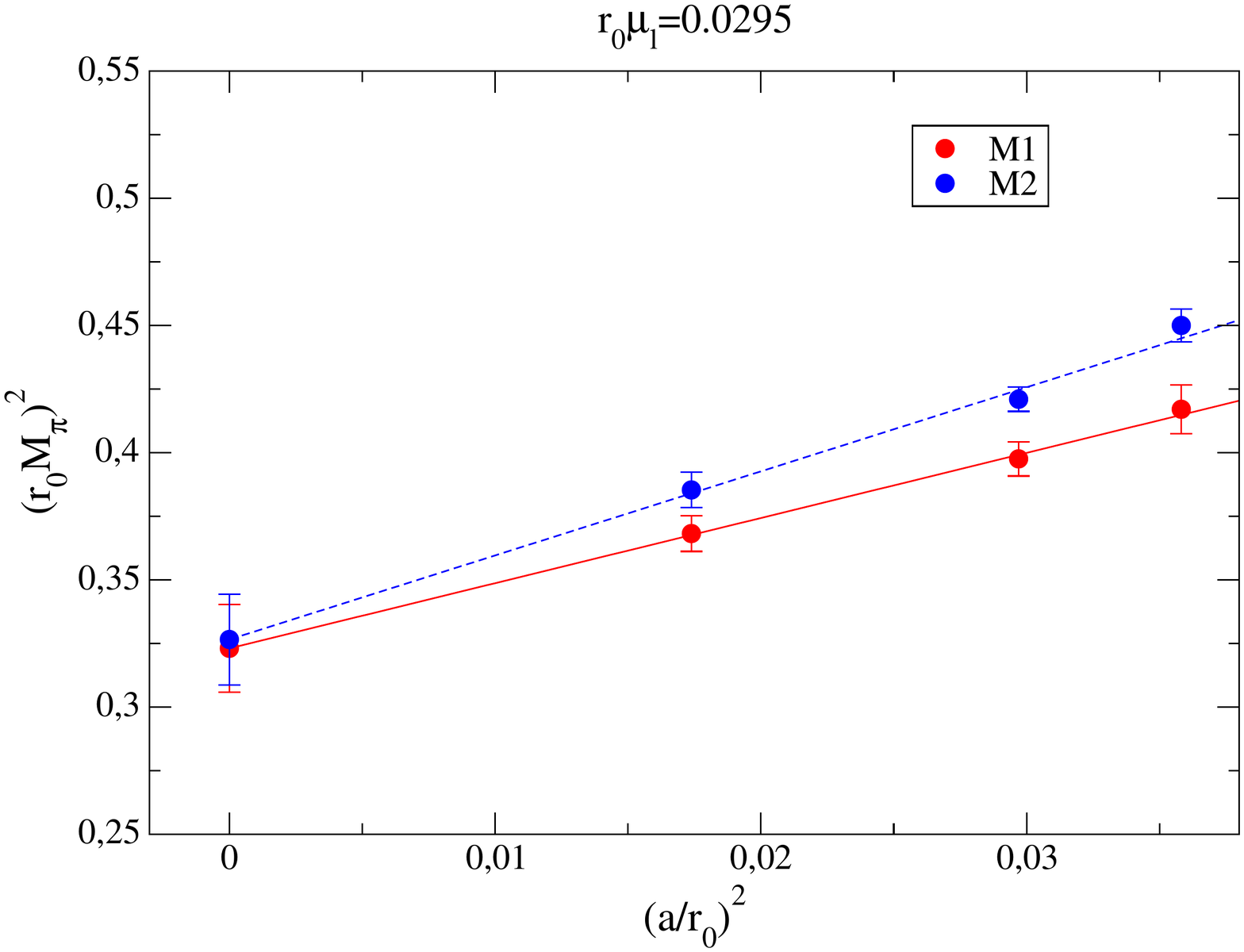}
\end{center}
\vspace*{-0.5cm}
\caption{\it Scaling of the squared pion mass computed at a fixed value of the renormalized light quark mass. The $\rm M1$ and $\rm M2$ determinations for $Z_P$ lead to compatible results in the continuum limit.}
\label{fig:ZP_M1M2}
\end{figure}

The method M2 consists in fitting the chirally extrapolated RC estimators to a  constant in the reduced momentum interval, $\tilde{p}^2 \in [11.5, 14.0]$ GeV$^2$, for all the three values of $\beta$. 
Since the momentum interval is kept constant while varying $\beta$, the ${\cal{O}}(a^2)$ artifacts occurring in the RCs of the method M2 will be removed once the continuum limit of the physical quantities of interest is taken, as shown in Fig.~\ref{fig:ZP_M1M2} in the case of the squared pion mass. 

\begin{figure}[!htb]
\begin{center}
\includegraphics[width=0.8\textwidth]{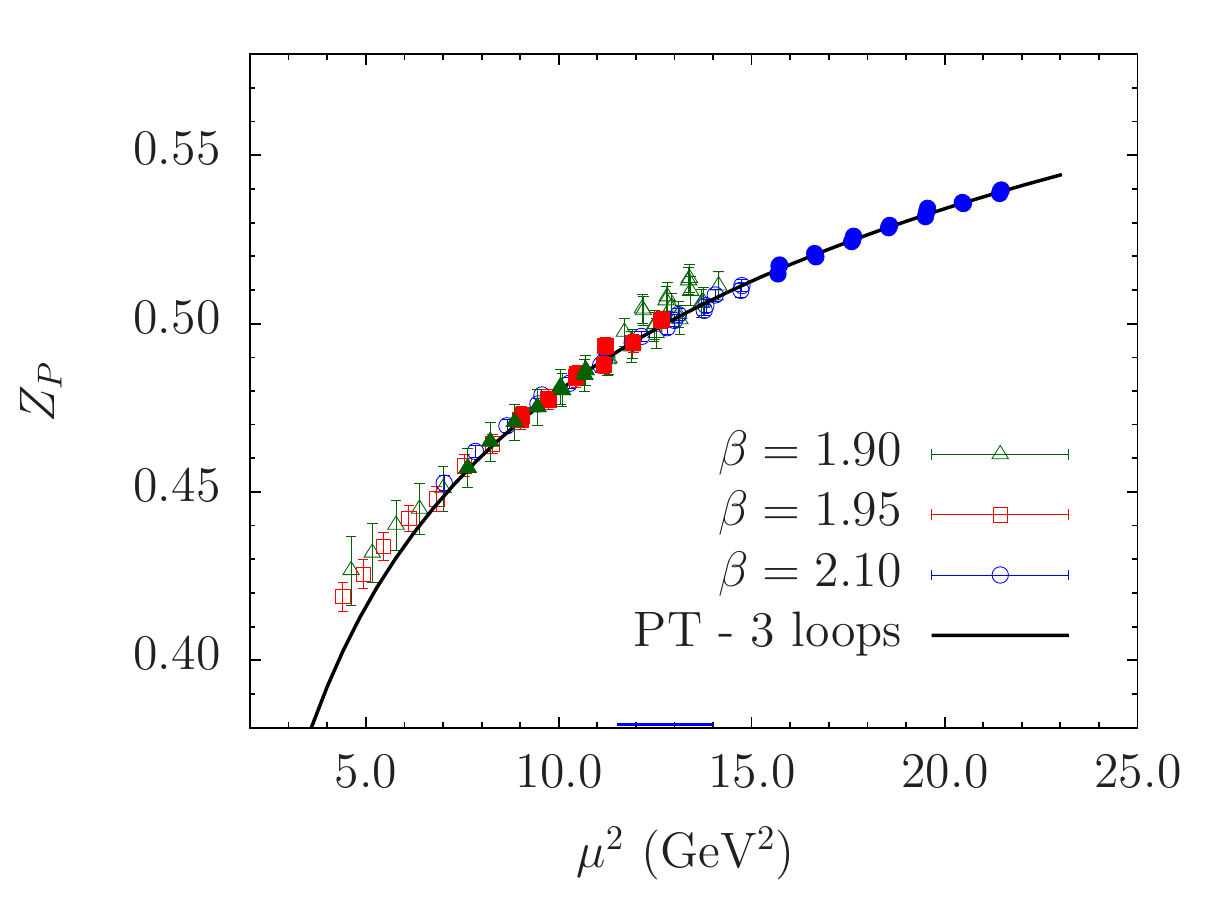}
\end{center}
\vspace*{-0.5cm}
\caption{\it Comparison of the evolution of the RC $Z_P$ determined non-perturbatively with the one obtained in perturbation theory at three loops \cite{Chetyrkin:1999pq}, as a function of the renormalization scale $\mu$. At each lattice coupling the full markers correspond to momenta $(a \tilde{p})^2$ in the range $[1.5, 2.2]$ used in the method M1, whereas the blue line on the x-axis identifies the range of the momenta $[11.5, 14.0]$ GeV$^2$ adopted in the method M2. }
\label{fig:ZP_PT}
\end{figure}

Finally, in Fig.~\ref{fig:ZP_PT} the scale evolution of the RC $Z_P$ determined non-perturbatively is compared with the one predicted by perturbation theory at three loops \cite{Chetyrkin:1999pq}, using $N_f = 4$ and $\Lambda_{QCD} = 296 \mev$ \cite{PDG}.
Notice that at each lattice coupling the full markers correspond to momenta $(a \tilde{p})^2$ in the range $[1.5, 2.2]$ used in the method M1, whereas the blue line on the x-axis identifies the range of the momenta $[11.5, 14.0]$ GeV$^2$ adopted in the method M2.
It can be seen that within the percent level of accuracy our lattice data match the perturbative evolution at three loops for scales above $\simeq 2.5 \gev$, providing also evidence that higher order perturbative contributions are not relevant for describing the renormalization scale dependence of the RC $Z_P$ in the region of momenta explored in this work.

\end{document}